\documentclass[10pt,superscriptaddress,tightenlines,twocolumn,amsmath,amssymb,aps, pra]{revtex4-2}
\UseRawInputEncoding
\usepackage{graphicx}
\usepackage{dcolumn}
\usepackage{bm}
\usepackage{color}
\usepackage{amsmath}
\usepackage{amssymb}
\usepackage{subfigure}
\usepackage{braket}
\usepackage{indentfirst}
\usepackage{mathrsfs}

\usepackage{amssymb}
\usepackage{amsmath}
\usepackage{amsfonts}
\usepackage{graphicx}
\usepackage{makecell}
\usepackage{hyperref}
\usepackage{makecell}
\usepackage{color}%
\usepackage{lmodern}
\usepackage{lineno}

\usepackage{xcolor}
\colorlet{RED}{red}
\colorlet{BLACK}{black}

\providecommand{\U}[1]{\protect\rule{.1in}{.1in}}

\hypersetup{colorlinks=true, linkcolor=red, citecolor=blue, urlcolor=black}
\newcommand{\kb}[1]{\ket{#1}\hspace{-1mm} \bra{#1}} 
\newcommand\Omicron{O}
\newcommand{\tabincell}[2]{\begin{tabular}{@{}#1@{}}#2\end{tabular}}
\begin{document}

\title{Finite-key analysis for coherent-one-way quantum key distribution}
\author{Ming-Yang Li}
\author{Xiao-Yu Cao}
\author{Yuan-Mei Xie}
\affiliation{National Laboratory of Solid State Microstructures and School of Physics, Collaborative Innovation Center of Advanced Microstructures, Nanjing University, Nanjing 210093, China}
\author{Hua-Lei Yin}\email{hlyin@ruc.edu.cn}
\affiliation{Department of Physics and Beijing Key Laboratory of Opto-electronic Functional Materials and Micro-nano Devices, Key Laboratory of Quantum State Construction and Manipulation (Ministry of Education), Renmin University of China, Beijing 100872, China}
\affiliation{National Laboratory of Solid State Microstructures and School of Physics, Collaborative Innovation Center of Advanced Microstructures, Nanjing University, Nanjing 210093, China}
\author{Zeng-Bing Chen}\email{zbchen@nju.edu.cn}
\affiliation{National Laboratory of Solid State Microstructures and School of Physics, Collaborative Innovation Center of Advanced Microstructures, Nanjing University, Nanjing 210093, China}

\date{\today}

\begin{abstract}
Coherent-one-way (COW) quantum key distribution (QKD) is a significant communication protocol that has been implemented experimentally and deployed in practical products due to its simple equipment requirements. However, existing security analyses of COW-QKD either provide a short transmission distance or lack immunity against coherent attacks in the finite-key regime. In this paper, we present a tight finite-key security analysis within the universally composable framework for a variant of COW-QKD, which has been proven to extend the secure transmission distance in the asymptotic case. We combine the quantum leftover hash lemma and entropic uncertainty relation to derive the key rate formula. When estimating statistical parameters, we use the recently proposed Kato's inequality to ensure security against coherent attacks and achieve a higher key rate. Our paper confirms the security and feasibility of COW-QKD for practical application and lays the foundation for further theoretical study and experimental implementation.
\end{abstract}
 
\maketitle

\section{\label{introduction}Introduction}

Quantum theory has been playing a significant role in the field of communications, leading to the development of primitives such as quantum repeaters~\cite{duan2001long,azuma2015all,li2023all}, quantum conference key agreement~\cite{chen_multi_2007,fu2015long,zhao_phase_2020,cao_coherent_2021,li_finite_2021,fletcher_continuous_2022,li_breakingqcka_2023}, quantum secret sharing~\cite{hillery_quantum_1999,cleve_how_1999,wei_experimental_2013,gu_differential_2021,williams2019quantum,shen_experimental_2023,de2020experimental,li_breakingqss_2023}, and quantum digital signatures~\cite{dunjko2014quantum,yin2016practical,qin2022quantum,yin2023experimental}. Among these primitives, quantum key distribution (QKD)~\cite{bennett_quantum_2014,ekert_quantum_1991} has received considerable attention due to its ability to provide two remote users with a secret key with unconditional security guaranteed by the laws of quantum mechanics. Since the first QKD protocol, the Bennett-Brassard 1984 protocol~\cite{bennett_quantum_2014} was proposed, various QKD schemes have been developed~\cite{scarani_security_2009,xu_secure_2020,pirandola_advances_2020} to improve its practicality. Among these developments, measurement-device-independent QKD~\cite{lo_measurement-device-independent_2012,braunstein_side-channel-free_2012} is of vital importance for its immunity against one of the most threatening attacks, detector attacks~\cite{lydersen_hacking_2010}, which enable experimental operations over a long distance~\cite{yin_measurement-device-independent_2016,zhou_making_2016}. However,  due to channel loss, the key rates of most QKD protocols are bounded by the secret-key capacity of repeaterless QKD~\cite{pirandola_direct_2009,takeoka_fundamental_2014,pirandola_fundamental_2017,das_unversal_2021}. A protocol called the twin-field QKD~\cite{lucamarini_overcoming_2018} and its variants~\cite{ma_phase-matching_2018,wang_twin-field_2018,yin_measurement-device-independent_2019,lin_simple_2018,cui_twin-field_2019,curty_simple_2019}, which are based on single-photon interference instead of two-photon interference, break this bound and increase the secure distance to 833 km~\cite{wang_twin-field_2022} and 1002 km~\cite{liu2023experimental} experimentally. Moreover, the recently proposed asynchronous measurement-device-independent QKD~\cite{xie_breaking_2022,zeng2022mode} (also named mode-pairing QKD) has become a practical approach for long-distance quantum communication systems~\cite{PhysRevLett.130.030801,zhou_experimental_2023,bai_asynchronous_2023,xie_advantages_2023} because it breaks the linear bound with its simple experimental implementation compared with twin-field QKD. The photon number splitting attack~\cite{brassard_limitations_2000} is another critical limitation of practical QKD that has been overcome by several means like decoy-state methods~\cite{hwang_quantum_2003,wang_beating_2005,lo_decoy_2005}, nonorthogonal coding methods~\cite{scarani_quantum_2004,tamaki_unconditionally_2006,yin_security_2016}, strong reference methods~\cite{koashi_unconditional_2004}, and distributed-phase-reference methods~\cite{inoue_differential_2002,inoue_differential-phase-shift_2003,stucki_fast_2005,sasaki_practical_2014}, including differential-phase-shift (DPS) QKD and coherent-one-way (COW) QKD.

DPS protocol~\cite{inoue_differential_2002,inoue_differential-phase-shift_2003} is becoming more significant for its excellent key rate performance achieved by the simple setup of equipment. The experimental progress~\cite{Takesue_differential_2005,Diamanti_100km_2006,takesue_quantum_2007,Sasaki_field_2011} shows the status of DPS-QKD as a promising protocol for realizing the quantum communication process in the real world. Theoretically, long-term security analyses of DPS-QKD have been proposed to establish a solid foundation to guarantee its unconditional security in reality. Assuming a single photon to be in each of the blocks, the analysis in Ref.~\cite{wen_unconditional_2009} provided security proof of DPS-QKD, and this impractical assumption was changed to use a blockwise phase-randomized coherent photon source in later developments~\cite{tamaki_2012_unconditional,Mizutani_information_2018}. Furthermore, recently proposed proofs~\cite{mizutani_quantum_2019,Mizutani_quantum_2020,Endo_line_2022} give more practical analyses, removing the requirement of a special photon source and covering more general cases. Finally, Refs.~\cite{mizutani_finite_2023,sandfuchs_2023_security} provide information-theoretic secure analyses to show the practicability of DPS-QKD in the finite-key regime, which builds a complete theoretic scheme for this protocol. We note that the security proof in Ref.~\cite{mizutani_finite_2023} results in a key rate that scales in the order of $\Omicron(\eta^{2})$ without relativistic constraint and has immunity against coherent attacks, which is of vital importance in realistic implementation.

COW-QKD~\cite{stucki_fast_2005} is another type of distributed-phase-reference protocol that has been implemented in practical quantum information processing~\cite{Peev_2009} with its easily achievable experimental requirements~\cite{stucki_continuous_2009,stucki_high_2009,walenta_fast_2014,korzh_provably_2015,sibson_chip-based_2017,sibson_integrated_2017,roberts_modulator-free_2017,dai_pass-block_2020}, which are similar to those of DPS-QKD. Contrary to the DPS-QKD, the security proof for the COW protocol remains incomplete, primarily due to the absence of a finite-key secure analysis that simultaneously offers robust key rate performance and security against coherent attacks. Typically, the security of COW-QKD is proven by measuring the interference visibility to estimate information leakage~\cite{branciard_upper_2008,moroder_security_2012}. However, when considering the zero-error attack~\cite{gonzalez-payo_upper_2020,trenyi_zero-error_2021}, which enables eavesdropping by Eve without introducing bit errors, COW-QKD is insecure if its key rate scales as $O(\eta)$~\cite{wang_characterising_2019}, which is the scale of the key rate used in many COW-QKD experiments~\cite{walenta_fast_2014,sibson_chip-based_2017,sibson_integrated_2017,roberts_modulator-free_2017,dai_pass-block_2020,de_marco_real-time_2021}. In Ref.~\cite{Lavie-improved-2022}, the authors introduced an innovative method for calculating the key rate of a variant of COW-QKD, resulting in an improved key rate in high-loss channels. However, the security of this protocol in the finite-key regime, particularly its immunity to coherent attacks, has yet to be proven. This is a crucial step in ensuring its practicality in real-world environments. In summary, the lack of a finite-key analysis for COW-QKD that offers both a high key rate and security against coherent attacks remains a significant challenge in enhancing the practicality of this technology. Recently, a security proof for COW-QKD was proposed~\cite{gao_simple_2022} based on an innovative practical implementation that retains the simplicity of the original version. By estimating the upper bound on the phase error rate instead of measuring the visibility of interference, it was shown that the secure 
transmission distance can be over 100 km, and an analytic formula for the key rate was provided. \textcolor{black}{Nevertheless, a practical QKD protocol only involves finite resources, which means only a finite number of states are sent, leading to statistical fluctuations between observed values and expected values. Consequently, before we promote this protocol into reality, finite key analysis must be completed to lay the theoretical foundation. The uncertainty relation of smooth entropies~\cite{tomamichel_uncertainty_2011} has been utilized to prove the finite-key security of the BB84 protocol~\cite{tomamichel_tight_2012} with composable security~\cite{Müller-Quade_2009} against general attacks. This entropic uncertainty relation framework has been further extended to other finite-key cases of QKD, even with imperfect light sources~\cite{wang_tight_2016,wang_finite_2019}. This demonstrates its robust capability to underpin finite-key security analysis for various protocols.}

In this paper, we extend the security proof in Ref.~\cite{gao_simple_2022} to the finite-key domain with composable security~\cite{Müller-Quade_2009} to demonstrate its real-world applicability. We employ the quantum leftover hash lemma~\cite{renner_security_2008} and the entropic uncertainty relation~\cite{tomamichel_uncertainty_2011,tomamichel_tight_2012} to derive a formula for the secure key length in the finite-key regime. When dealing with correlated random variables, we apply Kato's inequality~\cite{kato_concentration_2020} to estimate statistical fluctuations, ensuring security against coherent attacks and resulting in a higher key rate compared to Azuma's inequality~\cite{azuma_weighted_1967}. We simulate the performance of the key rate under different conditions, such as varying values of misalignment error and different choices of basis, to demonstrate the flexibility of our protocol. The simulation results and comparison with existing analyses and another similar protocol confirm the advantages of our approach. \textcolor{black}{Additionally, our protocol is employed to show the exceptional capability of Kato's inequality in providing significantly tighter bounds when addressing events with an extremely low probability of occurrence. The comparison of key rates with previous COW variants and a summary of differences in many aspects are presented as well, serving to clearly highlight the unique advantages of our protocol.}

This paper is organized as follows: Section~\ref{protocol} \textcolor{black}{introduces the assumptions on devices and our COW-QKD protocol scheme.} Section~\ref{formula} presents details of the key rate calculation. Numerical simulations of key rate performance under different conditions \textcolor{black}{and some comparisons} are shown in Sec.~\ref{simulation}, and we conclude in Sec.~\ref{conclusion}.

\section{\label{protocol}\textcolor{black}{Assumptions and protocol descriptions}}

\subsection{\label{assumutions}\textcolor{black}{Assumptions on devices}}

\textcolor{black}{For the completeness of this paper, the assumptions on the devices of sender Alice and receiver Bob are introduced here before we describe our variant of the COW-QKD. In this protocol, Alice encodes a random bit with a quantum state consisting of two pulses sent in adjacent time windows and extracts a string of secret bits from these states together with Bob.}

\begin{figure}[t]
    \centering
    \includegraphics[width=8.5cm]{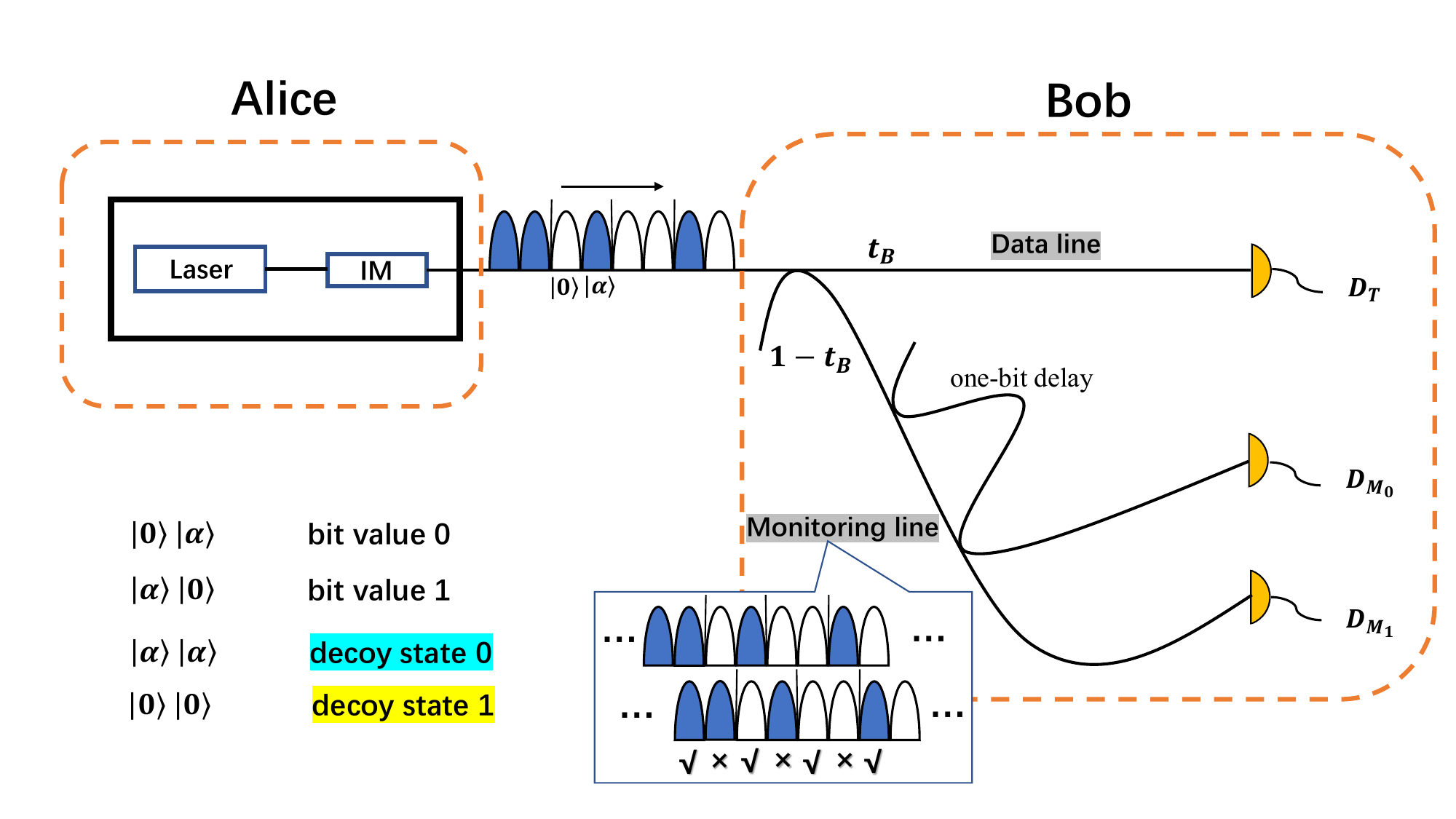}
    \caption{Experimental implementation of COW-QKD protocol in this paper. With an intensity modulator (IM), Alice can prepare quantum states $\ket{0}$ and $\ket{\alpha}$ in each time window experimentally to randomly sends a sequence of pulses that consists of states $\ket{0}_{2k-1}\ket{\alpha}_{2k}$, $\ket{\alpha}_{2k-1}\ket{0}_{2k}$, $\ket{\alpha}_{2k-1}\ket{\alpha}_{2k}$, and $\ket{0}_{2k-1}\ket{0}_{2k}$ to Bob. After passively distributing these states into the data line or the monitoring line with a beam splitter of transmittance $t_B$, Bob records the detector's click in each round. $D_T$, $D_{M_0}$, and $D_{M_1}$ are single-photon detectors. Compared to the original version of COW QKD, our protocol adds state $\ket{0}_{2k-1}\ket{0}_{2k}$ as another decoy state, which maintains the requirements for experimental equipment. \textcolor{black}{We note that on the monitoring line, Bob is only required to record the clicks that occur within specific time windows. These clicks are a result of interference involving two pulses from the same round, as illustrated in the provided figure.}}
    \label{scheme figure}
\end{figure}

\subsubsection{\label{assumptions_Alice}\textcolor{black}{Assumptions on Alice's devices}}
\textcolor{black}{The assumptions on Alice's devices are presented below:}
\begin{enumerate}
    \item 
    \textcolor{black}{Alice employs her sending equipment, which includes a mode-locked continuous-wave laser and an intensity modulator, to create weak coherent pulses. Alternatively, she can completely block the output to generate a vacuum state;}
    \item 
    \textcolor{black}{In our variant of COW-QKD, Alice randomly selects her initial bit string and encodes each bit into a two-pulse state. Additionally, she randomly determines which two adjacent time windows will be used to transmit decoy states. Consequently, the probability of the emitted state being a weak coherent pulse or a vacuum state in each time window is independent of the states sent previously.}
\end{enumerate}

\subsubsection{\label{assumptions_Bob}\textcolor{black}{Assumptions on Bob's devices}}
\textcolor{black}{The assumptions on Bob's devices are summarised as follows:}
\begin{enumerate}
    \item 
    \textcolor{black}{Bob's detection devices receive optical pulses transmitted through a quantum channel with a transmittance of $\eta$. These incoming states are then divided into a data line or a monitoring line using a beam splitter with a transmittance of $t_B$. Alternatively, an optical switch can be used in place of the beam splitter, allowing Bob to actively distribute the quantum states;}
    \item
    \textcolor{black}{On the data line, the quantum states are directly transmitted to a single-photon detector which measures the arrival time of the pulses. On the monitoring line, Bob utilizes an asymmetric Mach-Zehnder interferometer, which includes a one-bit delay and two single-photon detectors, to record which detector clicks within certain time windows. We note that all the single-photon detectors are threshold detectors, designed to simply determine the presence or absence of a photon.}
    \item
    \textcolor{black}{We assume the detection efficiency $\eta_d$ and dark-count rate $p_d$ of each detector to be the same and reasonable values of them are employed in Sec.~\ref{simulation} to numerically present the performance of our protocol.}
\end{enumerate}

\subsection{Detailed steps}
In COW-QKD protocol, sender Alice uses two-pulse states $\ket{0_k}=\ket{0}_{2k-1}\ket{\alpha}_{2k}$ and $\ket{1_k}=\ket{\alpha}_{2k-1}\ket{0}_{2k}$ at two time windows $2k-1$ and $2k$ ($k=1,2,\cdots,N$) to encode logic bits 0 and 1 in the $k$-th round respectively. \textcolor{black}{Here we use $\ket{0}$ to denote the vacuum state and $\ket{\alpha}$ to denote the coherent state whose mean photon number is $\mu=|\alpha|^2$.} As shown in Fig.~\ref{scheme figure}, the COW-QKD scheme used in this paper takes both two-pulse coherent state $\ket{\alpha}_{2k-1}\ket{\alpha}_{2k}$ and two-pulse vacuum state $\ket{0}_{2k-1}\ket{0}_{2k}$ as decoy states to estimate the phase error rate instead of using visibility to reflect the broken coherence.

\par The detailed steps of this scheme are:
\begin{enumerate}
    \item 
    Alice randomly sends a sequence of pulses that consists of states $\ket{0}_{2k-1}\ket{\alpha}_{2k}$, $\ket{\alpha}_{2k-1}\ket{0}_{2k}$, $\ket{\alpha}_{2k-1}\ket{\alpha}_{2k}$, and $\ket{0}_{2k-1}\ket{0}_{2k}$ with probability $p_z$, $p_z$, $p_{d_1}$, and $p_{d_2}$, respectively, to Bob where $p_z=\frac{1}{2}(1-p_{d_1}-p_{d_2})$. She records her choice of sending in each round. This step is repeated for $N$ rounds so we have $k=1,2,\cdots,N$.
    \item 
    Bob uses a beam splitter of transmittance $t_B$ to passively distribute incoming states into the data line or the monitoring line. On the data line, he measures the click time of each signal to determine which logic bit Alice encodes in this round and gets the raw key. On the monitoring line, he records which detector clicks in each round. \textcolor{black}{As illustrated in Fig.~\ref{scheme figure}, the clicks that are recorded are specifically those resulting from interference involving two pulses from the same round. Any other clicks should be disregarded.} Here we note that if multiple detectors click in one round, Bob records one of these detector \textcolor{black}{clicks} randomly.
    \item 
    Bob announces in which round he records a click on the data line. Alice only keeps her logic bits in those rounds and discards the rest to get the raw key.
    \item 
    Bob announces his click records of the monitoring line. Alice calculates the following click counts: $n_{0\alpha}^{M_i}$, $n_{\alpha 0}^{M_i}$, $n_{\alpha \alpha}^{M_i}$, and $n_{00}^{M_i}$ ($i=0$ or $1$). $n_{u}^{M_i}(u=0\alpha,\alpha 0,\alpha\alpha,00)$ are the click counts of states  $\ket{0}_{2k-1}\ket{\alpha}_{2k}$, $\ket{\alpha}_{2k-1}\ket{0}_{2k}$, $\ket{\alpha}_{2k-1}\ket{\alpha}_{2k}$, and $\ket{0}_{2k-1}\ket{0}_{2k}$, respectively, where the superscript $M_i$ refers to the clicking detectors on the monitoring line. By applying Kato's inequality, she can estimate the upper bound on phase error rate $\overline{E_p}$. The bit error rate $E_z$ can be calculated by revealing some bits from the raw key. If either $\overline{E_p}$ or $E_z$ exceeds the preset values, the protocol aborts.
    \item 
    After an error correction step is performed, at most $\rm{leak}_{\rm{EC}}$ bits of information are revealed. Then Alice and Bob verify whether the error correction step succeeds and perform privacy amplification to get the final key string.
\end{enumerate}

\section{\label{formula}The Key-length Formula}

\subsection{Security definition}
Before we present the security proof in the finite-key regime, we introduce the universally composable framework of QKD~\cite{Müller-Quade_2009}. Typically, performing a QKD protocol either generates a pair of bit strings $\hat{S}_A$ and $\hat{S}_B$ for Alice and Bob, respectively, or aborts so $\hat{S}_A=\hat{S}_B=\varnothing$. A secure QKD protocol must satisfy the two criteria below.
\par The first is the correctness criterion which is met if two bit strings are the same, i.e., $\hat{S}_A=\hat{S}_B$. In practical experiments, however, as it is not always possible to perfectly satisfy the correctness criterion, a small degree of error is typically allowed. Instead, we require that the probability of the two bit strings not being identical does not exceed a predetermined value, denoted as $\varepsilon_{\rm{cor}}$, In this case, we say that the protocol is $\varepsilon_{\rm{cor}}$-correct.
\par The second is the secrecy criterion which is met if there is no correlation between the system of the eavesdropper Eve and the bit strings of Alice. We assume the orthonormal basis which consists of Alice's quantum system and corresponds to each possible bit string of Alice to be $\{\ket{s}\}_s$. The secrecy criterion requires the joint quantum state of Alice and Eve to be $\rho_{\rm{AE}}= \rho_{\rm{AE}}^{\rm{ideal}} \equiv U_{\rm{A}} \otimes \rho_{\rm{E}}$, where $U_{\rm{A}}=\sum_s \frac{1}{|\mathcal{S}|}|s\rangle\langle s|$ is a uniform mixture which indicates that the probability of generating each possible bit string of Alice is uniformly distributed, and $\rho_{\rm{E}}$ is Eve's system, which does not correlate with Alice's system. However, it is not always possible to perfectly satisfy this criterion in practice. This means that a small deviation between the actual joint quantum state of Alice and Eve and the ideal state is permissible. The trace distance measures the difference and we say the protocol is $\varepsilon_{\rm{sec}}$-secret if the trace distance between the actual joint quantum state $\rho_{\rm{AE}}$ and the ideal state $\rho_{\rm{AE}}^{\rm{ideal}}$ does not exceed $\Delta$, i.e., 
\begin{equation}
    \frac{1}{2}\Vert \rho_{\rm{AE}}-\rho_{\rm{AE}}^{\rm{ideal}} \Vert_1 \le \Delta,
\end{equation} 
and$(1-p_{\rm{abort}})\Delta \le \varepsilon_{\rm{sec}}$, where $p_{\rm{abort}}$ is the probability for aborting this protocol and $\Vert \cdot \Vert_1$ denotes the trace norm.
\par Finally, a protocol is $\varepsilon_{\rm{s}}$-secure if it is both $\varepsilon_{\rm{cor}}$-correct and $\varepsilon_{\rm{sec}}$-secret with $\varepsilon_{\rm{cor}}+\varepsilon_{\rm{sec}} \le \varepsilon_{\rm{s}}$.

\subsection{\label{security}Security proof}
\par Here, a virtual entanglement-based protocol is introduced to obtain the secure key rate in the finite-key regime, which is based on the virtual entanglement-based protocol in Ref.~~\cite{gao_simple_2022}. To simplify the presentation, we ignore the label $k$ and express the state sent in the $k$-th round as $\ket{0_z}$ and $\ket{1_z}$. Let $\ket{0_x}=(\ket{0_z}+\ket{1_z})/\sqrt{N^+}$ and $\ket{1_x}=(\ket{0_z}-\ket{1_z})/\sqrt{N^-}$ be the logic bits 0 and 1 in the X basis, where $N^{\pm}=2(1\pm e^{-\mu})$ are the normalization factors. In the virtual entanglement-based protocol, Alice prepares $K$ pairs of the entangled state
\begin{equation}
    \begin{aligned}
        \ket{\phi} &= \frac{1}{\sqrt{2}}(\ket{+z}_A\ket{0_z}_{A^{\prime}}+\ket{-z}_A\ket{1_z}_{A^{\prime}})\\
        &=\frac{\sqrt{N^+}}{2}\ket{+x}_A\ket{0_x}_{A^{\prime}}+\frac{\sqrt{N^-}}{2}\ket{-x}_A\ket{1_x}_{A^{\prime}},
    \end{aligned}
\end{equation}
where $\ket{\pm x}$ and $\ket{\pm z}$ are the eigenstates of the Pauli matrices $X$ and $Z$, respectively, and subscripts $A$ and $A^{\prime}$ denote different quantum systems possessed by Alice. Then Alice measures the qubits in the system $A$ randomly in the Pauli $X$ or $Z$ basis to obtain the raw key $\hat{X}_A$ from the $X$ basis and $\hat{Z}_A$ from the $Z$ basis. Bob's experimental implementation is the same as the practical COW-QKD. He obtains his raw key $\hat{Z}_B$ of the $Z$ basis on the data line by measuring the click time just like the original protocol. He also records a bit value 0(1) in the $X$ basis when detector $D_{M_0}(D_{M_1})$ on the monitoring line clicks to obtain the raw key $\hat{X}_B$. $\hat{Z}_A$ and $\hat{Z}_B$ are used to extract the final key so the error-correction step and error-verification step are performed to them. If these steps succeed, Alice and Bob obtain the same bit string which we denote as $\hat{Z}$. All the information that the eavesdropper Eve possesses up to the error-correction step and error-verification step is denoted as $E^{\prime}$. The smooth min-entropy $H_{\rm{min}}^{\varepsilon}(\hat{Z}|E^{\prime})$ characterizes the mean probability that Eve can guess $\hat{Z}$ successfully with all information she owns using the optimal strategy~\cite{konig_operational_2009}. The smooth max-entropy $H_{\rm{max}}^{\varepsilon}(\hat{Z}_A|\hat{Z}_B)$ quantifies the number of bits required to reconstruct $\hat{Z}_A$ from $\hat{Z}_B$~\cite{renes_one-shot_2012}. 
\par According to the quantum leftover hashing lemma~\cite{renner_security_2008}, a $\Delta$-secret key of length $l$ can be extracted from $\hat{Z}$ when a random $\rm{universal}_2$ hash function to $\hat{Z}$, where parameter $\Delta$ satisfies 
\begin{equation}
    \Delta = 2\varepsilon+\frac{1}{2}\sqrt{2^{l-H_{\rm{min}}^{\varepsilon}(\hat{Z}|E^{\prime})}}.
\end{equation}
Letting $\varepsilon_0=\frac{1}{2}\sqrt{2^{l-H_{\rm{min}}^{\varepsilon}(\hat{Z}|E^{\prime})}}$, the length $l$ of the secret key is~\cite{yin_finite-key_2019}
\begin{equation}
    l=H_{\rm{min}}^{\varepsilon}(\hat{Z}|E^{\prime})-2\log_2\left(\frac{1}{2\varepsilon_0}\right).
\end{equation}
A chain-rule inequality for these smooth entropies is used to describe the error-correction step and error-verification step. That is,
\begin{equation}
    \begin{aligned}
        H_{\rm{min}}^{\varepsilon}(\hat{Z}|E^{\prime}) &\ge H_{\rm{min}}^{\varepsilon}(\hat{Z}_A|E) - H_{\rm{max}}^{\varepsilon}(\hat{Z}_A|\hat{Z}_B)\\
        &=H_{\rm{min}}^{\varepsilon}(\hat{Z}_A|E)-\rm{leak}_{\rm{EC}}-\log_2\left(\frac{2}{\varepsilon_{\rm{cor}}}\right),
    \end{aligned}
\end{equation}
where $\rm{leak}_{\rm{EC}}$ and $\log_2(\frac{2}{\varepsilon_{\rm{cor}}})$ are the numbers of bits that are revealed during the error-correction and error-verification procedure, respectively, to generate a $\varepsilon_{\rm{cor}}$-correct key, and $\hat{E}$ denotes all the information that Eve possesses before the error-correction step and error-verification step. The lower bound on the smooth min-entropy can be obtained by the entropic uncertainty relation~\cite{tomamichel_uncertainty_2011}. We denote the binary Shannon entropy as $h(x)=-x\log_2x-(1-x)\log_2(1-x)$. Let $\hat{X}_A^{\prime}$ and $\hat{X}_B^{\prime}$ be the bit strings that Alice and Bob would have obtained if Alice had measured in the $X$ basis, which is actually measured in the $Z$ basis in the virtual protocol. So, we have $H_{\rm{max}}^{\varepsilon}(\hat{X}_A^{\prime}|\hat{X}_B^{\prime}) \le n_z h(E_x)$, where $n_z$ is the size of $\hat{Z}_A$ and $E_x$ is the bit error rate in the $X$ basis. By exploiting the the entropic uncertainty relation, we have
\begin{equation}
    H_{\rm{min}}^{\varepsilon}(\hat{Z}_A|E) \ge n_z-H_{\rm{max}}^{\varepsilon}(\hat{X}_A^{\prime}|\hat{X}_B^{\prime}) \ge n_z[1-h(E_x)],
\end{equation}
and the final key length is
\begin{equation}
    l \ge  n_z[1-h(E_x)]-\rm{leak}_{\rm{EC}}-\log_2\left(\frac{2}{\varepsilon_{\rm{cor}}}\right)-2\log_2\left(\frac{1}{2\varepsilon_0}\right).
\end{equation}

\subsection{Phase error rate}
The phase error rate formula is derived by the same method in Ref.~\cite{gao_simple_2022}. For the completeness of this paper, a brief deduction is presented here. We consider a prepare-and-measure protocol which is equivalent to the entanglement-based protocol in Sec.~\ref{security}. In this protocol, when Alice prepares her optical signals, she randomly chooses the Z or X basis. If she chooses the Z basis, she prepares states $\ket{0_z}$ and $\ket{1_z}$ with the same probability. If she chooses the X basis, she prepares states $\ket{0_x}$ and $\ket{1_x}$ with probability $\frac{N^+}{4}$ and $\frac{N^-}{4}$, respectively. She sends her states to Bob, and Bob uses the same implementation as the practical protocol to measure these states in the $Z$ basis (data line) or in the $X$ basis (monitoring line) distributed by a beam splitter.
\par It is obvious that the density matrices of the X and Z basis are the same. That is,
\begin{equation}
    \begin{aligned}
        \rho &= (\kb{0_z}+\kb{1_z})/2\\
        &=(N^+\kb{0_x}+N^-\kb{1_x})/4.
    \end{aligned}
    \label{density matrices}
\end{equation} 
Therefore, the bit error rate of the X basis can be obtained as follows:
\begin{equation}
    \begin{aligned}
        E_x&=\frac{N^+Q_{0_x}^{M_1}+N^-Q_{1_x}^{M_0}}{N^+(Q_{0_x}^{M_0}+Q_{0_x}^{M_1})+N^-(Q_{1_x}^{M_0}+Q_{1_x}^{M_1})}\\
        &=\frac{N^+(Q_{0_x}^{M_1}-Q_{0_x}^{M_0})+2(Q_{0_z}^{M_0}+Q_{1_z}^{M_0})}{2(Q_{0_z}^{M_0}+Q_{0_z}^{M_1}+Q_{1_z}^{M_0}+Q_{1_z}^{M_1})},
    \end{aligned}
    \label{BitErrorOfX}
\end{equation}
which is equal to the bit error rate in the virtual entanglement-based protocol, where $Q_{s_{x(z)}}^{M_i}$ refers to the gain of the event which Alice prepares state $\ket{s_{x(z)}} (s=0,1) $ and Bob get a click with detector $D_{M_i} (i=0,1) $ on the monitoring line. The relation $\frac{N^-}{4}Q_{1x}^{M_i}+\frac{N^+}{4}Q_{0x}^{M_i}=\frac{1}{2}(Q_{1z}^{M_i}+Q_{0z}^{M_i})$ is used in the second equation, which can be obtained from Eq.~(\ref{density matrices}). Because the density matrices of the Z basis and X basis are the same, the eavesdropper Eve cannot distinguish whether the prepare-and-measure protocol or the practical COW-QKD protocol is actually performed by Alice and Bob. The phase error rate in the practical COW-QKD protocol is equal to the bit error rate of the X basis in the prepare-and-measure protocol.
\par In practical  COW-QKD protocol, states $\ket{0_x}$ and $\ket{1_x}$ are not sent, so we can not calculate $Q_{0_x}^{M_1}$ and $Q_{0_x}^{M_0}$ directly. The decoy states $\ket{\alpha}_{2k-1}\ket{\alpha}_{2k}$ and $\ket{0}_{2k-1}\ket{0}_{2k}$ are used to estimate $\overline{Q_{0_x}^{M_1}}$ and $\underline{Q_{0_x}^{M_0}}$, where $\overline{O}$ and $\underline{O}$ are the upper and lower bounds on value $O$, respectively. The expressions are

\begin{equation}
\begin{aligned}
    \overline{Q_{0x}^{M_1}} &= \frac{1}{N^+}\left(e^{\frac{\mu}{2}}\sqrt{Q_{\alpha \alpha}^{M_1}}+e^{-\frac{\mu}{2}}\sqrt{Q_{00}^{M_1}}\right)^2\\
    &+ \frac{N^-}{N^+}\left( \frac{e^{\mu}N^-}{4}+e^{\mu} \sqrt{Q_{\alpha \alpha}^{M_1}}+\sqrt{Q_{00}^{M_1}}\right)
\end{aligned}
    \label{gain M1}
\end{equation}
and
\begin{equation}
\begin{aligned}
    \underline{Q_{0x}^{M_0}} &= \frac{1}{N^+}\left(e^{\mu}Q_{\alpha \alpha}^{M_0}+e^{-\mu}Q_{00}^{M_0}-2\sqrt{Q_{00}^{M_0}Q_{\alpha \alpha}^{M_0}}\right)\\
    &-\frac{N^-}{N^+}\left( e^{\mu} \sqrt{Q_{\alpha \alpha}^{M_0}}+\sqrt{Q_{00}^{M_0}}\right),
\end{aligned}
    \label{gain M0}
\end{equation}
where $Q_w^{M_i} (w=\alpha\alpha,00)$ denotes the gain of the event which Alice sends state $\ket{\alpha}\ket{\alpha}$ or $\ket{0}\ket{0}$, respectively, where we also omit the subscripts of states $\ket{\alpha}_{2k-1}\ket{\alpha}_{2k}$ and $\ket{0}_{2k-1}\ket{0}_{2k}$ and Bob gets a click with detector $D_{M_i} (i=0,1)$. The details of how to obtain Eqs.~(\ref{gain M1}) and~(\ref{gain M0}) can be found in Ref.~\cite{gao_simple_2022}.

\subsection{Statistical fluctuations}
\textcolor{black}{Given the impact of finite-key effects, it is crucial to ensure the security of our protocol within the finite-key regime if we want to facilitate the practical application of this technology. This involves taking into account the statistical fluctuations between observed and expected values and estimating the lower bound of the final key length using a concentration inequality~\cite{azuma_weighted_1967,kato_concentration_2020}.}
\par Typically, Azuma's inequality~\cite{azuma_weighted_1967} is applied to convert observed values to the upper or lower bound on corresponding expected values and vice versa. As shown in Ref.~\cite{mizutani_finite_2023}, it can be concluded that a loose bound will be obtained when using Azuma's inequality to estimate the statistical fluctuations of events that occur with a very small probability. Specifically, the estimation of the gains of decoy states \textcolor{black}{$\ket{0}_{2k-1}\ket{0}_{2k}$} is loose when using Azuma's inequality. Instead, we use a concentration inequality named Kato's inequality~\cite{kato_concentration_2020} to make our estimation tighter so a higher key rate can be obtained. Here we introduce the general form of Kato's inequality which has been employed in some finite key analyses~\cite{curras-lorenzo_tight_2021,mizutani_finite_2023}. Let $n_1,n_2,\cdots,n_k$ be a sequence of random variables which satisfies $0\le n_i \le 1,(i=1,2,\cdots,k)$. Let $\Gamma_i=\sum_{u=1}^i n_u$ and $f_i$ be the $\sigma$-algebra generated by $\{ n_1,n_2,\cdots,n_k \}$, which is called the natural filtration of this sequence of random variables. For any $k,a \in \mathbb{R}$ and any $b$ s.t. $b \ge |a|$, according to Kato's inequality we have that
\begin{equation}
  \begin{aligned}
    Pr&\left[\sum_{u=1}^kE(n_u|f_{u-1})-\Gamma_k \ge \left[ b+a\left( \frac{2\Gamma_k}{k}-1 \right) \right]\sqrt{k} \right] \\
    &\le \exp \left[ \frac{-2(b^2-a^2)}{(1+\frac{4a}{3\sqrt{k}})^2} \right] ,
  \end{aligned}
    \label{kato1}
\end{equation}
where $E(\cdot)$ refers to the expected value. Another form of Kato's inequality can be derived by replacing $n_i \rightarrow 1-n_i$ and $a \rightarrow -a$, which is
\begin{equation}
  \begin{aligned}
    Pr&\left[\Gamma_k-\sum_{u=1}^kE(n_u|f_{u-1}) \ge \left[ b+a\left( \frac{2\Gamma_k}{k}-1 \right) \right]\sqrt{k} \right] \\
    &\le \exp \left[ \frac{-2(b^2-a^2)}{(1-\frac{4a}{3\sqrt{k}})^2} \right] .
  \end{aligned} 
    \label{kato2}
\end{equation} The details of how to use Kato's inequality to accomplish parameter estimation tasks are shown in the Appendix~\ref{appendix}. In the description below, we let $O^*$ be the expected value of $O$. 
\par After performing the COW-QKD protocol, observed values $n_{\alpha \alpha}^{M_i}$ and $n_{00}^{M_i}$ ($i=0,1$) are obtained, which stand for the total click counts of detector $M_i$ when state $\ket{0}\ket{0}$ or $\ket{\alpha}\ket{\alpha}$ is sent, respectively. $Q_{s_{z}}^{M_i}(s=0,1)$ and the click count of detector $D_T$, which we express as $n_z$, can be directly calculated as well. First, we use these four observed values to estimate their upper bounds on corresponding expected values by Kato's inequality as follows:
\begin{equation}
        n_{w}^{M_i*} \le \overline{n_{w}^{M_i*}}=n_{w}^{M_i} +\Delta_{w}^{M_i}   ,
\end{equation}
where $w=00,\alpha \alpha$ and $i=0,1$. The statistical fluctuation parameters here are obtained in the way presented in the Appendix~\ref{appendix}. Similarly, two lower bounds $\underline{n_{w}^{M_0*}}$ ($w=00,\alpha \alpha$) need to be calculated as follows:
\begin{equation}
        n_{w}^{M_0*} \ge \underline{n_{w}^{M_0*}}=n_{w}^{M_0} -\Delta_{w}^{M_0\prime}.
\end{equation}
We set the failure probability for estimating each of the six bounds above to be $\varepsilon_1$. The total number of rounds performed is set to be $N$. So, we can denote the number of state  $\ket{\alpha}\ket{\alpha}$ sent by Alice as $N_{\alpha\alpha}=N\times p_{d_1}$ and the number of state $\ket{0}\ket{0}$ as  $N_{00}=N\times p_{d_2}$. We calculate the upper and lower bounds on gains of each event as follows:
\begin{equation}
      \overline{Q_{w}^{M_i*}}=\overline{n_{w}^{M_i*}}/n_{w},
\end{equation} 
\begin{equation}
  \underline{Q_{w}^{M_0*}}=\underline{n_{w}^{M_0*}}/n_{w}.
\end{equation}
Then, by applying Eq.~(\ref{gain M1}) and~(\ref{gain M0}) in the expected case, we obtain the expected values as follows:

\begin{equation}
\begin{aligned}
    \overline{Q_{0x}^{M_1*}} &= \frac{1}{N^+}\left(e^{\frac{\mu}{2}}\sqrt{\overline{Q_{\alpha \alpha}^{M_1*}}}+e^{-\frac{\mu}{2}}\sqrt{\overline{Q_{00}^{M_1*}}}\right)^2\\
    &+ \frac{N^-}{N^+}\left( \frac{e^{\mu}N^-}{4}+e^{\mu} \sqrt{\overline{Q_{\alpha \alpha}^{M_1*}}}+\sqrt{\overline{Q_{00}^{M_1*}}}\right),
\end{aligned}
    \label{gain M1 ex}
\end{equation}
\begin{equation}
\begin{aligned}
    \underline{Q_{0x}^{M_0*}} &= \frac{1}{N^+}\left(e^{\mu}\underline{Q_{\alpha \alpha}^{M_0*}}+e^{-\mu}\underline{Q_{00}^{M_0*}} -2\sqrt{\overline{Q_{00}^{M_0*}}\times\overline{Q_{\alpha \alpha}^{M_0*}}}\right)\\
    &-\frac{N^-}{N^+}\left( e^{\mu} \sqrt{\overline{Q_{\alpha \alpha}^{M_0*}}}+\sqrt{\overline{Q_{00}^{M_0*}}}\right).
\end{aligned}
    \label{gain M0 ex}
\end{equation}
Finally, by applying Eq.~(\ref{BitErrorOfX}) and considering the bit error rate in the X basis to be equal to the phase error rate in the Z basis in the expected value case~\cite{zhou_making_2016}, the expected upper bound on the phase error rate is
\begin{equation}
    \overline{E_{p}^*}=\overline{E_{x}^*}=\frac{N^+(\overline{Q_{0x}^{M_1*}}-\underline{Q_{0x}^{M_0*}})+2(Q_{0_z}^{M_0}+Q_{1_z}^{M_0})}{2(Q_{0_z}^{M_0}+Q_{0_z}^{M_1}+Q_{1_z}^{M_0}+Q_{1_z}^{M_1})}.
    \label{phase error rate}
\end{equation}
We use Kato's inequality again but in a different form which is explained in the Appendix\ref{appendix} to calculate the upper bound on phase error rate in the observed value case. The expected number of clicks corresponding to phase errors is $\overline{n_p^*}=N\times\overline{E_{p}^*}$. So, the upper bound on the observed value is
\begin{equation}
    n_p \le \overline{n_p} = \overline{n_p^*} + \Delta_p,
    \label{Ep}
\end{equation}
where $\Delta_p=\sqrt{\frac{1}{2}n_{z}\ln\varepsilon_2^{-1}}$ and $\varepsilon_2$ is the failure probability for estimating $\overline{n_p}$. So, the upper bound on the phase error rate is
\begin{equation}
    \overline{E_p}=\overline{n_p}/n_z.
\end{equation}

\subsection{Composable security}
\par Considering the failure probability for estimating the statistical fluctuations described in Sec, the practical protocol has a total secrecy of $\varepsilon_{\rm{sec}}=2\varepsilon + \varepsilon_0+6\varepsilon_1+\varepsilon_2$, where we take $\varepsilon=\varepsilon_0=\varepsilon_1=\varepsilon_2=\varepsilon_{\rm{sec}}/10$. So, the final key length is denoted as
\begin{equation}
    l \ge n_z[1-h(\overline{E_p})]-\rm{leak}_{\rm{EC}}-\log_2\left(\frac{2}{\varepsilon_{\rm{cor}}}\right)-2\log_2\left(\frac{5}{\varepsilon_{\rm{sec}}}\right),
\end{equation}
and the COW-QKD protocol in this paper is $\varepsilon_{\rm{s}}$-secret, where $\varepsilon_{\rm{s}}=\varepsilon_{\rm{cor}}+\varepsilon_{\rm{sec}}$.

\section{\label{simulation}numerical simulation\\ and discussion}

To numerically simulate the key rate performance of our protocol in the finite-key regime, we assume that the dark-count rate is $p_d=2\times10^{-8}$ and the efficiency of the photon detectors is $\eta_d=70\%$. The number of bits that are revealed in the error-correction step $\rm{leak}_{\rm{EC}}$ is $f n_z h(E_z)$, where the correction efficiency $f$ is set to 1.1. The transmittance of the optical fiber with length $L$ is expressed by $\eta=10^{-0.016L}$. For finite-key analysis, the security bounds of correctness and secrecy are fixed to  $\varepsilon_{\rm{cor}}=10^{-15}$ and $\varepsilon_{\rm{sec}}=10^{-10}$. Other experimental parameters such as the mean photon number $\mu=|\alpha|^2$ and the transmittance of the beam-splitter used to distribute incoming states are decided by an optimization algorithm.

\begin{figure}[t]
    \centering
    \includegraphics[width=8.5cm]{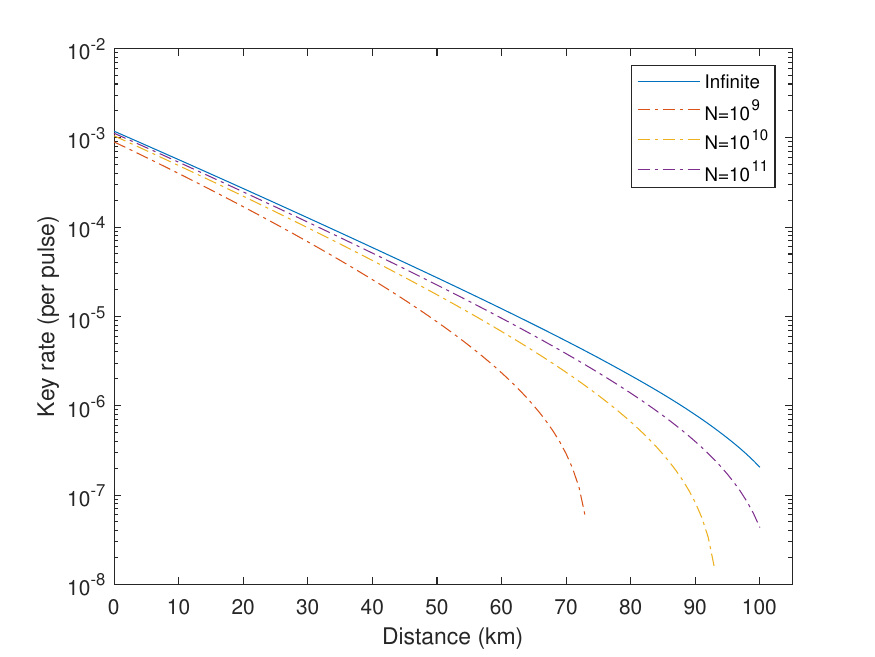}
    \caption{Secret key rate with different values of the total number of rounds, $N=10^9,10^{10},$ and $10^{11}$, using passive basis choice. The misalignment error $e_d$ is set to $1\%$. When $N=10^{11}$, which is reasonable in experimental implementation, the key rate is quite close to the performance in the asymptotic case. The key rate performance also shows that the security of our protocol ensures a secure distance exceeding 100km for COW-QKD in practical implementation.}
    \label{passive}
\end{figure}

We present the performance of the key rate of COW-QKD with different total numbers of rounds compared with the key rate of infinite limit~\cite{gao_simple_2022}, where the misalignment error rate is fixed to $e_d=1\%$. As shown in Fig.~\Ref{passive}, we can conclude that if the state-sending step of our protocol is repeated for $N=10^{11}$ rounds, the key rate is close to that of infinite limit, showing the practicality of our protocol in the finite-key regime. When choosing $N=10^{11}$, a 3-Mbit key can be obtained through 34 km fiber by Alice and Bob if they run our protocol with a laser operating at 1 GHZ for only 30 seconds, which presents the superiority of this protocol in short-distance communication. The results also demonstrate that our security analysis guarantees an unconditionally secure communication range exceeding 100km for COW-QKD, given its straightforward experimental setup.

\begin{figure}[t]
  \centering
  \subfigure{
  \includegraphics[width=8.5cm]{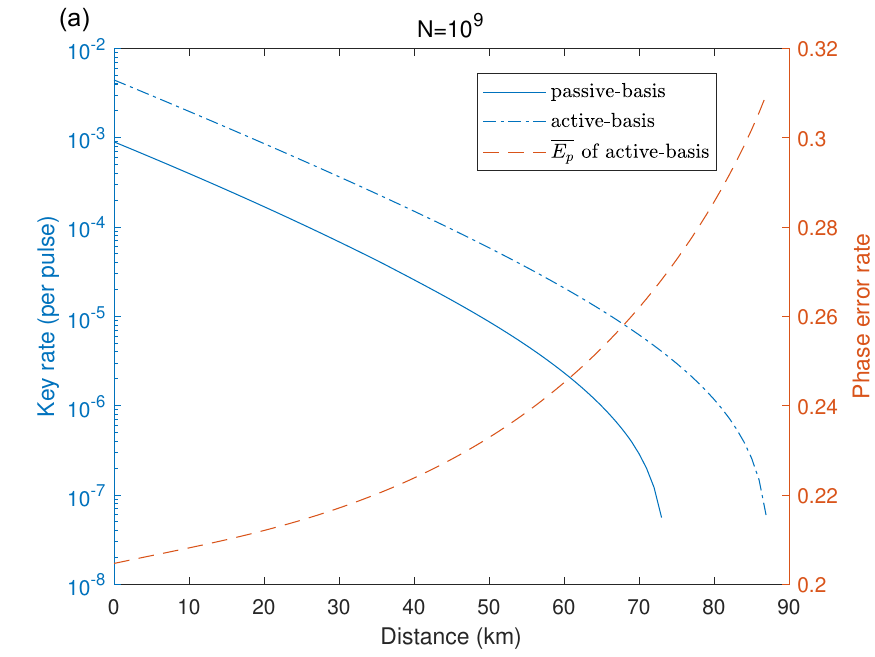}}
  \subfigure{
  \includegraphics[width=8.5cm]{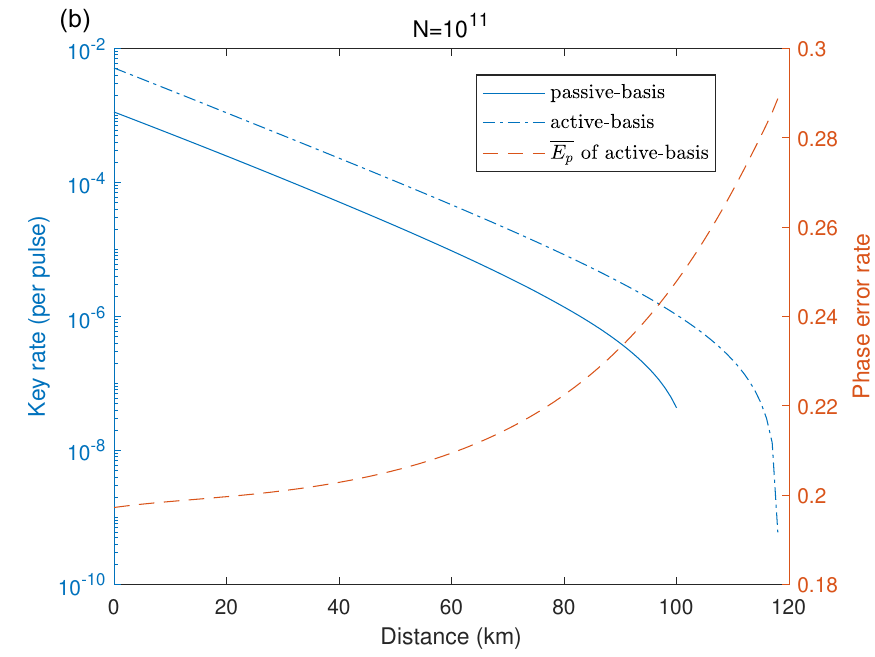}}
  
  \caption{Comparison of key rates using passive basis and active basis when $N=10^9$ and $N=10^{11}$. The upper bound on the phase error rate $\overline{E_p}$ using active basis choice is presented by the dashed red line. The misalignment error $e_d$ is set to $1\%$. (a) The total number of rounds is $N=10^9$. (b) The total number of rounds is $N=10^{11}$.}
  \label{PassiveVsActive} 
  
  \end{figure}

\begin{figure}[t]
\centering
\subfigure{
\includegraphics[width=8.5cm]{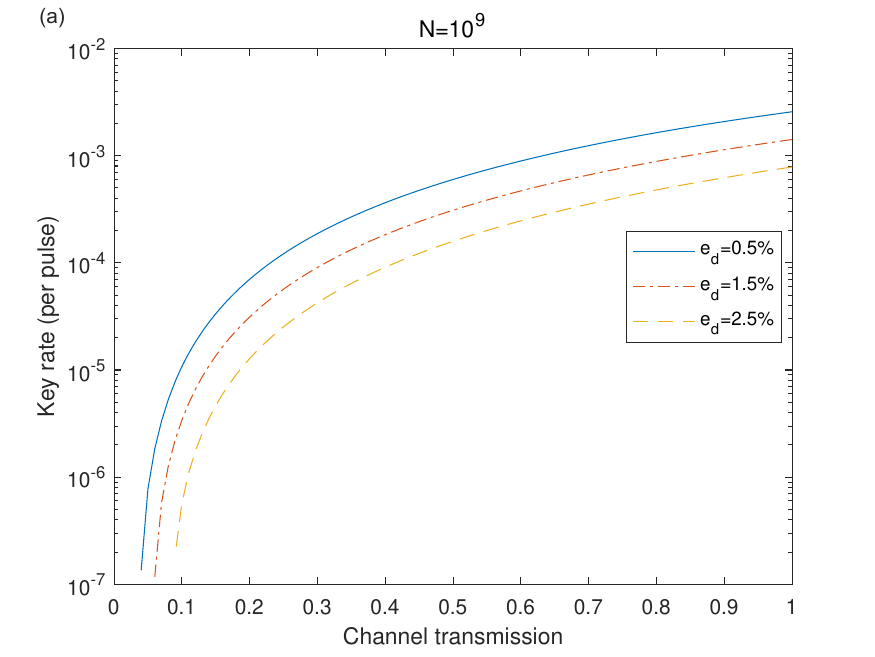}}
\subfigure{
\includegraphics[width=8.5cm]{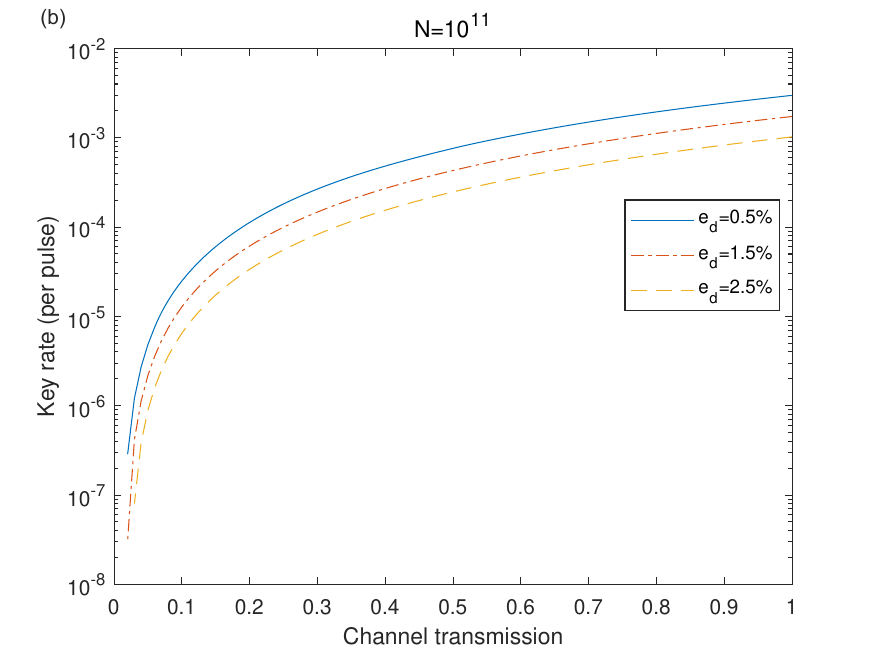}}

\caption{Secret key rate simulation in the finite-key case with different values of misalignment error when using passive basis. When a large misalignment error is employed, the key rates do not drop significantly, which is an important advantage in constructing quantum information systems. (a) The total number of rounds is $N=10^9$. (b) The total number of rounds is $N=10^{11}$.}
\label{DifferentError} 
\end{figure}

We demonstrate the flexibility of our protocol by presenting its key rate performance under different conditions. First, the beam splitter used to passively distribute optical pulses into the data or monitoring lines can be replaced by an optical switch that actively divides incoming states into different lines. This is referred to as the passive and active basis choice, respectively. A comparison of the key rate between passive and active basis choice is presented in Fig.~\ref{PassiveVsActive}, along with the estimated upper bound on the phase error rate $\overline{E_p}$ when using an active basis, demonstrating the applicability of our analysis with an active basis. Our protocol also exhibits robustness when faced with varying values of the misalignment error rate, as shown in Fig.~\ref{DifferentError}. The results show that even with a large misalignment error, the key rates are not significantly affected. This indicates the practicality of our protocol in constructing quantum communication systems under different experimental conditions.

\begin{figure}[t]
    \centering
    \includegraphics[width=8.5cm]{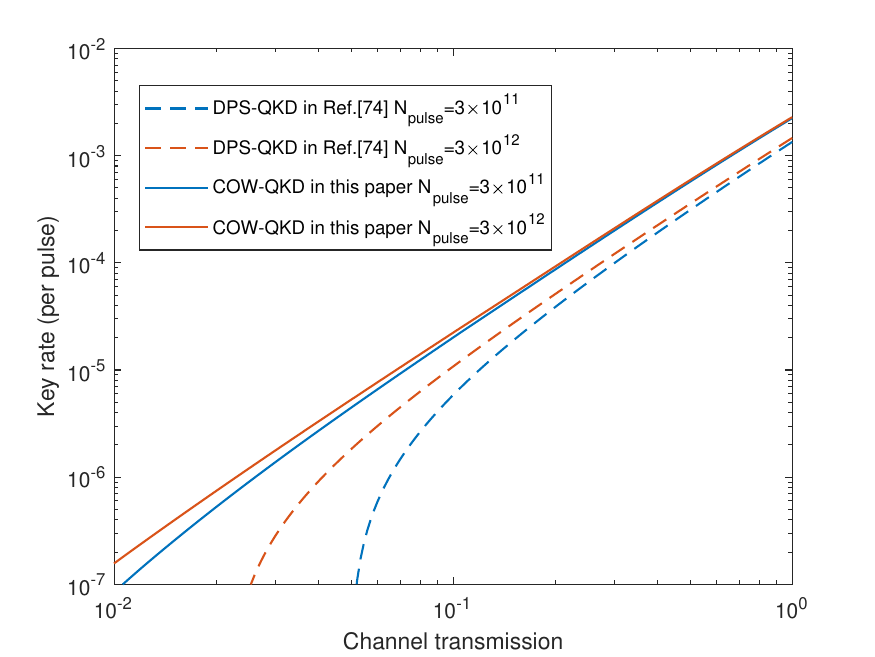}
    \caption{Comparison of key rates of our protocol and the DPS-QKD protocol in Ref.~\cite{mizutani_finite_2023}. Two protocols have similar experimental settings and both belong to the distributed-phase-reference QKD protocols. The bit error rate is $1\%$ and the correction efficiency $f$ is 1.16. For convenience, the security bounds of both correctness and secrecy are fixed to $2^{-28}$ as in Ref.~\cite{mizutani_finite_2023}. We compare the key rates when the numbers of pulses are $N_{\rm{pulse}}=3\times10^{11}$ and $N_{\rm{pulse}}=3\times10^{12}$ and conclude that our protocol has advantages on the secure transmission distance and key rate performance. }
    \label{comparison}
\end{figure}

We compare our protocol with DPS-QKD~\cite{mizutani_quantum_2019}, whose equipment requirements are similar to that of COW-QKD. The finite-key security analysis of it in Ref.~\cite{mizutani_finite_2023} shows a tighter bound on phase error rate can be obtained by Kato's inequality. To simulate under the same experimental conditions, we follow the choice in Ref.~\cite{mizutani_finite_2023} and fix the security bounds of both correctness and secrecy to $2^{-28}$ to get a total secrecy of $2^{-27} \approx 10^{-8.1}$. The misalignment error $e_d$ is set to 0.01 and the dark-count rate is 0, so we can ensure the bit error rate is $1\%$ as chosen in Ref.~\cite{mizutani_finite_2023}. The correction efficiency $f$ is 1.16, and we have simulated the key rate under varying values of overall channel transmittance, taking into account both optical fiber loss and detection efficiency. We note that in our COW-QKD protocol, Alice sends two pulses in each round, whereas the DPS-QKD protocol requires three pulses. For comparison, we need to ensure that the total number of pulses $N_{\rm pulse}$ is the same for both protocols. We compare the key rates of the two protocols when $N_{\rm pulse}$ is set to $3\times10^{11}$ and $3\times10^{12}$, so one of the results for DPS-QKD is consistent with that reported in Ref.~\cite{mizutani_finite_2023}. The simulation is shown in Fig.~\ref{comparison}. The results reveal that the key rates of our protocol are significantly higher than those reported in Ref.~\cite{mizutani_finite_2023}.

\begin{figure}[t]
\centering
\subfigure{
\includegraphics[width=8.5cm]{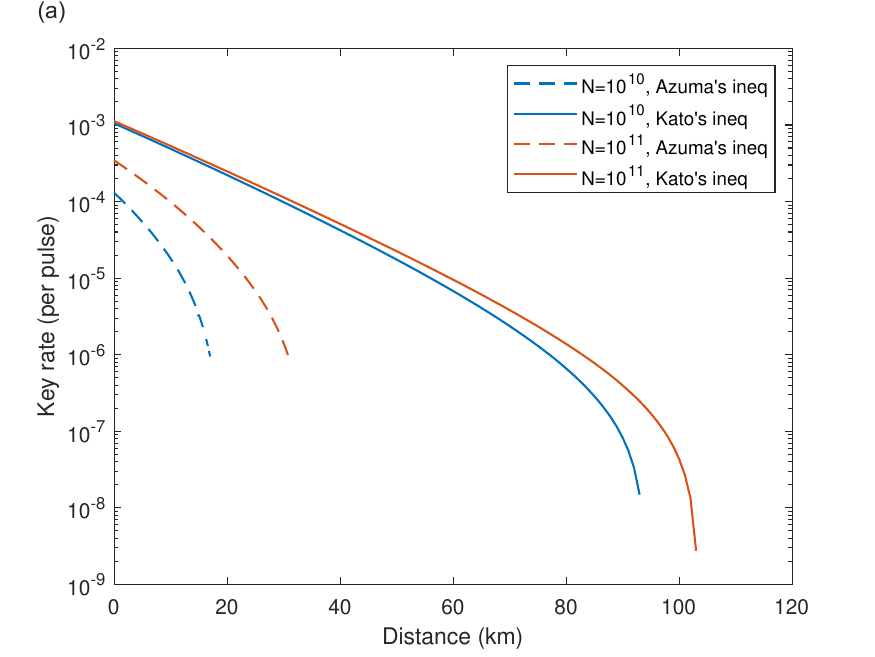}}
\subfigure{
\includegraphics[width=8.5cm]{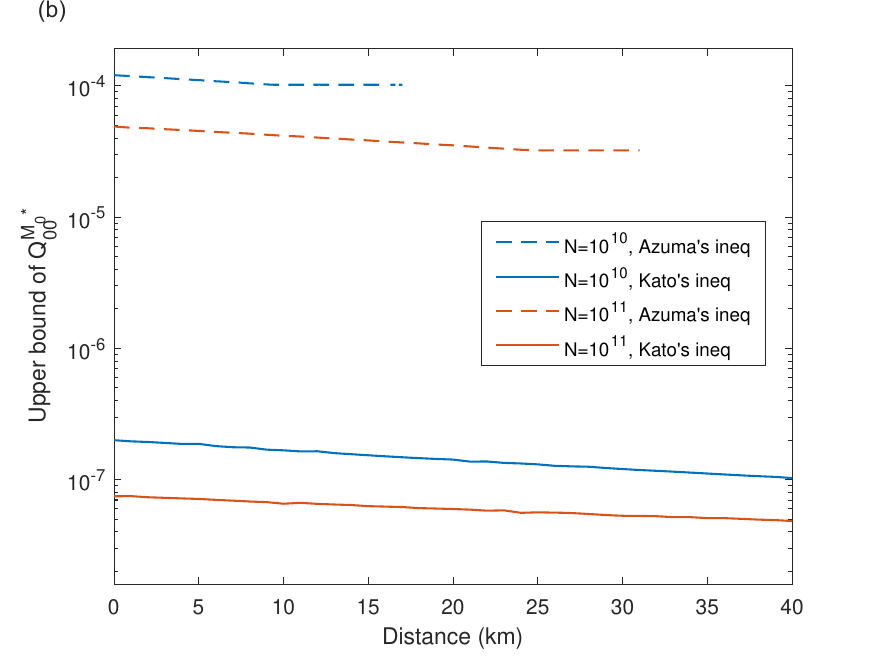}}

\caption{\textcolor{black}{(a) Finite secret key rate simulation of our protocol using Azuma's or Kato's inequalities to estimate the upper bound on $Q_{00}^{M_0*}$ with $N=10^{10}$ and $N=10^{11}$. The misalignment error $e_d$ is set to $1\%$. It can be concluded that significantly higher key rates are obtained by utilizing Kato's inequality. (b) The detailed estimated values of $\overline{Q_{00}^{M_0*}}$ when applying Azuma's or Kato's inequalities. Statistical fluctuations dominate the estimation of the upper bound, leading to significantly higher values. Consequently, higher phase error rates are calculated as per Eqs.(\ref{gain M1 ex})-(\ref{phase error rate}) and result in lower key rates.}}
\label{KatovsAzuma} 
\end{figure}

\textcolor{black}{Following the approach proposed in Ref.~\cite{mizutani_finite_2023}, our protocol demonstrates an enhancement of Kato's inequality over Azuma's inequality as well. In our protocol, the gain of the decoy states $\ket{0}_{2k-1}\ket{0}_{2k}$, represented as $Q_{00}^{M_i}$, is entirely attributed to the non-zero dark count rate, which is approximately on the scale of $10^{-8}$. However, the fluctuations estimated by Azuma's inequality between observed and expected values exhibit a linear dependence on $\sqrt{N_{00}}$. These statistical fluctuations dominate the estimation of the upper limit on $Q_{00}^{M_i*}$, resulting in a decrease in $\underline{Q_{0x}^{M_0*}}$ and an increase in $\overline{Q_{0x}^{M_1*}}$ according to Eqs.~(\ref{gain M1 ex}) and (\ref{gain M0 ex}). This gives rise to a higher phase error rate as per Eq.~(\ref{phase error rate}), consequently leading to a diminished key rate. To compare the results, we conduct a numerical simulation of our protocol's key rates, using Azuma's or Kato's inequalities to estimate the upper bound on $Q_{00}^{M_i*}$. The results are depicted in Fig.~\ref{KatovsAzuma} together with a detailed comparison of the upper bounds on $Q_{00}^{M_0*}$ obtained by two inequalities.}

\begin{figure}[t]
\centering
\includegraphics[width=8.5cm]{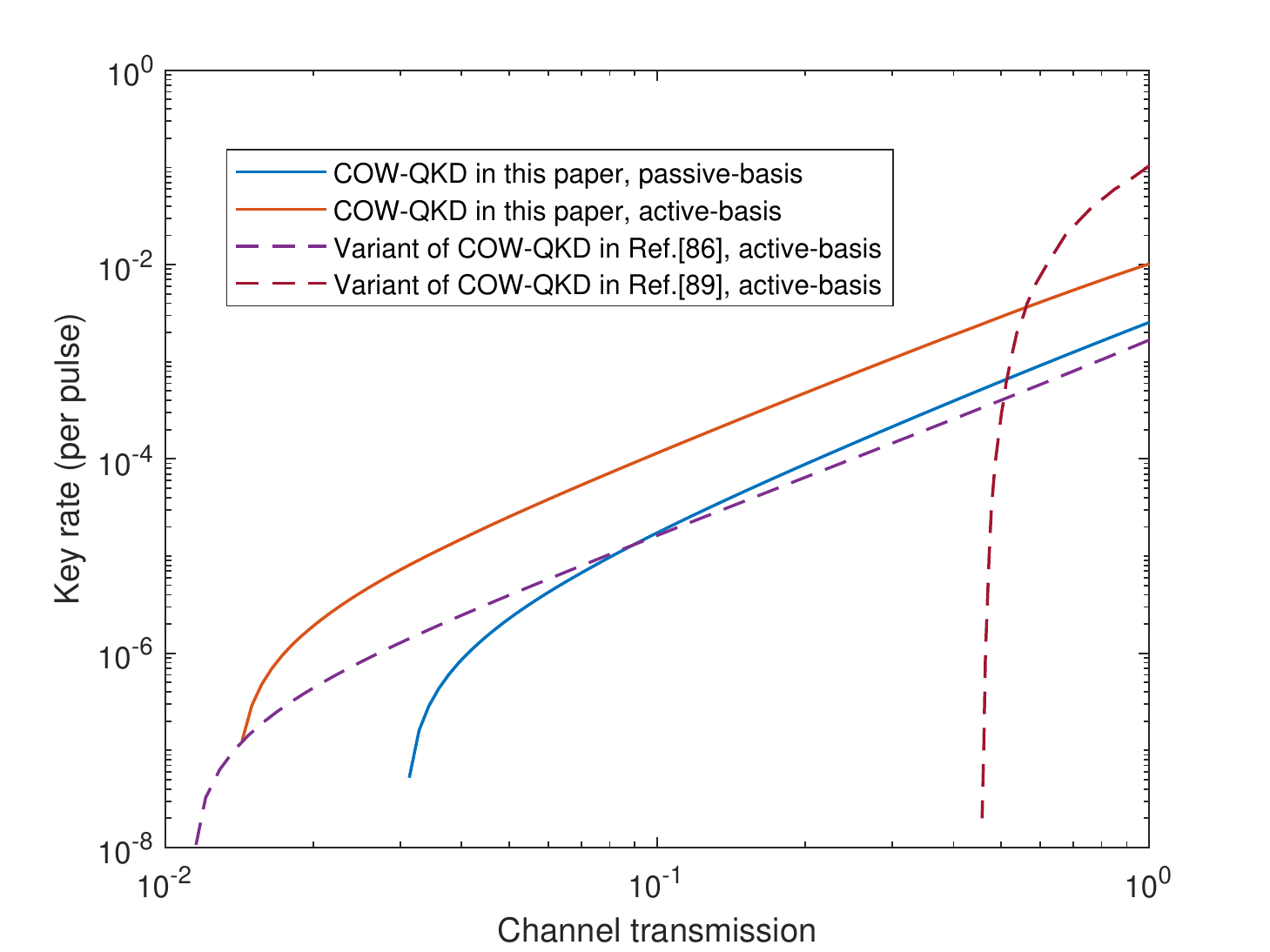}
\caption{\textcolor{black}{Secret key rates of our protocol compared with previous variants in Refs.~\cite{moroder_security_2012,wang_characterising_2019}. The dark count rate $p_d$ is fixed to $10^{-7}$ and the efficiency of detectors is set to $99\%$. The experimental requirements for the variant in Ref.~\cite{moroder_security_2012} are more complex compared to the original protocol. However, both the variant in Ref.~\cite{wang_characterising_2019} and our protocol maintain the simplicity of the setup. Under certain channel transmission values, our protocol with passive basis choice can still outperform the previous variants that use active basis choice in terms of key rate performance.}}
\label{COWcomparison} 
\end{figure}

\textcolor{black}{Compared with other security analyses of previous variants of COW-QKD, our protocol has notable advantages because its key rate performance is better when considering the highest security standard, i.e., the security against both zero-error attack and coherent attacks. For the completeness of this paper, in Fig.~\ref{COWcomparison} we compare the key rates in the asymptotic case of our COW protocol with the variants proposed in Refs.~\cite{moroder_security_2012,wang_characterising_2019} that have the immunity against these two attacks. To fairly compare, the parameter choices in Refs.~\cite{moroder_security_2012,wang_characterising_2019} are adopted, which fix the dark count rate $p_d$ to $10^{-7}$ and set the detection efficiency $\eta_d$ to $99\%$. In our numerical simulation of the variant presented in Ref.~\cite{moroder_security_2012}, we opt for a scenario where each three-signal block, comprising six optical pulses, shares the same phase. This modification to the experimental setup has inevitably altered the simplicity of the original protocol. Both of these previous variants are simulated with an active basis choice. It can be concluded that our protocol not only outperforms in terms of key rates and transmission distance but also preserves the original simplicity of the COW-QKD setting.}

\begin{table*}[t]\color{black}
\renewcommand\arraystretch{1.5}
\caption{\label{TabelComparison}The differences between our protocol and several variants of COW-QKD. Here, state $\ket{0}\ket{\alpha}$ is abbreviated expression of state $\ket{0}_{2k-1}\ket{\alpha}_{2k}$ and other states have similar meanings. States $\ket{0}\ket{0}\ket{\alpha}$,$\ket{0}\ket{\alpha}\ket{0}$, and $\ket{0}\ket{\beta}\ket{\beta}$ are three-pulse states used in Ref.~\cite{Lavie-improved-2022}.  EUR stands for entropic uncertainty relation and SDP stands for semidefinite programming techniques.}
\begin{ruledtabular}
\begin{tabular}{lcccccc}
 & \tabincell{c}{\bf \small Branciard \emph {et~al.}\\ \bf \small(2008) \cite{branciard_upper_2008}} &\tabincell{c}{\bf \small  Moroder \emph {et~al.}\\ \bf \small (2012) \cite{moroder_security_2012}} & \tabincell{c}{\bf \small Korzh \emph {et~al.}\\ \bf \small(2015) \cite{korzh_provably_2015}}  &\tabincell{c}{\bf \small Wang \emph {et~al.}\\ \bf\small (2019) \cite{wang_characterising_2019}}&\tabincell{c}{\bf \small Lavie \emph {et~al.}\\ \bf \small (2022) \cite{Lavie-improved-2022}} & \tabincell{c}{\\ \bf \small This work}\\ \hline
 Signal states&$\ket{0}\ket{\alpha}$,$\ket{\alpha}\ket{0}$ &$\ket{0}\ket{\alpha}$,$\ket{\alpha}\ket{0}$\footnotemark[1]&$\ket{0}\ket{\alpha}$,$\ket{\alpha}\ket{0}$ & $\ket{0}\ket{\alpha}$,$\ket{\alpha}\ket{0}$ & $\ket{0}\ket{0}\ket{\alpha}$,$\ket{0}\ket{\alpha}\ket{0}$ &$\ket{0}\ket{\alpha}$,$\ket{\alpha}\ket{0}$\\
 Decoy states&$\ket{\alpha}\ket{\alpha}$
 &$\ket{\alpha}\ket{\alpha}$\footnotemark[1]&$\ket{\alpha}\ket{\alpha}$&$\ket{\alpha}\ket{\alpha}$&$\ket{0}\ket{\beta}\ket{\beta}$&$\ket{\alpha}\ket{\alpha}$,$\ket{0}\ket{0}$\\
 Attack model&\tabincell{c}{Collective\footnotemark[2]}
 &\tabincell{c}{Coherent}&\tabincell{c}{Collective}&\tabincell{c}{Coherent}&\tabincell{c}{Collective}&\tabincell{c}{Coherent}\\
  
 Key rate&$\Omicron(\eta)$
 &$\Omicron(\eta^2)$&$\Omicron(\eta)$&$\Omicron(\eta^2)$&$\Omicron(\eta^2)$&$\Omicron(\eta^2)$\\
Security framework& EUR&SDP&EUR&SDP&SDP&EUR\\
\tabincell{l}{Against zero-error attack}& NO&YES&NO&YES&YES&YES\\
 \tabincell{l}{Finite feasibility} &NO&NO&YES&NO&NO&YES
\end{tabular}
\end{ruledtabular}
\footnotetext[1]{This protocol needs three states to be sent in a group with the same phase, making the setup more complicated.}
\footnotetext[2]{This protocol only considers restricted types of collective attacks.}
\end{table*}

\textcolor{black}{Moreover, we have summarized the differences with several previous works on COW-QKD in aspects such as experimental setting and key rate performance in Table~\Ref{TabelComparison}. As previously discussed, our protocol exhibits superior key rate performance compared to variants~\cite{moroder_security_2012,wang_characterising_2019} that share an equivalent security level. A method for calculating the lower bound on the secure key rate of a COW variant was proposed in Ref.~\cite{Lavie-improved-2022} recently, which shows significantly improved key rate performance. However, the absence of finite-key regime analysis and immunity against coherent attacks pose significant challenges to the practical implementation of this theoretical protocol. Earlier works such as Refs.~\cite{branciard_upper_2008,korzh_provably_2015} presented security proofs that could achieve high key rates scaling as $\Omicron(\eta)$. However, these variants have been demonstrated to be insecure when confronting zero-error attacks\cite{gonzalez-payo_upper_2020,trenyi_zero-error_2021}. Since our security proof provides the analytical formulas of the key rate, the extension of analyzing performance with finite key length can be completed as presented in Sec.~\ref{formula}. With the help of Kato's inequality, our analysis gives a tight bound on the key rate and guarantees security against coherent attacks, establishing foundations for further practical applications.}

\section{\label{conclusion}Conclusion}
In this paper, we present a finite-key analysis for the COW-QKD protocol proposed in Ref.~\cite{gao_simple_2022}. We apply the quantum leftover hashing lemma and entropic uncertainty relation to derive an analytic formula for the key length. When dealing with correlated random variables, we use Kato's inequality to ensure security against coherent attacks and achieve a higher key rate. By considering the failure probabilities for estimating statistical fluctuations between observed and expected values, we complete the security proof within the universally composable framework. Our finite-key analysis shows that the key transmission distance can exceed 100 km in specific cases, providing a feasible approach for the secure implementation of quantum communication processes. In short-distance communication, the numerical simulation in Fig.~\Ref{passive} has shown that our protocol can generate a 3-Mbit secret key over 34km fiber by running this protocol for only 30 seconds with a photon source operating at 1 GHZ repetition rate. We also present numerical simulations of key rates under different conditions, demonstrating the practicality and flexibility of our protocol. Compared to the finite-key analysis of DPS-QKD in Ref.~\cite{mizutani_finite_2023}, our protocol obtains a significantly higher key rate with almost the same experimental setup. \textcolor{black}{Additionally, we show that our COW-QKD protocol can also be used to show the superiority of Kato's inequality when estimating events with a small probability of occurrence. Furthermore, we present the comparison of key rates with previous COW variants in the asymptotic scenario and summarize the differences in many aspects to clearly illustrate the distinct advantages of our protocol.} In conclusion, our protocol lays a theoretical foundation for applying COW-QKD in real-world scenarios by offering both a high key rate and unconditional security against coherent attacks and completes the intact security proof for this protocol. Our protocol may be employed in future quantum communication with minuscule devices like chips and quantum information networks due to the simple experimental setup and excellent key rate performance.

\begin{acknowledgments}
  This work is supported by the National Natural Science Foundation of China (No.12274223), the Natural Science Foundation of Jiangsu Province (No. BK20211145), the Fundamental Research Funds for the Central Universities (No. 020414380182), and the Program for Innovative Talents and Entrepreneurs in Jiangsu (No. JSSCRC2021484).
\end{acknowledgments}

\appendix*

\section{\label{appendix}Kato's inequality}
Kato's inequality~\cite{kato_concentration_2020} is used to deal with the correlated random variables in this paper when estimating parameters. Here we introduce how to use Kato's inequality to complete the estimation in the main text.

\par We can use Eq.~(\ref{kato1}) to estimate the upper bounds on expected values from the corresponding observed values. In the estimation of QKD protocols, the random variables $n_i,i=1,2,\cdots,k$, which indicate whether the detector clicks in the $i$-th round, respectively, are Bernoulli random variables. If the detector clicks in the $u$-th round, $n_u=1$, and if it doesn't, click $n_u=0$. So, we have $E(n_u|f_{u-1})=\Pr(n_u=1|f_{u-1})$. $\Gamma_k$ is an observed value that denotes the total number 
of detector click during k rounds. 

To get the tightest bound, one should choose the optimal values for $a$ and $b$ to minimize the deviation $\left[ b+a\left( \frac{2\Gamma_k}{k}-1 \right) \right]\sqrt{k} $ by solving an optimization problem. To demonstrate, we let $\varepsilon_a$ be the failure probability for estimating the upper bounds, i.e., $\exp\left[ \frac{-2(b^2-a^2)}{(1+\frac{4a}{3\sqrt{k}})^2} \right]=\varepsilon_a$, and the optimization problem which is denoted as:
$
        \min_{a,b\ge |a|} \left[ b+a\left( \frac{2\Gamma_k}{k}-1 \right) \right]\sqrt{k};
$.

This is solved by Ref.~\cite{curras-lorenzo_tight_2021} and the solutions are
\begin{widetext}
\begin{equation}
    \begin{aligned}
        a_1&=a_1(\Gamma_k,k,\varepsilon_a)\\
         &=\frac{3\left( 72\sqrt{k}\Gamma_k(k-\Gamma_k)\ln \varepsilon_a-16k^{3/2}\ln^2\varepsilon_a+9\sqrt{2}(k-2\Gamma_k)\sqrt{-k^2\ln \varepsilon_a(9\Gamma_k(k-\Gamma_k)-2k\ln \varepsilon_a)} \right)}{4(9k-8\ln \varepsilon_a)(9\Gamma_k(k-\Gamma_k)-2k\ln \varepsilon_a)},
    \end{aligned}
    \label{a1}
\end{equation}
\begin{equation}
    b_1=b_1(a_1,k,\varepsilon_a)=\frac{\sqrt{18a_1^2k-(16a_1^2+24a_1\sqrt{k}+9k)\ln \varepsilon_a}}{3\sqrt{2k}}.
    \label{b1}
\end{equation}
\end{widetext}
With the fixed values $a_1$ and $b_1$, we get the upper bound on expected value as follows according to Eq.~(\ref{kato1}):
\begin{equation}
    \Gamma_k^* \le \overline{\Gamma_k^*} = \Gamma_k + \Delta_1(a_1,b_1,k,\Gamma_k),
\end{equation}
where $\Delta_1(a_1,b_1,k,\Gamma_k) = \left[ b_1+a_1\left(\frac{2\Gamma_k}{k}-1 \right)  \right]\sqrt{k}$ and we use the expected value $\Gamma_k^*$ to denote $\sum_{u=1}^kE(n_u|f_{u-1})$.
\par Similarly, Eq.~(\ref{kato2}) can be applied to estimate the lower bound on expected values, where 
we need to solve another optimization problem. That is, $\min_{a,b\ge |a|} \left[ b+a\left( \frac{2\Gamma_k}{k}-1 \right) \right]\sqrt{k}$, where $\exp\left[ \frac{-2(b^2-a^2)}{(1-\frac{4a}{3\sqrt{k}})^2} \right]=\varepsilon_a$.

The solutions are
\begin{widetext}
\begin{equation}
    \begin{aligned}
        a_2&=a_2(\Gamma_k,k,\varepsilon_a)\\
         &=-\frac{3\left( 72\sqrt{k}\Gamma_k(k-\Gamma_k)\ln \varepsilon_a-16k^{3/2}\ln^2\varepsilon_a-9\sqrt{2}(k-2\Gamma_k)\sqrt{-k^2\ln \varepsilon_a(9\Gamma_k(k-\Gamma_k)-2k\ln \varepsilon_a)} \right)}{4(9k-8\ln \varepsilon_a)(9\Gamma_k(k-\Gamma_k)-2k\ln \varepsilon_a)},
    \end{aligned}
    \label{a2}
\end{equation}
\begin{equation}
    b_2=b_2(a_2,k,\varepsilon_a)=\frac{\sqrt{18a_2^2k-(16a_2^2-24a_2\sqrt{k}+9k)\ln \varepsilon_a}}{3\sqrt{2k}},
    \label{b2}
\end{equation}
\end{widetext}
and we get the lower bound
\begin{equation}
    \Gamma_k^* \ge \underline{\Gamma_k^*} = \Gamma_k - \Delta_2(a_2,b_2,k,\Gamma_k),
\end{equation}
where $\Delta_2(a_2,b_2,k,\Gamma_k) = \left[ b_2+a_2\left(\frac{2\Gamma_k}{k}-1 \right)  \right]\sqrt{k}$. With the methods above, the estimations of $\overline{n_{w}^{M_i*}}$ and $\underline{n_{w}^{M_0*}}$ in the main text can be done, where $w=00,\alpha \alpha$ and $i=0,1$. The failure probability for each estimation is $\varepsilon_a$, which is considered when explaining the composable security of our protocol.
\par When converting expected values to observed values, Kato's inequality is available as well. However, to get specific values of the optimal $a_i,b_i$ and the deviation $\Delta_i$ where $i=1,2$, the observed value $\Gamma_k$ needs to be employed, which is not known. So, we follow the method used in Ref.~~\cite{curras-lorenzo_tight_2021}. Let $a=0$ and set the failure probabilities in Eqs.~(\ref{kato1}) and (\ref{kato2}) to be $\varepsilon_b$. We obtain the inequalities below:
\begin{equation}
    \sum_{u=1}^kE(n_u|f_{u-1}) \le \Gamma_k +\Delta,
\end{equation}
\begin{equation}
    \sum_{u=1}^kE(n_u|f_{u-1}) \ge \Gamma_k -\Delta,
\end{equation}
where $\Delta=\sqrt{\frac{1}{2}k\ln \varepsilon_b^{-1}}$. This is how the estimation procedure of Eq.~(\ref{Ep}) is done.

\begin{thebibliography}{104}%
\makeatletter
\providecommand \@ifxundefined [1]{%
 \@ifx{#1\undefined}
}%
\providecommand \@ifnum [1]{%
 \ifnum #1\expandafter \@firstoftwo
 \else \expandafter \@secondoftwo
 \fi
}%
\providecommand \@ifx [1]{%
 \ifx #1\expandafter \@firstoftwo
 \else \expandafter \@secondoftwo
 \fi
}%
\providecommand \natexlab [1]{#1}%
\providecommand \enquote  [1]{``#1''}%
\providecommand \bibnamefont  [1]{#1}%
\providecommand \bibfnamefont [1]{#1}%
\providecommand \citenamefont [1]{#1}%
\providecommand \href@noop [0]{\@secondoftwo}%
\providecommand \href [0]{\begingroup \@sanitize@url \@href}%
\providecommand \@href[1]{\@@startlink{#1}\@@href}%
\providecommand \@@href[1]{\endgroup#1\@@endlink}%
\providecommand \@sanitize@url [0]{\catcode `\\12\catcode `\$12\catcode
  `\&12\catcode `\#12\catcode `\^12\catcode `\_12\catcode `\%12\relax}%
\providecommand \@@startlink[1]{}%
\providecommand \@@endlink[0]{}%
\providecommand \url  [0]{\begingroup\@sanitize@url \@url }%
\providecommand \@url [1]{\endgroup\@href {#1}{\urlprefix }}%
\providecommand \urlprefix  [0]{URL }%
\providecommand \Eprint [0]{\href }%
\providecommand \doibase [0]{https://doi.org/}%
\providecommand \selectlanguage [0]{\@gobble}%
\providecommand \bibinfo  [0]{\@secondoftwo}%
\providecommand \bibfield  [0]{\@secondoftwo}%
\providecommand \translation [1]{[#1]}%
\providecommand \BibitemOpen [0]{}%
\providecommand \bibitemStop [0]{}%
\providecommand \bibitemNoStop [0]{.\EOS\space}%
\providecommand \EOS [0]{\spacefactor3000\relax}%
\providecommand \BibitemShut  [1]{\csname bibitem#1\endcsname}%
\let\auto@bib@innerbib\@empty
\bibitem [{\citenamefont {Duan}\ \emph {et~al.}(2001)\citenamefont {Duan},
  \citenamefont {Lukin}, \citenamefont {Cirac},\ and\ \citenamefont
  {Zoller}}]{duan2001long}%
  \BibitemOpen
  \bibfield  {author} {\bibinfo {author} {\bibfnamefont {L.-M.}\ \bibnamefont
  {Duan}}, \bibinfo {author} {\bibfnamefont {M.~D.}\ \bibnamefont {Lukin}},
  \bibinfo {author} {\bibfnamefont {J.~I.}\ \bibnamefont {Cirac}},\ and\
  \bibinfo {author} {\bibfnamefont {P.}~\bibnamefont {Zoller}},\ }\bibfield
  {title} {\bibinfo {title} {Long-distance quantum communication with atomic
  ensembles and linear optics},\ }\href@noop {} {\bibfield  {journal} {\bibinfo
   {journal} {Nature}\ }\textbf {\bibinfo {volume} {414}},\ \bibinfo {pages}
  {413} (\bibinfo {year} {2001})}\BibitemShut {NoStop}%
\bibitem [{\citenamefont {Azuma}\ \emph {et~al.}(2015)\citenamefont {Azuma},
  \citenamefont {Tamaki},\ and\ \citenamefont {Lo}}]{azuma2015all}%
  \BibitemOpen
  \bibfield  {author} {\bibinfo {author} {\bibfnamefont {K.}~\bibnamefont
  {Azuma}}, \bibinfo {author} {\bibfnamefont {K.}~\bibnamefont {Tamaki}},\ and\
  \bibinfo {author} {\bibfnamefont {H.-K.}\ \bibnamefont {Lo}},\ }\bibfield
  {title} {\bibinfo {title} {All-photonic quantum repeaters},\ }\href@noop {}
  {\bibfield  {journal} {\bibinfo  {journal} {Nat. Commun.}\ }\textbf {\bibinfo
  {volume} {6}},\ \bibinfo {pages} {6787} (\bibinfo {year} {2015})}\BibitemShut
  {NoStop}%
\bibitem [{\citenamefont {Li}\ \emph {et~al.}(2023{\natexlab{a}})\citenamefont
  {Li}, \citenamefont {Fu}, \citenamefont {Liu}, \citenamefont {Xie},
  \citenamefont {Li}, \citenamefont {Zhou}, \citenamefont {Yin},\ and\
  \citenamefont {Chen}}]{li2023all}%
  \BibitemOpen
  \bibfield  {author} {\bibinfo {author} {\bibfnamefont {C.-L.}\ \bibnamefont
  {Li}}, \bibinfo {author} {\bibfnamefont {Y.}~\bibnamefont {Fu}}, \bibinfo
  {author} {\bibfnamefont {W.-B.}\ \bibnamefont {Liu}}, \bibinfo {author}
  {\bibfnamefont {Y.-M.}\ \bibnamefont {Xie}}, \bibinfo {author} {\bibfnamefont
  {B.-H.}\ \bibnamefont {Li}}, \bibinfo {author} {\bibfnamefont {M.-G.}\
  \bibnamefont {Zhou}}, \bibinfo {author} {\bibfnamefont {H.-L.}\ \bibnamefont
  {Yin}},\ and\ \bibinfo {author} {\bibfnamefont {Z.-B.}\ \bibnamefont
  {Chen}},\ }\bibfield  {title} {\bibinfo {title} {All-photonic quantum
  repeater for multipartite entanglement generation},\ }\href@noop {}
  {\bibfield  {journal} {\bibinfo  {journal} {Opt. Lett.}\ }\textbf {\bibinfo
  {volume} {48}},\ \bibinfo {pages} {1244} (\bibinfo {year}
  {2023}{\natexlab{a}})}\BibitemShut {NoStop}%
\bibitem [{\citenamefont {Chen}\ and\ \citenamefont
  {Lo}(2007)}]{chen_multi_2007}%
  \BibitemOpen
  \bibfield  {author} {\bibinfo {author} {\bibfnamefont {K.}~\bibnamefont
  {Chen}}\ and\ \bibinfo {author} {\bibfnamefont {H.-K.}\ \bibnamefont {Lo}},\
  }\bibfield  {title} {\bibinfo {title} {Multi-partite quantum cryptographic
  protocols with noisy GHZ states},\ }\href@noop {} {\bibfield  {journal}
  {\bibinfo  {journal} {Quantum Inf. Comput.}\ }\textbf {\bibinfo {volume}
  {7}},\ \bibinfo {pages} {689} (\bibinfo {year} {2007})}\BibitemShut {NoStop}%
\bibitem [{\citenamefont {Fu}\ \emph {et~al.}(2015)\citenamefont {Fu},
  \citenamefont {Yin}, \citenamefont {Chen},\ and\ \citenamefont
  {Chen}}]{fu2015long}%
  \BibitemOpen
  \bibfield  {author} {\bibinfo {author} {\bibfnamefont {Y.}~\bibnamefont
  {Fu}}, \bibinfo {author} {\bibfnamefont {H.-L.}\ \bibnamefont {Yin}},
  \bibinfo {author} {\bibfnamefont {T.-Y.}\ \bibnamefont {Chen}},\ and\
  \bibinfo {author} {\bibfnamefont {Z.-B.}\ \bibnamefont {Chen}},\ }\bibfield
  {title} {\bibinfo {title} {Long-distance measurement-device-independent
  multiparty quantum communication},\ }\href@noop {} {\bibfield  {journal}
  {\bibinfo  {journal} {Phys. Rev. Lett.}\ }\textbf {\bibinfo {volume} {114}},\
  \bibinfo {pages} {090501} (\bibinfo {year} {2015})}\BibitemShut {NoStop}%
\bibitem [{\citenamefont {Zhao}\ \emph {et~al.}(2020)\citenamefont {Zhao},
  \citenamefont {Zeng}, \citenamefont {Cao}, \citenamefont {Xu}, \citenamefont
  {Zhen}, \citenamefont {Ma}, \citenamefont {Li}, \citenamefont {Liu},\ and\
  \citenamefont {Chen}}]{zhao_phase_2020}%
  \BibitemOpen
  \bibfield  {author} {\bibinfo {author} {\bibfnamefont {S.}~\bibnamefont
  {Zhao}}, \bibinfo {author} {\bibfnamefont {P.}~\bibnamefont {Zeng}}, \bibinfo
  {author} {\bibfnamefont {W.-F.}\ \bibnamefont {Cao}}, \bibinfo {author}
  {\bibfnamefont {X.-Y.}\ \bibnamefont {Xu}}, \bibinfo {author} {\bibfnamefont
  {Y.-Z.}\ \bibnamefont {Zhen}}, \bibinfo {author} {\bibfnamefont
  {X.}~\bibnamefont {Ma}}, \bibinfo {author} {\bibfnamefont {L.}~\bibnamefont
  {Li}}, \bibinfo {author} {\bibfnamefont {N.-L.}\ \bibnamefont {Liu}},\ and\
  \bibinfo {author} {\bibfnamefont {K.}~\bibnamefont {Chen}},\ }\bibfield
  {title} {\bibinfo {title} {Phase-matching quantum cryptographic
  conferencing},\ }\href {https://doi.org/10.1103/PhysRevApplied.14.024010}
  {\bibfield  {journal} {\bibinfo  {journal} {Phys. Rev. Appl.}\ }\textbf
  {\bibinfo {volume} {14}},\ \bibinfo {pages} {024010} (\bibinfo {year}
  {2020})}\BibitemShut {NoStop}%
\bibitem [{\citenamefont {Cao}\ \emph {et~al.}(2021)\citenamefont {Cao},
  \citenamefont {Gu}, \citenamefont {Lu}, \citenamefont {Yin},\ and\
  \citenamefont {Chen}}]{cao_coherent_2021}%
  \BibitemOpen
  \bibfield  {author} {\bibinfo {author} {\bibfnamefont {X.-Y.}\ \bibnamefont
  {Cao}}, \bibinfo {author} {\bibfnamefont {J.}~\bibnamefont {Gu}}, \bibinfo
  {author} {\bibfnamefont {Y.-S.}\ \bibnamefont {Lu}}, \bibinfo {author}
  {\bibfnamefont {H.-L.}\ \bibnamefont {Yin}},\ and\ \bibinfo {author}
  {\bibfnamefont {Z.-B.}\ \bibnamefont {Chen}},\ }\bibfield  {title} {\bibinfo
  {title} {Coherent one-way quantum conference key agreement based on twin
  field},\ }\href {https://doi.org/10.1088/1367-2630/abef98} {\bibfield
  {journal} {\bibinfo  {journal} {New J. Phys.}\ }\textbf {\bibinfo {volume}
  {23}},\ \bibinfo {pages} {043002} (\bibinfo {year} {2021})}\BibitemShut
  {NoStop}%
\bibitem [{\citenamefont {Li}\ \emph {et~al.}(2021)\citenamefont {Li},
  \citenamefont {Cao}, \citenamefont {Li}, \citenamefont {Weng}, \citenamefont
  {Gu}, \citenamefont {Yin},\ and\ \citenamefont {Chen}}]{li_finite_2021}%
  \BibitemOpen
  \bibfield  {author} {\bibinfo {author} {\bibfnamefont {Z.}~\bibnamefont
  {Li}}, \bibinfo {author} {\bibfnamefont {X.-Y.}\ \bibnamefont {Cao}},
  \bibinfo {author} {\bibfnamefont {C.-L.}\ \bibnamefont {Li}}, \bibinfo
  {author} {\bibfnamefont {C.-X.}\ \bibnamefont {Weng}}, \bibinfo {author}
  {\bibfnamefont {J.}~\bibnamefont {Gu}}, \bibinfo {author} {\bibfnamefont
  {H.-L.}\ \bibnamefont {Yin}},\ and\ \bibinfo {author} {\bibfnamefont {Z.-B.}\
  \bibnamefont {Chen}},\ }\bibfield  {title} {\bibinfo {title} {Finite-key
  analysis for quantum conference key agreement with asymmetric channels},\
  }\href@noop {} {\bibfield  {journal} {\bibinfo  {journal} {Quantum Sci.
  Technol}\ }\textbf {\bibinfo {volume} {6}},\ \bibinfo {pages} {045019}
  (\bibinfo {year} {2021})}\BibitemShut {NoStop}%
\bibitem [{\citenamefont {Fletcher}\ and\ \citenamefont
  {Pirandola}(2022)}]{fletcher_continuous_2022}%
  \BibitemOpen
  \bibfield  {author} {\bibinfo {author} {\bibfnamefont {A.~I.}\ \bibnamefont
  {Fletcher}}\ and\ \bibinfo {author} {\bibfnamefont {S.}~\bibnamefont
  {Pirandola}},\ }\bibfield  {title} {\bibinfo {title} {Continuous variable
  measurement device independent quantum conferencing with postselection},\
  }\href@noop {} {\bibfield  {journal} {\bibinfo  {journal} {Sci. Rep.}\
  }\textbf {\bibinfo {volume} {12}},\ \bibinfo {pages} {17329} (\bibinfo {year}
  {2022})}\BibitemShut {NoStop}%
\bibitem [{\citenamefont {Li}\ \emph {et~al.}(2023{\natexlab{b}})\citenamefont
  {Li}, \citenamefont {Fu}, \citenamefont {Liu}, \citenamefont {Xie},
  \citenamefont {Li}, \citenamefont {Zhou}, \citenamefont {Yin},\ and\
  \citenamefont {Chen}}]{li_breakingqcka_2023}%
  \BibitemOpen
  \bibfield  {author} {\bibinfo {author} {\bibfnamefont {C.-L.}\ \bibnamefont
  {Li}}, \bibinfo {author} {\bibfnamefont {Y.}~\bibnamefont {Fu}}, \bibinfo
  {author} {\bibfnamefont {W.-B.}\ \bibnamefont {Liu}}, \bibinfo {author}
  {\bibfnamefont {Y.-M.}\ \bibnamefont {Xie}}, \bibinfo {author} {\bibfnamefont
  {B.-H.}\ \bibnamefont {Li}}, \bibinfo {author} {\bibfnamefont {M.-G.}\
  \bibnamefont {Zhou}}, \bibinfo {author} {\bibfnamefont {H.-L.}\ \bibnamefont
  {Yin}},\ and\ \bibinfo {author} {\bibfnamefont {Z.-B.}\ \bibnamefont
  {Chen}},\ }\bibfield  {title} {\bibinfo {title} {Breaking universal
  limitations on quantum conference key agreement without quantum memory},\
  }\href {https://doi.org/10.1038/s42005-023-01238-5} {\bibfield  {journal}
  {\bibinfo  {journal} {Commun. Phys.}\ }\textbf {\bibinfo {volume} {6}},\
  \bibinfo {pages} {122} (\bibinfo {year} {2023}{\natexlab{b}})}\BibitemShut
  {NoStop}%
\bibitem [{\citenamefont {Hillery}\ \emph {et~al.}(1999)\citenamefont
  {Hillery}, \citenamefont {Bu\ifmmode~\check{z}\else \v{z}\fi{}ek},\ and\
  \citenamefont {Berthiaume}}]{hillery_quantum_1999}%
  \BibitemOpen
  \bibfield  {author} {\bibinfo {author} {\bibfnamefont {M.}~\bibnamefont
  {Hillery}}, \bibinfo {author} {\bibfnamefont {V.}~\bibnamefont
  {Bu\ifmmode~\check{z}\else \v{z}\fi{}ek}},\ and\ \bibinfo {author}
  {\bibfnamefont {A.}~\bibnamefont {Berthiaume}},\ }\bibfield  {title}
  {\bibinfo {title} {Quantum secret sharing},\ }\href
  {https://doi.org/10.1103/PhysRevA.59.1829} {\bibfield  {journal} {\bibinfo
  {journal} {Phys. Rev. A}\ }\textbf {\bibinfo {volume} {59}},\ \bibinfo
  {pages} {1829} (\bibinfo {year} {1999})}\BibitemShut {NoStop}%
\bibitem [{\citenamefont {Cleve}\ \emph {et~al.}(1999)\citenamefont {Cleve},
  \citenamefont {Gottesman},\ and\ \citenamefont {Lo}}]{cleve_how_1999}%
  \BibitemOpen
  \bibfield  {author} {\bibinfo {author} {\bibfnamefont {R.}~\bibnamefont
  {Cleve}}, \bibinfo {author} {\bibfnamefont {D.}~\bibnamefont {Gottesman}},\
  and\ \bibinfo {author} {\bibfnamefont {H.-K.}\ \bibnamefont {Lo}},\
  }\bibfield  {title} {\bibinfo {title} {How to share a quantum secret},\
  }\href {https://doi.org/10.1103/PhysRevLett.83.648} {\bibfield  {journal}
  {\bibinfo  {journal} {Phys. Rev. Lett.}\ }\textbf {\bibinfo {volume} {83}},\
  \bibinfo {pages} {648} (\bibinfo {year} {1999})}\BibitemShut {NoStop}%
\bibitem [{\citenamefont {Wei}\ \emph {et~al.}(2013)\citenamefont {Wei},
  \citenamefont {Ma},\ and\ \citenamefont {Yang}}]{wei_experimental_2013}%
  \BibitemOpen
  \bibfield  {author} {\bibinfo {author} {\bibfnamefont {K.-J.}\ \bibnamefont
  {Wei}}, \bibinfo {author} {\bibfnamefont {H.-Q.}\ \bibnamefont {Ma}},\ and\
  \bibinfo {author} {\bibfnamefont {J.-H.}\ \bibnamefont {Yang}},\ }\bibfield
  {title} {\bibinfo {title} {Experimental circular quantum secret sharing over
  telecom fiber network},\ }\href {https://doi.org/10.1364/OE.21.016663}
  {\bibfield  {journal} {\bibinfo  {journal} {Opt. Express}\ }\textbf {\bibinfo
  {volume} {21}},\ \bibinfo {pages} {16663} (\bibinfo {year}
  {2013})}\BibitemShut {NoStop}%
\bibitem [{\citenamefont {Gu}\ \emph {et~al.}(2021)\citenamefont {Gu},
  \citenamefont {Cao}, \citenamefont {Yin},\ and\ \citenamefont
  {Chen}}]{gu_differential_2021}%
  \BibitemOpen
  \bibfield  {author} {\bibinfo {author} {\bibfnamefont {J.}~\bibnamefont
  {Gu}}, \bibinfo {author} {\bibfnamefont {X.-Y.}\ \bibnamefont {Cao}},
  \bibinfo {author} {\bibfnamefont {H.-L.}\ \bibnamefont {Yin}},\ and\ \bibinfo
  {author} {\bibfnamefont {Z.-B.}\ \bibnamefont {Chen}},\ }\bibfield  {title}
  {\bibinfo {title} {Differential phase shift quantum secret sharing using a
  twin field},\ }\href {https://doi.org/10.1364/OE.417856} {\bibfield
  {journal} {\bibinfo  {journal} {Opt. Express}\ }\textbf {\bibinfo {volume}
  {29}},\ \bibinfo {pages} {9165} (\bibinfo {year} {2021})}\BibitemShut
  {NoStop}%
\bibitem [{\citenamefont {Williams}\ \emph {et~al.}(2019)\citenamefont
  {Williams}, \citenamefont {Lukens}, \citenamefont {Peters}, \citenamefont
  {Qi},\ and\ \citenamefont {Grice}}]{williams2019quantum}%
  \BibitemOpen
  \bibfield  {author} {\bibinfo {author} {\bibfnamefont {B.~P.}\ \bibnamefont
  {Williams}}, \bibinfo {author} {\bibfnamefont {J.~M.}\ \bibnamefont
  {Lukens}}, \bibinfo {author} {\bibfnamefont {N.~A.}\ \bibnamefont {Peters}},
  \bibinfo {author} {\bibfnamefont {B.}~\bibnamefont {Qi}},\ and\ \bibinfo
  {author} {\bibfnamefont {W.~P.}\ \bibnamefont {Grice}},\ }\bibfield  {title}
  {\bibinfo {title} {Quantum secret sharing with polarization-entangled photon
  pairs},\ }\href@noop {} {\bibfield  {journal} {\bibinfo  {journal} {Phys.
  Rev. A}\ }\textbf {\bibinfo {volume} {99}},\ \bibinfo {pages} {062311}
  (\bibinfo {year} {2019})}\BibitemShut {NoStop}%
\bibitem [{\citenamefont {Shen}\ \emph {et~al.}(2023)\citenamefont {Shen},
  \citenamefont {Cao}, \citenamefont {Wang}, \citenamefont {Fu}, \citenamefont
  {Gu}, \citenamefont {Liu}, \citenamefont {Weng}, \citenamefont {Yin},\ and\
  \citenamefont {Chen}}]{shen_experimental_2023}%
  \BibitemOpen
  \bibfield  {author} {\bibinfo {author} {\bibfnamefont {A.}~\bibnamefont
  {Shen}}, \bibinfo {author} {\bibfnamefont {X.-Y.}\ \bibnamefont {Cao}},
  \bibinfo {author} {\bibfnamefont {Y.}~\bibnamefont {Wang}}, \bibinfo {author}
  {\bibfnamefont {Y.}~\bibnamefont {Fu}}, \bibinfo {author} {\bibfnamefont
  {J.}~\bibnamefont {Gu}}, \bibinfo {author} {\bibfnamefont {W.-B.}\
  \bibnamefont {Liu}}, \bibinfo {author} {\bibfnamefont {C.-X.}\ \bibnamefont
  {Weng}}, \bibinfo {author} {\bibfnamefont {H.-L.}\ \bibnamefont {Yin}},\ and\
  \bibinfo {author} {\bibfnamefont {Z.-B.}\ \bibnamefont {Chen}},\ }\bibfield
  {title} {\bibinfo {title} {Experimental quantum secret sharing based on phase
  encoding of coherent states},\ }\href
  {https://doi.org/10.1007/s11433-023-2105-7} {\bibfield  {journal} {\bibinfo
  {journal} {Sci. China-Phys. Mech. Astron.}\ }\textbf {\bibinfo {volume}
  {66}},\ \bibinfo {pages} {260311} (\bibinfo {year} {2023})}\BibitemShut
  {NoStop}%
\bibitem [{\citenamefont {De~Oliveira}\ \emph {et~al.}(2020)\citenamefont
  {De~Oliveira}, \citenamefont {Nape}, \citenamefont {Pinnell}, \citenamefont
  {TabeBordbar},\ and\ \citenamefont {Forbes}}]{de2020experimental}%
  \BibitemOpen
  \bibfield  {author} {\bibinfo {author} {\bibfnamefont {M.}~\bibnamefont
  {De~Oliveira}}, \bibinfo {author} {\bibfnamefont {I.}~\bibnamefont {Nape}},
  \bibinfo {author} {\bibfnamefont {J.}~\bibnamefont {Pinnell}}, \bibinfo
  {author} {\bibfnamefont {N.}~\bibnamefont {TabeBordbar}},\ and\ \bibinfo
  {author} {\bibfnamefont {A.}~\bibnamefont {Forbes}},\ }\bibfield  {title}
  {\bibinfo {title} {Experimental high-dimensional quantum secret sharing with
  spin-orbit-structured photons},\ }\href@noop {} {\bibfield  {journal}
  {\bibinfo  {journal} {Phys. Rev. A}\ }\textbf {\bibinfo {volume} {101}},\
  \bibinfo {pages} {042303} (\bibinfo {year} {2020})}\BibitemShut {NoStop}%
\bibitem [{\citenamefont {Li}\ \emph {et~al.}(2023{\natexlab{c}})\citenamefont
  {Li}, \citenamefont {Fu}, \citenamefont {Liu}, \citenamefont {Xie},
  \citenamefont {Li}, \citenamefont {Zhou}, \citenamefont {Yin},\ and\
  \citenamefont {Chen}}]{li_breakingqss_2023}%
  \BibitemOpen
  \bibfield  {author} {\bibinfo {author} {\bibfnamefont {C.-L.}\ \bibnamefont
  {Li}}, \bibinfo {author} {\bibfnamefont {Y.}~\bibnamefont {Fu}}, \bibinfo
  {author} {\bibfnamefont {W.-B.}\ \bibnamefont {Liu}}, \bibinfo {author}
  {\bibfnamefont {Y.-M.}\ \bibnamefont {Xie}}, \bibinfo {author} {\bibfnamefont
  {B.-H.}\ \bibnamefont {Li}}, \bibinfo {author} {\bibfnamefont {M.-G.}\
  \bibnamefont {Zhou}}, \bibinfo {author} {\bibfnamefont {H.-L.}\ \bibnamefont
  {Yin}},\ and\ \bibinfo {author} {\bibfnamefont {Z.-B.}\ \bibnamefont
  {Chen}},\ }\bibfield  {title} {\bibinfo {title} {Breaking the rate-distance
  limitation of measurement-device-independent quantum secret sharing},\ }\href
  {https://doi.org/10.1103/PhysRevResearch.5.033077} {\bibfield  {journal}
  {\bibinfo  {journal} {Phys. Rev. Res.}\ }\textbf {\bibinfo {volume} {5}},\
  \bibinfo {pages} {033077} (\bibinfo {year} {2023}{\natexlab{c}})}\BibitemShut
  {NoStop}%
\bibitem [{\citenamefont {Dunjko}\ \emph {et~al.}(2014)\citenamefont {Dunjko},
  \citenamefont {Wallden},\ and\ \citenamefont
  {Andersson}}]{dunjko2014quantum}%
  \BibitemOpen
  \bibfield  {author} {\bibinfo {author} {\bibfnamefont {V.}~\bibnamefont
  {Dunjko}}, \bibinfo {author} {\bibfnamefont {P.}~\bibnamefont {Wallden}},\
  and\ \bibinfo {author} {\bibfnamefont {E.}~\bibnamefont {Andersson}},\
  }\bibfield  {title} {\bibinfo {title} {Quantum digital signatures without
  quantum memory},\ }\href@noop {} {\bibfield  {journal} {\bibinfo  {journal}
  {Phys. Rev. Lett.}\ }\textbf {\bibinfo {volume} {112}},\ \bibinfo {pages}
  {040502} (\bibinfo {year} {2014})}\BibitemShut {NoStop}%
\bibitem [{\citenamefont {Yin}\ \emph {et~al.}(2016{\natexlab{a}})\citenamefont
  {Yin}, \citenamefont {Fu},\ and\ \citenamefont {Chen}}]{yin2016practical}%
  \BibitemOpen
  \bibfield  {author} {\bibinfo {author} {\bibfnamefont {H.-L.}\ \bibnamefont
  {Yin}}, \bibinfo {author} {\bibfnamefont {Y.}~\bibnamefont {Fu}},\ and\
  \bibinfo {author} {\bibfnamefont {Z.-B.}\ \bibnamefont {Chen}},\ }\bibfield
  {title} {\bibinfo {title} {Practical quantum digital signature},\ }\href@noop
  {} {\bibfield  {journal} {\bibinfo  {journal} {Phys. Rev. A}\ }\textbf
  {\bibinfo {volume} {93}},\ \bibinfo {pages} {032316} (\bibinfo {year}
  {2016}{\natexlab{a}})}\BibitemShut {NoStop}%
\bibitem [{\citenamefont {Qin}\ \emph {et~al.}(2022)\citenamefont {Qin},
  \citenamefont {Jiang}, \citenamefont {Yu},\ and\ \citenamefont
  {Wang}}]{qin2022quantum}%
  \BibitemOpen
  \bibfield  {author} {\bibinfo {author} {\bibfnamefont {J.-Q.}\ \bibnamefont
  {Qin}}, \bibinfo {author} {\bibfnamefont {C.}~\bibnamefont {Jiang}}, \bibinfo
  {author} {\bibfnamefont {Y.-L.}\ \bibnamefont {Yu}},\ and\ \bibinfo {author}
  {\bibfnamefont {X.-B.}\ \bibnamefont {Wang}},\ }\bibfield  {title} {\bibinfo
  {title} {Quantum digital signatures with random pairing},\ }\href@noop {}
  {\bibfield  {journal} {\bibinfo  {journal} {Phys. Rev. Appl.}\ }\textbf
  {\bibinfo {volume} {17}},\ \bibinfo {pages} {044047} (\bibinfo {year}
  {2022})}\BibitemShut {NoStop}%
\bibitem [{\citenamefont {Yin}\ \emph {et~al.}(2023)\citenamefont {Yin},
  \citenamefont {Fu}, \citenamefont {Li}, \citenamefont {Weng}, \citenamefont
  {Li}, \citenamefont {Gu}, \citenamefont {Lu}, \citenamefont {Huang},\ and\
  \citenamefont {Chen}}]{yin2023experimental}%
  \BibitemOpen
  \bibfield  {author} {\bibinfo {author} {\bibfnamefont {H.-L.}\ \bibnamefont
  {Yin}}, \bibinfo {author} {\bibfnamefont {Y.}~\bibnamefont {Fu}}, \bibinfo
  {author} {\bibfnamefont {C.-L.}\ \bibnamefont {Li}}, \bibinfo {author}
  {\bibfnamefont {C.-X.}\ \bibnamefont {Weng}}, \bibinfo {author}
  {\bibfnamefont {B.-H.}\ \bibnamefont {Li}}, \bibinfo {author} {\bibfnamefont
  {J.}~\bibnamefont {Gu}}, \bibinfo {author} {\bibfnamefont {Y.-S.}\
  \bibnamefont {Lu}}, \bibinfo {author} {\bibfnamefont {S.}~\bibnamefont
  {Huang}},\ and\ \bibinfo {author} {\bibfnamefont {Z.-B.}\ \bibnamefont
  {Chen}},\ }\bibfield  {title} {\bibinfo {title} {Experimental quantum secure
  network with digital signatures and encryption},\ }\href@noop {} {\bibfield
  {journal} {\bibinfo  {journal} {Natl. Sci. Rev.}\ }\textbf {\bibinfo {volume}
  {10}},\ \bibinfo {pages} {nwac228} (\bibinfo {year} {2023})}\BibitemShut
  {NoStop}%
\bibitem [{\citenamefont {Bennett}\ and\ \citenamefont
  {Brassard}(2014)}]{bennett_quantum_2014}%
  \BibitemOpen
  \bibfield  {author} {\bibinfo {author} {\bibfnamefont {C.~H.}\ \bibnamefont
  {Bennett}}\ and\ \bibinfo {author} {\bibfnamefont {G.}~\bibnamefont
  {Brassard}},\ }\bibfield  {title} {\bibinfo {title} {Quantum cryptography:
  {Public} key distribution and coin tossing},\ }\href
  {https://doi.org/10.1016/j.tcs.2014.05.025} {\bibfield  {journal} {\bibinfo
  {journal} {Theor Comput Sci}\ }\textbf {\bibinfo {volume} {560}},\ \bibinfo
  {pages} {7} (\bibinfo {year} {2014})}\BibitemShut {NoStop}%
\bibitem [{\citenamefont {Ekert}(1991)}]{ekert_quantum_1991}%
  \BibitemOpen
  \bibfield  {author} {\bibinfo {author} {\bibfnamefont {A.~K.}\ \bibnamefont
  {Ekert}},\ }\bibfield  {title} {\bibinfo {title} {Quantum cryptography based
  on bell's theorem},\ }\href {https://doi.org/10.1103/PhysRevLett.67.661}
  {\bibfield  {journal} {\bibinfo  {journal} {Phys. Rev. Lett.}\ }\textbf
  {\bibinfo {volume} {67}},\ \bibinfo {pages} {661} (\bibinfo {year}
  {1991})}\BibitemShut {NoStop}%
\bibitem [{\citenamefont {Scarani}\ \emph {et~al.}(2009)\citenamefont
  {Scarani}, \citenamefont {Bechmann-Pasquinucci}, \citenamefont {Cerf},
  \citenamefont {Du\~sek}, \citenamefont {L\"utkenhaus},\ and\ \citenamefont
  {Peev}}]{scarani_security_2009}%
  \BibitemOpen
  \bibfield  {author} {\bibinfo {author} {\bibfnamefont {V.}~\bibnamefont
  {Scarani}}, \bibinfo {author} {\bibfnamefont {H.}~\bibnamefont
  {Bechmann-Pasquinucci}}, \bibinfo {author} {\bibfnamefont {N.~J.}\
  \bibnamefont {Cerf}}, \bibinfo {author} {\bibfnamefont {M.}~\bibnamefont
  {Du\~sek}}, \bibinfo {author} {\bibfnamefont {N.}~\bibnamefont
  {L\"utkenhaus}},\ and\ \bibinfo {author} {\bibfnamefont {M.}~\bibnamefont
  {Peev}},\ }\bibfield  {title} {\bibinfo {title} {The security of practical
  quantum key distribution},\ }\href
  {https://doi.org/10.1103/RevModPhys.81.1301} {\bibfield  {journal} {\bibinfo
  {journal} {Rev. Mod. Phys.}\ }\textbf {\bibinfo {volume} {81}},\ \bibinfo
  {pages} {1301} (\bibinfo {year} {2009})}\BibitemShut {NoStop}%
\bibitem [{\citenamefont {Xu}\ \emph {et~al.}(2020)\citenamefont {Xu},
  \citenamefont {Ma}, \citenamefont {Zhang}, \citenamefont {Lo},\ and\
  \citenamefont {Pan}}]{xu_secure_2020}%
  \BibitemOpen
  \bibfield  {author} {\bibinfo {author} {\bibfnamefont {F.}~\bibnamefont
  {Xu}}, \bibinfo {author} {\bibfnamefont {X.}~\bibnamefont {Ma}}, \bibinfo
  {author} {\bibfnamefont {Q.}~\bibnamefont {Zhang}}, \bibinfo {author}
  {\bibfnamefont {H.-K.}\ \bibnamefont {Lo}},\ and\ \bibinfo {author}
  {\bibfnamefont {J.-W.}\ \bibnamefont {Pan}},\ }\bibfield  {title} {\bibinfo
  {title} {Secure quantum key distribution with realistic devices},\ }\href
  {https://doi.org/10.1103/RevModPhys.92.025002} {\bibfield  {journal}
  {\bibinfo  {journal} {Rev. Mod. Phys.}\ }\textbf {\bibinfo {volume} {92}},\
  \bibinfo {pages} {025002} (\bibinfo {year} {2020})}\BibitemShut {NoStop}%
\bibitem [{\citenamefont {Pirandola}\ \emph {et~al.}(2020)\citenamefont
  {Pirandola}, \citenamefont {Andersen}, \citenamefont {Banchi}, \citenamefont
  {Berta}, \citenamefont {Bunandar}, \citenamefont {Colbeck}, \citenamefont
  {Englund}, \citenamefont {Gehring}, \citenamefont {Lupo}, \citenamefont
  {Ottaviani}, \citenamefont {Pereira}, \citenamefont {Razavi}, \citenamefont
  {Shamsul~Shaari}, \citenamefont {Tomamichel}, \citenamefont {Usenko},
  \citenamefont {Vallone}, \citenamefont {Villoresi},\ and\ \citenamefont
  {Wallden}}]{pirandola_advances_2020}%
  \BibitemOpen
  \bibfield  {author} {\bibinfo {author} {\bibfnamefont {S.}~\bibnamefont
  {Pirandola}}, \bibinfo {author} {\bibfnamefont {U.~L.}\ \bibnamefont
  {Andersen}}, \bibinfo {author} {\bibfnamefont {L.}~\bibnamefont {Banchi}},
  \bibinfo {author} {\bibfnamefont {M.}~\bibnamefont {Berta}}, \bibinfo
  {author} {\bibfnamefont {D.}~\bibnamefont {Bunandar}}, \bibinfo {author}
  {\bibfnamefont {R.}~\bibnamefont {Colbeck}}, \bibinfo {author} {\bibfnamefont
  {D.}~\bibnamefont {Englund}}, \bibinfo {author} {\bibfnamefont
  {T.}~\bibnamefont {Gehring}}, \bibinfo {author} {\bibfnamefont
  {C.}~\bibnamefont {Lupo}}, \bibinfo {author} {\bibfnamefont {C.}~\bibnamefont
  {Ottaviani}}, \bibinfo {author} {\bibfnamefont {J.~L.}\ \bibnamefont
  {Pereira}}, \bibinfo {author} {\bibfnamefont {M.}~\bibnamefont {Razavi}},
  \bibinfo {author} {\bibfnamefont {J.}~\bibnamefont {Shamsul~Shaari}},
  \bibinfo {author} {\bibfnamefont {M.}~\bibnamefont {Tomamichel}}, \bibinfo
  {author} {\bibfnamefont {V.~C.}\ \bibnamefont {Usenko}}, \bibinfo {author}
  {\bibfnamefont {G.}~\bibnamefont {Vallone}}, \bibinfo {author} {\bibfnamefont
  {P.}~\bibnamefont {Villoresi}},\ and\ \bibinfo {author} {\bibfnamefont
  {P.}~\bibnamefont {Wallden}},\ }\bibfield  {title} {\bibinfo {title}
  {Advances in quantum cryptography},\ }\href
  {https://doi.org/10.1364/AOP.361502} {\bibfield  {journal} {\bibinfo
  {journal} {Adv. Opt. Photonics}\ }\textbf {\bibinfo {volume} {12}},\ \bibinfo
  {pages} {1012} (\bibinfo {year} {2020})}\BibitemShut {NoStop}%
\bibitem [{\citenamefont {Lo}\ \emph {et~al.}(2012)\citenamefont {Lo},
  \citenamefont {Curty},\ and\ \citenamefont
  {Qi}}]{lo_measurement-device-independent_2012}%
  \BibitemOpen
  \bibfield  {author} {\bibinfo {author} {\bibfnamefont {H.-K.}\ \bibnamefont
  {Lo}}, \bibinfo {author} {\bibfnamefont {M.}~\bibnamefont {Curty}},\ and\
  \bibinfo {author} {\bibfnamefont {B.}~\bibnamefont {Qi}},\ }\bibfield
  {title} {\bibinfo {title} {Measurement-{device}-{independent} {quantum} {key}
  {distribution}},\ }\href {https://doi.org/10.1103/PhysRevLett.108.130503}
  {\bibfield  {journal} {\bibinfo  {journal} {Phys. Rev. Lett.}\ }\textbf
  {\bibinfo {volume} {108}},\ \bibinfo {pages} {130503} (\bibinfo {year}
  {2012})}\BibitemShut {NoStop}%
\bibitem [{\citenamefont {Braunstein}\ and\ \citenamefont
  {Pirandola}(2012)}]{braunstein_side-channel-free_2012}%
  \BibitemOpen
  \bibfield  {author} {\bibinfo {author} {\bibfnamefont {S.~L.}\ \bibnamefont
  {Braunstein}}\ and\ \bibinfo {author} {\bibfnamefont {S.}~\bibnamefont
  {Pirandola}},\ }\bibfield  {title} {\bibinfo {title} {Side-{channel}-{free}
  {quantum} {key} {distribution}},\ }\href
  {https://doi.org/10.1103/PhysRevLett.108.130502} {\bibfield  {journal}
  {\bibinfo  {journal} {Phys. Rev. Lett.}\ }\textbf {\bibinfo {volume} {108}},\
  \bibinfo {pages} {130502} (\bibinfo {year} {2012})}\BibitemShut {NoStop}%
\bibitem [{\citenamefont {Lydersen}\ \emph {et~al.}(2010)\citenamefont
  {Lydersen}, \citenamefont {Wiechers}, \citenamefont {Wittmann}, \citenamefont
  {Elser}, \citenamefont {Skaar},\ and\ \citenamefont
  {Makarov}}]{lydersen_hacking_2010}%
  \BibitemOpen
  \bibfield  {author} {\bibinfo {author} {\bibfnamefont {L.}~\bibnamefont
  {Lydersen}}, \bibinfo {author} {\bibfnamefont {C.}~\bibnamefont {Wiechers}},
  \bibinfo {author} {\bibfnamefont {C.}~\bibnamefont {Wittmann}}, \bibinfo
  {author} {\bibfnamefont {D.}~\bibnamefont {Elser}}, \bibinfo {author}
  {\bibfnamefont {J.}~\bibnamefont {Skaar}},\ and\ \bibinfo {author}
  {\bibfnamefont {V.}~\bibnamefont {Makarov}},\ }\bibfield  {title} {\bibinfo
  {title} {Hacking commercial quantum cryptography systems by tailored bright
  illumination},\ }\href {https://doi.org/10.1038/nphoton.2010.214} {\bibfield
  {journal} {\bibinfo  {journal} {Nat. Photonics}\ }\textbf {\bibinfo {volume}
  {4}},\ \bibinfo {pages} {686} (\bibinfo {year} {2010})}\BibitemShut {NoStop}%
\bibitem [{\citenamefont {Yin}\ \emph {et~al.}(2016{\natexlab{b}})\citenamefont
  {Yin}, \citenamefont {Chen}, \citenamefont {Yu}, \citenamefont {Liu},
  \citenamefont {You}, \citenamefont {Zhou}, \citenamefont {Chen},
  \citenamefont {Mao}, \citenamefont {Huang}, \citenamefont {Zhang},
  \citenamefont {Chen}, \citenamefont {Li}, \citenamefont {Nolan},
  \citenamefont {Zhou}, \citenamefont {Jiang}, \citenamefont {Wang},
  \citenamefont {Zhang}, \citenamefont {Wang},\ and\ \citenamefont
  {Pan}}]{yin_measurement-device-independent_2016}%
  \BibitemOpen
  \bibfield  {author} {\bibinfo {author} {\bibfnamefont {H.-L.}\ \bibnamefont
  {Yin}}, \bibinfo {author} {\bibfnamefont {T.-Y.}\ \bibnamefont {Chen}},
  \bibinfo {author} {\bibfnamefont {Z.-W.}\ \bibnamefont {Yu}}, \bibinfo
  {author} {\bibfnamefont {H.}~\bibnamefont {Liu}}, \bibinfo {author}
  {\bibfnamefont {L.-X.}\ \bibnamefont {You}}, \bibinfo {author} {\bibfnamefont
  {Y.-H.}\ \bibnamefont {Zhou}}, \bibinfo {author} {\bibfnamefont {S.-J.}\
  \bibnamefont {Chen}}, \bibinfo {author} {\bibfnamefont {Y.}~\bibnamefont
  {Mao}}, \bibinfo {author} {\bibfnamefont {M.-Q.}\ \bibnamefont {Huang}},
  \bibinfo {author} {\bibfnamefont {W.-J.}\ \bibnamefont {Zhang}}, \bibinfo
  {author} {\bibfnamefont {H.}~\bibnamefont {Chen}}, \bibinfo {author}
  {\bibfnamefont {M.~J.}\ \bibnamefont {Li}}, \bibinfo {author} {\bibfnamefont
  {D.}~\bibnamefont {Nolan}}, \bibinfo {author} {\bibfnamefont
  {F.}~\bibnamefont {Zhou}}, \bibinfo {author} {\bibfnamefont {X.}~\bibnamefont
  {Jiang}}, \bibinfo {author} {\bibfnamefont {Z.}~\bibnamefont {Wang}},
  \bibinfo {author} {\bibfnamefont {Q.}~\bibnamefont {Zhang}}, \bibinfo
  {author} {\bibfnamefont {X.-B.}\ \bibnamefont {Wang}},\ and\ \bibinfo
  {author} {\bibfnamefont {J.-W.}\ \bibnamefont {Pan}},\ }\bibfield  {title}
  {\bibinfo {title} {Measurement-{device}-{independent} {quantum} {key}
  {distribution} {over} a 404 km {optical} {fiber}},\ }\href
  {https://doi.org/10.1103/PhysRevLett.117.190501} {\bibfield  {journal}
  {\bibinfo  {journal} {Phys. Rev. Lett.}\ }\textbf {\bibinfo {volume} {117}},\
  \bibinfo {pages} {190501} (\bibinfo {year} {2016}{\natexlab{b}})}\BibitemShut
  {NoStop}%
\bibitem [{\citenamefont {Zhou}\ \emph {et~al.}(2016)\citenamefont {Zhou},
  \citenamefont {Yu},\ and\ \citenamefont {Wang}}]{zhou_making_2016}%
  \BibitemOpen
  \bibfield  {author} {\bibinfo {author} {\bibfnamefont {Y.-H.}\ \bibnamefont
  {Zhou}}, \bibinfo {author} {\bibfnamefont {Z.-W.}\ \bibnamefont {Yu}},\ and\
  \bibinfo {author} {\bibfnamefont {X.-B.}\ \bibnamefont {Wang}},\ }\bibfield
  {title} {\bibinfo {title} {Making the decoy-state
  measurement-device-independent quantum key distribution practically useful},\
  }\href {https://doi.org/10.1103/PhysRevA.93.042324} {\bibfield  {journal}
  {\bibinfo  {journal} {Phys. Rev. A}\ }\textbf {\bibinfo {volume} {93}},\
  \bibinfo {pages} {042324} (\bibinfo {year} {2016})}\BibitemShut {NoStop}%
\bibitem [{\citenamefont {Pirandola}\ \emph {et~al.}(2009)\citenamefont
  {Pirandola}, \citenamefont {Garc\'ia-Patr\'on}, \citenamefont {Braunstein},\
  and\ \citenamefont {Lloyd}}]{pirandola_direct_2009}%
  \BibitemOpen
  \bibfield  {author} {\bibinfo {author} {\bibfnamefont {S.}~\bibnamefont
  {Pirandola}}, \bibinfo {author} {\bibfnamefont {R.}~\bibnamefont
  {Garc\'ia-Patr\'on}}, \bibinfo {author} {\bibfnamefont {S.~L.}\ \bibnamefont
  {Braunstein}},\ and\ \bibinfo {author} {\bibfnamefont {S.}~\bibnamefont
  {Lloyd}},\ }\bibfield  {title} {\bibinfo {title} {Direct and {reverse}
  {secret}-{key} {capacities} of a {quantum} {channel}},\ }\href
  {https://doi.org/10.1103/PhysRevLett.102.050503} {\bibfield  {journal}
  {\bibinfo  {journal} {Phys. Rev. Lett.}\ }\textbf {\bibinfo {volume} {102}},\
  \bibinfo {pages} {050503} (\bibinfo {year} {2009})}\BibitemShut {NoStop}%
\bibitem [{\citenamefont {Takeoka}\ \emph {et~al.}(2014)\citenamefont
  {Takeoka}, \citenamefont {Guha},\ and\ \citenamefont
  {Wilde}}]{takeoka_fundamental_2014}%
  \BibitemOpen
  \bibfield  {author} {\bibinfo {author} {\bibfnamefont {M.}~\bibnamefont
  {Takeoka}}, \bibinfo {author} {\bibfnamefont {S.}~\bibnamefont {Guha}},\ and\
  \bibinfo {author} {\bibfnamefont {M.~M.}\ \bibnamefont {Wilde}},\ }\bibfield
  {title} {\bibinfo {title} {Fundamental rate-loss tradeoff for optical quantum
  key distribution},\ }\href {https://doi.org/10.1038/ncomms6235} {\bibfield
  {journal} {\bibinfo  {journal} {Nat. Commun.}\ }\textbf {\bibinfo {volume}
  {5}},\ \bibinfo {pages} {5235} (\bibinfo {year} {2014})}\BibitemShut
  {NoStop}%
\bibitem [{\citenamefont {Pirandola}\ \emph {et~al.}(2017)\citenamefont
  {Pirandola}, \citenamefont {Laurenza}, \citenamefont {Ottaviani},\ and\
  \citenamefont {Banchi}}]{pirandola_fundamental_2017}%
  \BibitemOpen
  \bibfield  {author} {\bibinfo {author} {\bibfnamefont {S.}~\bibnamefont
  {Pirandola}}, \bibinfo {author} {\bibfnamefont {R.}~\bibnamefont {Laurenza}},
  \bibinfo {author} {\bibfnamefont {C.}~\bibnamefont {Ottaviani}},\ and\
  \bibinfo {author} {\bibfnamefont {L.}~\bibnamefont {Banchi}},\ }\bibfield
  {title} {\bibinfo {title} {Fundamental limits of repeaterless quantum
  communications},\ }\href {https://doi.org/10.1038/ncomms15043} {\bibfield
  {journal} {\bibinfo  {journal} {Nat. Commun.}\ }\textbf {\bibinfo {volume}
  {8}},\ \bibinfo {pages} {15043} (\bibinfo {year} {2017})}\BibitemShut
  {NoStop}%
\bibitem [{\citenamefont {Das}\ \emph {et~al.}(2021)\citenamefont {Das},
  \citenamefont {B\"auml}, \citenamefont {Winczewski},\ and\ \citenamefont
  {Horodecki}}]{das_unversal_2021}%
  \BibitemOpen
  \bibfield  {author} {\bibinfo {author} {\bibfnamefont {S.}~\bibnamefont
  {Das}}, \bibinfo {author} {\bibfnamefont {S.}~\bibnamefont {B\"auml}},
  \bibinfo {author} {\bibfnamefont {M.}~\bibnamefont {Winczewski}},\ and\
  \bibinfo {author} {\bibfnamefont {K.}~\bibnamefont {Horodecki}},\ }\bibfield
  {title} {\bibinfo {title} {Universal limitations on quantum key distribution
  over a network},\ }\href {https://doi.org/10.1103/PhysRevX.11.041016}
  {\bibfield  {journal} {\bibinfo  {journal} {Phys. Rev. X}\ }\textbf {\bibinfo
  {volume} {11}},\ \bibinfo {pages} {041016} (\bibinfo {year}
  {2021})}\BibitemShut {NoStop}%
\bibitem [{\citenamefont {Lucamarini}\ \emph {et~al.}(2018)\citenamefont
  {Lucamarini}, \citenamefont {Yuan}, \citenamefont {Dynes},\ and\
  \citenamefont {Shields}}]{lucamarini_overcoming_2018}%
  \BibitemOpen
  \bibfield  {author} {\bibinfo {author} {\bibfnamefont {M.}~\bibnamefont
  {Lucamarini}}, \bibinfo {author} {\bibfnamefont {Z.~L.}\ \bibnamefont
  {Yuan}}, \bibinfo {author} {\bibfnamefont {J.~F.}\ \bibnamefont {Dynes}},\
  and\ \bibinfo {author} {\bibfnamefont {A.~J.}\ \bibnamefont {Shields}},\
  }\bibfield  {title} {\bibinfo {title} {Overcoming the rate-distance limit of
  quantum key distribution without quantum repeaters},\ }\href@noop {}
  {\bibfield  {journal} {\bibinfo  {journal} {Nature}\ }\textbf {\bibinfo
  {volume} {557}},\ \bibinfo {pages} {400} (\bibinfo {year}
  {2018})}\BibitemShut {NoStop}%
\bibitem [{\citenamefont {Ma}\ \emph {et~al.}(2018)\citenamefont {Ma},
  \citenamefont {Zeng},\ and\ \citenamefont {Zhou}}]{ma_phase-matching_2018}%
  \BibitemOpen
  \bibfield  {author} {\bibinfo {author} {\bibfnamefont {X.}~\bibnamefont
  {Ma}}, \bibinfo {author} {\bibfnamefont {P.}~\bibnamefont {Zeng}},\ and\
  \bibinfo {author} {\bibfnamefont {H.}~\bibnamefont {Zhou}},\ }\bibfield
  {title} {\bibinfo {title} {Phase-{matching} {quantum} {key} {distribution}},\
  }\href {https://doi.org/10.1103/PhysRevX.8.031043} {\bibfield  {journal}
  {\bibinfo  {journal} {Phys. Rev. X}\ }\textbf {\bibinfo {volume} {8}},\
  \bibinfo {pages} {031043} (\bibinfo {year} {2018})}\BibitemShut {NoStop}%
\bibitem [{\citenamefont {Wang}\ \emph {et~al.}(2018)\citenamefont {Wang},
  \citenamefont {Yu},\ and\ \citenamefont {Hu}}]{wang_twin-field_2018}%
  \BibitemOpen
  \bibfield  {author} {\bibinfo {author} {\bibfnamefont {X.-B.}\ \bibnamefont
  {Wang}}, \bibinfo {author} {\bibfnamefont {Z.-W.}\ \bibnamefont {Yu}},\ and\
  \bibinfo {author} {\bibfnamefont {X.-L.}\ \bibnamefont {Hu}},\ }\bibfield
  {title} {\bibinfo {title} {Twin-field quantum key distribution with large
  misalignment error},\ }\href {https://doi.org/10.1103/PhysRevA.98.062323}
  {\bibfield  {journal} {\bibinfo  {journal} {Phys. Rev. A}\ }\textbf {\bibinfo
  {volume} {98}},\ \bibinfo {pages} {062323} (\bibinfo {year}
  {2018})}\BibitemShut {NoStop}%
\bibitem [{\citenamefont {Yin}\ and\ \citenamefont
  {Fu}(2019)}]{yin_measurement-device-independent_2019}%
  \BibitemOpen
  \bibfield  {author} {\bibinfo {author} {\bibfnamefont {H.-L.}\ \bibnamefont
  {Yin}}\ and\ \bibinfo {author} {\bibfnamefont {Y.}~\bibnamefont {Fu}},\
  }\bibfield  {title} {\bibinfo {title} {Measurement-{device}-{independent}
  {twin}-{field} {quantum} {key} {distribution}},\ }\href
  {https://doi.org/10.1038/s41598-019-39454-1} {\bibfield  {journal} {\bibinfo
  {journal} {Sci. Rep.}\ }\textbf {\bibinfo {volume} {9}},\ \bibinfo {pages}
  {3045} (\bibinfo {year} {2019})}\BibitemShut {NoStop}%
\bibitem [{\citenamefont {Lin}\ and\ \citenamefont
  {L\"utkenhaus}(2018)}]{lin_simple_2018}%
  \BibitemOpen
  \bibfield  {author} {\bibinfo {author} {\bibfnamefont {J.}~\bibnamefont
  {Lin}}\ and\ \bibinfo {author} {\bibfnamefont {N.}~\bibnamefont
  {L\"utkenhaus}},\ }\bibfield  {title} {\bibinfo {title} {Simple security
  analysis of phase-matching measurement-device-independent quantum key
  distribution},\ }\href {https://doi.org/10.1103/PhysRevA.98.042332}
  {\bibfield  {journal} {\bibinfo  {journal} {Phys. Rev. A}\ }\textbf {\bibinfo
  {volume} {98}},\ \bibinfo {pages} {042332} (\bibinfo {year}
  {2018})}\BibitemShut {NoStop}%
\bibitem [{\citenamefont {Cui}\ \emph {et~al.}(2019)\citenamefont {Cui},
  \citenamefont {Yin}, \citenamefont {Wang}, \citenamefont {Chen},
  \citenamefont {Wang}, \citenamefont {Guo},\ and\ \citenamefont
  {Han}}]{cui_twin-field_2019}%
  \BibitemOpen
  \bibfield  {author} {\bibinfo {author} {\bibfnamefont {C.}~\bibnamefont
  {Cui}}, \bibinfo {author} {\bibfnamefont {Z.-Q.}\ \bibnamefont {Yin}},
  \bibinfo {author} {\bibfnamefont {R.}~\bibnamefont {Wang}}, \bibinfo {author}
  {\bibfnamefont {W.}~\bibnamefont {Chen}}, \bibinfo {author} {\bibfnamefont
  {S.}~\bibnamefont {Wang}}, \bibinfo {author} {\bibfnamefont {G.-C.}\
  \bibnamefont {Guo}},\ and\ \bibinfo {author} {\bibfnamefont {Z.-F.}\
  \bibnamefont {Han}},\ }\bibfield  {title} {\bibinfo {title} {Twin-{field}
  {quantum} {key} {distribution} without {phase} {postselection}},\ }\href
  {https://doi.org/10.1103/PhysRevApplied.11.034053} {\bibfield  {journal}
  {\bibinfo  {journal} {Phys. Rev. Appl.}\ }\textbf {\bibinfo {volume} {11}},\
  \bibinfo {pages} {034053} (\bibinfo {year} {2019})}\BibitemShut {NoStop}%
\bibitem [{\citenamefont {Curty}\ \emph {et~al.}(2019)\citenamefont {Curty},
  \citenamefont {Azuma},\ and\ \citenamefont {Lo}}]{curty_simple_2019}%
  \BibitemOpen
  \bibfield  {author} {\bibinfo {author} {\bibfnamefont {M.}~\bibnamefont
  {Curty}}, \bibinfo {author} {\bibfnamefont {K.}~\bibnamefont {Azuma}},\ and\
  \bibinfo {author} {\bibfnamefont {H.-K.}\ \bibnamefont {Lo}},\ }\bibfield
  {title} {\bibinfo {title} {Simple security proof of twin-field type quantum
  key distribution protocol},\ }\href
  {https://doi.org/10.1038/s41534-019-0175-6} {\bibfield  {journal} {\bibinfo
  {journal} {npj Quantum Inf.}\ }\textbf {\bibinfo {volume} {5}},\ \bibinfo
  {pages} {64} (\bibinfo {year} {2019})}\BibitemShut {NoStop}%
\bibitem [{\citenamefont {Wang}\ \emph {et~al.}(2022)\citenamefont {Wang},
  \citenamefont {Yin}, \citenamefont {He}, \citenamefont {Chen}, \citenamefont
  {Wang}, \citenamefont {Ye}, \citenamefont {Zhou}, \citenamefont {Fan-Yuan},
  \citenamefont {Wang}, \citenamefont {Chen}, \citenamefont {Zhu},
  \citenamefont {Morozov}, \citenamefont {Divochiy}, \citenamefont {Zhou},
  \citenamefont {Guo},\ and\ \citenamefont {Han}}]{wang_twin-field_2022}%
  \BibitemOpen
  \bibfield  {author} {\bibinfo {author} {\bibfnamefont {S.}~\bibnamefont
  {Wang}}, \bibinfo {author} {\bibfnamefont {Z.-Q.}\ \bibnamefont {Yin}},
  \bibinfo {author} {\bibfnamefont {D.-Y.}\ \bibnamefont {He}}, \bibinfo
  {author} {\bibfnamefont {W.}~\bibnamefont {Chen}}, \bibinfo {author}
  {\bibfnamefont {R.-Q.}\ \bibnamefont {Wang}}, \bibinfo {author}
  {\bibfnamefont {P.}~\bibnamefont {Ye}}, \bibinfo {author} {\bibfnamefont
  {Y.}~\bibnamefont {Zhou}}, \bibinfo {author} {\bibfnamefont {G.-J.}\
  \bibnamefont {Fan-Yuan}}, \bibinfo {author} {\bibfnamefont {F.-X.}\
  \bibnamefont {Wang}}, \bibinfo {author} {\bibfnamefont {W.}~\bibnamefont
  {Chen}}, \bibinfo {author} {\bibfnamefont {Y.-G.}\ \bibnamefont {Zhu}},
  \bibinfo {author} {\bibfnamefont {P.~V.}\ \bibnamefont {Morozov}}, \bibinfo
  {author} {\bibfnamefont {A.~V.}\ \bibnamefont {Divochiy}}, \bibinfo {author}
  {\bibfnamefont {Z.}~\bibnamefont {Zhou}}, \bibinfo {author} {\bibfnamefont
  {G.-C.}\ \bibnamefont {Guo}},\ and\ \bibinfo {author} {\bibfnamefont {Z.-F.}\
  \bibnamefont {Han}},\ }\bibfield  {title} {\bibinfo {title} {Twin-field
  quantum key distribution over 830-km fibre},\ }\href
  {https://doi.org/10.1038/s41566-021-00928-2} {\bibfield  {journal} {\bibinfo
  {journal} {Nat. Photonics}\ }\textbf {\bibinfo {volume} {16}},\ \bibinfo
  {pages} {154} (\bibinfo {year} {2022})}\BibitemShut {NoStop}%
\bibitem [{\citenamefont {Liu}\ \emph {et~al.}(2023)\citenamefont {Liu},
  \citenamefont {Zhang}, \citenamefont {Jiang}, \citenamefont {Chen},
  \citenamefont {Zhang}, \citenamefont {Pan}, \citenamefont {Ma}, \citenamefont
  {Dong}, \citenamefont {Xiong}, \citenamefont {Zhang}, \citenamefont {Li},
  \citenamefont {Wang}, \citenamefont {Wu}, \citenamefont {Chen}, \citenamefont
  {You}, \citenamefont {Wang}, \citenamefont {Zhang},\ and\ \citenamefont
  {Pan}}]{liu2023experimental}%
  \BibitemOpen
  \bibfield  {author} {\bibinfo {author} {\bibfnamefont {Y.}~\bibnamefont
  {Liu}}, \bibinfo {author} {\bibfnamefont {W.-J.}\ \bibnamefont {Zhang}},
  \bibinfo {author} {\bibfnamefont {C.}~\bibnamefont {Jiang}}, \bibinfo
  {author} {\bibfnamefont {J.-P.}\ \bibnamefont {Chen}}, \bibinfo {author}
  {\bibfnamefont {C.}~\bibnamefont {Zhang}}, \bibinfo {author} {\bibfnamefont
  {W.-X.}\ \bibnamefont {Pan}}, \bibinfo {author} {\bibfnamefont
  {D.}~\bibnamefont {Ma}}, \bibinfo {author} {\bibfnamefont {H.}~\bibnamefont
  {Dong}}, \bibinfo {author} {\bibfnamefont {J.-M.}\ \bibnamefont {Xiong}},
  \bibinfo {author} {\bibfnamefont {C.-J.}\ \bibnamefont {Zhang}}, \bibinfo
  {author} {\bibfnamefont {H.}~\bibnamefont {Li}}, \bibinfo {author}
  {\bibfnamefont {R.-C.}\ \bibnamefont {Wang}}, \bibinfo {author}
  {\bibfnamefont {J.}~\bibnamefont {Wu}}, \bibinfo {author} {\bibfnamefont
  {T.-Y.}\ \bibnamefont {Chen}}, \bibinfo {author} {\bibfnamefont
  {L.}~\bibnamefont {You}}, \bibinfo {author} {\bibfnamefont {X.-B.}\
  \bibnamefont {Wang}}, \bibinfo {author} {\bibfnamefont {Q.}~\bibnamefont
  {Zhang}},\ and\ \bibinfo {author} {\bibfnamefont {J.-W.}\ \bibnamefont
  {Pan}},\ }\bibfield  {title} {\bibinfo {title} {Experimental twin-field
  quantum key distribution over 1000 km fiber distance},\ }\href@noop {}
  {\bibfield  {journal} {\bibinfo  {journal} {Phys. Rev. Lett.}\ }\textbf
  {\bibinfo {volume} {130}},\ \bibinfo {pages} {210801} (\bibinfo {year}
  {2023})}\BibitemShut {NoStop}%
\bibitem [{\citenamefont {Xie}\ \emph {et~al.}(2022)\citenamefont {Xie},
  \citenamefont {Lu}, \citenamefont {Weng}, \citenamefont {Cao}, \citenamefont
  {Jia}, \citenamefont {Bao}, \citenamefont {Wang}, \citenamefont {Fu},
  \citenamefont {Yin},\ and\ \citenamefont {Chen}}]{xie_breaking_2022}%
  \BibitemOpen
  \bibfield  {author} {\bibinfo {author} {\bibfnamefont {Y.-M.}\ \bibnamefont
  {Xie}}, \bibinfo {author} {\bibfnamefont {Y.-S.}\ \bibnamefont {Lu}},
  \bibinfo {author} {\bibfnamefont {C.-X.}\ \bibnamefont {Weng}}, \bibinfo
  {author} {\bibfnamefont {X.-Y.}\ \bibnamefont {Cao}}, \bibinfo {author}
  {\bibfnamefont {Z.-Y.}\ \bibnamefont {Jia}}, \bibinfo {author} {\bibfnamefont
  {Y.}~\bibnamefont {Bao}}, \bibinfo {author} {\bibfnamefont {Y.}~\bibnamefont
  {Wang}}, \bibinfo {author} {\bibfnamefont {Y.}~\bibnamefont {Fu}}, \bibinfo
  {author} {\bibfnamefont {H.-L.}\ \bibnamefont {Yin}},\ and\ \bibinfo {author}
  {\bibfnamefont {Z.-B.}\ \bibnamefont {Chen}},\ }\bibfield  {title} {\bibinfo
  {title} {Breaking the {rate}-{loss} {bound} of {quantum} {key} {distribution}
  with {asynchronous} {two}-{photon} {interference}},\ }\href
  {https://doi.org/10.1103/PRXQuantum.3.020315} {\bibfield  {journal} {\bibinfo
   {journal} {PRX Quantum}\ }\textbf {\bibinfo {volume} {3}},\ \bibinfo {pages}
  {020315} (\bibinfo {year} {2022})}\BibitemShut {NoStop}%
\bibitem [{\citenamefont {Zeng}\ \emph {et~al.}(2022)\citenamefont {Zeng},
  \citenamefont {Zhou}, \citenamefont {Wu},\ and\ \citenamefont
  {Ma}}]{zeng2022mode}%
  \BibitemOpen
  \bibfield  {author} {\bibinfo {author} {\bibfnamefont {P.}~\bibnamefont
  {Zeng}}, \bibinfo {author} {\bibfnamefont {H.}~\bibnamefont {Zhou}}, \bibinfo
  {author} {\bibfnamefont {W.}~\bibnamefont {Wu}},\ and\ \bibinfo {author}
  {\bibfnamefont {X.}~\bibnamefont {Ma}},\ }\bibfield  {title} {\bibinfo
  {title} {Mode-pairing quantum key distribution},\ }\href@noop {} {\bibfield
  {journal} {\bibinfo  {journal} {Nat. Commun.}\ }\textbf {\bibinfo {volume}
  {13}},\ \bibinfo {pages} {3903} (\bibinfo {year} {2022})}\BibitemShut
  {NoStop}%
\bibitem [{\citenamefont {Zhu}\ \emph {et~al.}(2023)\citenamefont {Zhu},
  \citenamefont {Huang}, \citenamefont {Liu}, \citenamefont {Zeng},
  \citenamefont {Zou}, \citenamefont {Dai}, \citenamefont {Tang}, \citenamefont
  {Li}, \citenamefont {You}, \citenamefont {Wang}, \citenamefont {Chen},
  \citenamefont {Ma}, \citenamefont {Chen},\ and\ \citenamefont
  {Pan}}]{PhysRevLett.130.030801}%
  \BibitemOpen
  \bibfield  {author} {\bibinfo {author} {\bibfnamefont {H.-T.}\ \bibnamefont
  {Zhu}}, \bibinfo {author} {\bibfnamefont {Y.}~\bibnamefont {Huang}}, \bibinfo
  {author} {\bibfnamefont {H.}~\bibnamefont {Liu}}, \bibinfo {author}
  {\bibfnamefont {P.}~\bibnamefont {Zeng}}, \bibinfo {author} {\bibfnamefont
  {M.}~\bibnamefont {Zou}}, \bibinfo {author} {\bibfnamefont {Y.}~\bibnamefont
  {Dai}}, \bibinfo {author} {\bibfnamefont {S.}~\bibnamefont {Tang}}, \bibinfo
  {author} {\bibfnamefont {H.}~\bibnamefont {Li}}, \bibinfo {author}
  {\bibfnamefont {L.}~\bibnamefont {You}}, \bibinfo {author} {\bibfnamefont
  {Z.}~\bibnamefont {Wang}}, \bibinfo {author} {\bibfnamefont {Y.-A.}\
  \bibnamefont {Chen}}, \bibinfo {author} {\bibfnamefont {X.}~\bibnamefont
  {Ma}}, \bibinfo {author} {\bibfnamefont {T.-Y.}\ \bibnamefont {Chen}},\ and\
  \bibinfo {author} {\bibfnamefont {J.-W.}\ \bibnamefont {Pan}},\ }\bibfield
  {title} {\bibinfo {title} {Experimental mode-pairing
  measurement-device-independent quantum key distribution without global phase
  locking},\ }\href {https://doi.org/10.1103/PhysRevLett.130.030801} {\bibfield
   {journal} {\bibinfo  {journal} {Phys. Rev. Lett.}\ }\textbf {\bibinfo
  {volume} {130}},\ \bibinfo {pages} {030801} (\bibinfo {year}
  {2023})}\BibitemShut {NoStop}%
\bibitem [{\citenamefont {Zhou}\ \emph {et~al.}(2023)\citenamefont {Zhou},
  \citenamefont {Lin}, \citenamefont {Xie}, \citenamefont {Lu}, \citenamefont
  {Jing}, \citenamefont {Yin},\ and\ \citenamefont
  {Yuan}}]{zhou_experimental_2023}%
  \BibitemOpen
  \bibfield  {author} {\bibinfo {author} {\bibfnamefont {L.}~\bibnamefont
  {Zhou}}, \bibinfo {author} {\bibfnamefont {J.}~\bibnamefont {Lin}}, \bibinfo
  {author} {\bibfnamefont {Y.-M.}\ \bibnamefont {Xie}}, \bibinfo {author}
  {\bibfnamefont {Y.-S.}\ \bibnamefont {Lu}}, \bibinfo {author} {\bibfnamefont
  {Y.}~\bibnamefont {Jing}}, \bibinfo {author} {\bibfnamefont {H.-L.}\
  \bibnamefont {Yin}},\ and\ \bibinfo {author} {\bibfnamefont {Z.}~\bibnamefont
  {Yuan}},\ }\bibfield  {title} {\bibinfo {title} {Experimental {quantum}
  {communication} {overcomes} the {rate}-{loss} {limit} without {global}
  {phase} {tracking}},\ }\href {https://doi.org/10.1103/PhysRevLett.130.250801}
  {\bibfield  {journal} {\bibinfo  {journal} {Phys. Rev. Lett.}\ }\textbf
  {\bibinfo {volume} {130}},\ \bibinfo {pages} {250801} (\bibinfo {year}
  {2023})}\BibitemShut {NoStop}%
\bibitem [{\citenamefont {Bai}\ \emph {et~al.}(2023)\citenamefont {Bai},
  \citenamefont {Xie}, \citenamefont {Fu}, \citenamefont {Yin},\ and\
  \citenamefont {Chen}}]{bai_asynchronous_2023}%
  \BibitemOpen
  \bibfield  {author} {\bibinfo {author} {\bibfnamefont {J.-L.}\ \bibnamefont
  {Bai}}, \bibinfo {author} {\bibfnamefont {Y.-M.}\ \bibnamefont {Xie}},
  \bibinfo {author} {\bibfnamefont {Y.}~\bibnamefont {Fu}}, \bibinfo {author}
  {\bibfnamefont {H.-L.}\ \bibnamefont {Yin}},\ and\ \bibinfo {author}
  {\bibfnamefont {Z.-B.}\ \bibnamefont {Chen}},\ }\bibfield  {title} {\bibinfo
  {title} {Asynchronous measurement-device-independent quantum key distribution
  with hybrid source},\ }\href {https://doi.org/10.1364/OL.491511} {\bibfield
  {journal} {\bibinfo  {journal} {Opt. Lett.}\ }\textbf {\bibinfo {volume}
  {48}},\ \bibinfo {pages} {3551} (\bibinfo {year} {2023})}\BibitemShut
  {NoStop}%
\bibitem [{\citenamefont {Xie}\ \emph {et~al.}(2023)\citenamefont {Xie},
  \citenamefont {Bai}, \citenamefont {Lu}, \citenamefont {Weng}, \citenamefont
  {Yin},\ and\ \citenamefont {Chen}}]{xie_advantages_2023}%
  \BibitemOpen
  \bibfield  {author} {\bibinfo {author} {\bibfnamefont {Y.-M.}\ \bibnamefont
  {Xie}}, \bibinfo {author} {\bibfnamefont {J.-L.}\ \bibnamefont {Bai}},
  \bibinfo {author} {\bibfnamefont {Y.-S.}\ \bibnamefont {Lu}}, \bibinfo
  {author} {\bibfnamefont {C.-X.}\ \bibnamefont {Weng}}, \bibinfo {author}
  {\bibfnamefont {H.-L.}\ \bibnamefont {Yin}},\ and\ \bibinfo {author}
  {\bibfnamefont {Z.-B.}\ \bibnamefont {Chen}},\ }\bibfield  {title} {\bibinfo
  {title} {Advantages of {asynchronous} {measurement}-{device}-{independent}
  {quantum} {key} {distribution} in {intercity} {networks}},\ }\href
  {https://doi.org/10.1103/PhysRevApplied.19.054070} {\bibfield  {journal}
  {\bibinfo  {journal} {Phys. Rev. Appl.}\ }\textbf {\bibinfo {volume} {19}},\
  \bibinfo {pages} {054070} (\bibinfo {year} {2023})}\BibitemShut {NoStop}%
\bibitem [{\citenamefont {Brassard}\ \emph {et~al.}(2000)\citenamefont
  {Brassard}, \citenamefont {L\"utkenhaus}, \citenamefont {Mor},\ and\
  \citenamefont {Sanders}}]{brassard_limitations_2000}%
  \BibitemOpen
  \bibfield  {author} {\bibinfo {author} {\bibfnamefont {G.}~\bibnamefont
  {Brassard}}, \bibinfo {author} {\bibfnamefont {N.}~\bibnamefont
  {L\"utkenhaus}}, \bibinfo {author} {\bibfnamefont {T.}~\bibnamefont {Mor}},\
  and\ \bibinfo {author} {\bibfnamefont {B.~C.}\ \bibnamefont {Sanders}},\
  }\bibfield  {title} {\bibinfo {title} {Limitations on practical quantum
  cryptography},\ }\href {https://doi.org/10.1103/PhysRevLett.85.1330}
  {\bibfield  {journal} {\bibinfo  {journal} {Phys. Rev. Lett.}\ }\textbf
  {\bibinfo {volume} {85}},\ \bibinfo {pages} {1330} (\bibinfo {year}
  {2000})}\BibitemShut {NoStop}%
\bibitem [{\citenamefont {Hwang}(2003)}]{hwang_quantum_2003}%
  \BibitemOpen
  \bibfield  {author} {\bibinfo {author} {\bibfnamefont {W.-Y.}\ \bibnamefont
  {Hwang}},\ }\bibfield  {title} {\bibinfo {title} {Quantum {key}
  {distribution} with {high} {loss}: {toward} {global} {secure}
  {communication}},\ }\href {https://doi.org/10.1103/PhysRevLett.91.057901}
  {\bibfield  {journal} {\bibinfo  {journal} {Phys. Rev. Lett.}\ }\textbf
  {\bibinfo {volume} {91}},\ \bibinfo {pages} {057901} (\bibinfo {year}
  {2003})}\BibitemShut {NoStop}%
\bibitem [{\citenamefont {Wang}(2005)}]{wang_beating_2005}%
  \BibitemOpen
  \bibfield  {author} {\bibinfo {author} {\bibfnamefont {X.-B.}\ \bibnamefont
  {Wang}},\ }\bibfield  {title} {\bibinfo {title} {Beating the
  {photon}-{number}-{splitting} {attack} in {practical} {quantum}
  {cryptography}},\ }\href {https://doi.org/10.1103/PhysRevLett.94.230503}
  {\bibfield  {journal} {\bibinfo  {journal} {Phys. Rev. Lett.}\ }\textbf
  {\bibinfo {volume} {94}},\ \bibinfo {pages} {230503} (\bibinfo {year}
  {2005})}\BibitemShut {NoStop}%
\bibitem [{\citenamefont {Lo}\ \emph {et~al.}(2005)\citenamefont {Lo},
  \citenamefont {Ma},\ and\ \citenamefont {Chen}}]{lo_decoy_2005}%
  \BibitemOpen
  \bibfield  {author} {\bibinfo {author} {\bibfnamefont {H.-K.}\ \bibnamefont
  {Lo}}, \bibinfo {author} {\bibfnamefont {X.}~\bibnamefont {Ma}},\ and\
  \bibinfo {author} {\bibfnamefont {K.}~\bibnamefont {Chen}},\ }\bibfield
  {title} {\bibinfo {title} {Decoy state quantum key distribution},\ }\href
  {https://doi.org/10.1103/PhysRevLett.94.230504} {\bibfield  {journal}
  {\bibinfo  {journal} {Phys. Rev. Lett.}\ }\textbf {\bibinfo {volume} {94}},\
  \bibinfo {pages} {230504} (\bibinfo {year} {2005})}\BibitemShut {NoStop}%
\bibitem [{\citenamefont {Scarani}\ \emph {et~al.}(2004)\citenamefont
  {Scarani}, \citenamefont {Ac\'in}, \citenamefont {Ribordy},\ and\
  \citenamefont {Gisin}}]{scarani_quantum_2004}%
  \BibitemOpen
  \bibfield  {author} {\bibinfo {author} {\bibfnamefont {V.}~\bibnamefont
  {Scarani}}, \bibinfo {author} {\bibfnamefont {A.}~\bibnamefont {Ac\'in}},
  \bibinfo {author} {\bibfnamefont {G.}~\bibnamefont {Ribordy}},\ and\ \bibinfo
  {author} {\bibfnamefont {N.}~\bibnamefont {Gisin}},\ }\bibfield  {title}
  {\bibinfo {title} {Quantum cryptography protocols robust
against photon number splitting attacks for weak laser pulse implementations},\ }\href {https://doi.org/10.1103/PhysRevLett.92.057901}
  {\bibfield  {journal} {\bibinfo  {journal} {Phys. Rev. Lett.}\ }\textbf
  {\bibinfo {volume} {92}},\ \bibinfo {pages} {057901} (\bibinfo {year}
  {2004})}\BibitemShut {NoStop}%
\bibitem [{\citenamefont {Tamaki}\ and\ \citenamefont
  {Lo}(2006)}]{tamaki_unconditionally_2006}%
  \BibitemOpen
  \bibfield  {author} {\bibinfo {author} {\bibfnamefont {K.}~\bibnamefont
  {Tamaki}}\ and\ \bibinfo {author} {\bibfnamefont {H.-K.}\ \bibnamefont
  {Lo}},\ }\bibfield  {title} {\bibinfo {title} {Unconditionally secure key
  distillation from multiphotons},\ }\href
  {https://doi.org/10.1103/PhysRevA.73.010302} {\bibfield  {journal} {\bibinfo
  {journal} {Phys. Rev. A}\ }\textbf {\bibinfo {volume} {73}},\ \bibinfo
  {pages} {010302} (\bibinfo {year} {2006})}\BibitemShut {NoStop}%
\bibitem [{\citenamefont {Yin}\ \emph {et~al.}(2016{\natexlab{c}})\citenamefont
  {Yin}, \citenamefont {Fu}, \citenamefont {Mao},\ and\ \citenamefont
  {Chen}}]{yin_security_2016}%
  \BibitemOpen
  \bibfield  {author} {\bibinfo {author} {\bibfnamefont {H.-L.}\ \bibnamefont
  {Yin}}, \bibinfo {author} {\bibfnamefont {Y.}~\bibnamefont {Fu}}, \bibinfo
  {author} {\bibfnamefont {Y.}~\bibnamefont {Mao}},\ and\ \bibinfo {author}
  {\bibfnamefont {Z.-B.}\ \bibnamefont {Chen}},\ }\bibfield  {title} {\bibinfo
  {title} {Security of quantum key distribution with multiphoton components},\
  }\href {https://doi.org/10.1038/srep29482} {\bibfield  {journal} {\bibinfo
  {journal} {Sci. Rep.}\ }\textbf {\bibinfo {volume} {6}},\ \bibinfo {pages}
  {29482} (\bibinfo {year} {2016}{\natexlab{c}})}\BibitemShut {NoStop}%
\bibitem [{\citenamefont {Koashi}(2004)}]{koashi_unconditional_2004}%
  \BibitemOpen
  \bibfield  {author} {\bibinfo {author} {\bibfnamefont {M.}~\bibnamefont
  {Koashi}},\ }\bibfield  {title} {\bibinfo {title} {Unconditional security of coherent-state quantum key distribution with a strong phase-reference pulse},\ }\href
  {https://doi.org/10.1103/PhysRevLett.93.120501} {\bibfield  {journal}
  {\bibinfo  {journal} {Phys. Rev. Lett.}\ }\textbf {\bibinfo {volume} {93}},\
  \bibinfo {pages} {120501} (\bibinfo {year} {2004})}\BibitemShut {NoStop}%
\bibitem [{\citenamefont {Inoue}\ \emph {et~al.}(2002)\citenamefont {Inoue},
  \citenamefont {Waks},\ and\ \citenamefont
  {Yamamoto}}]{inoue_differential_2002}%
  \BibitemOpen
  \bibfield  {author} {\bibinfo {author} {\bibfnamefont {K.}~\bibnamefont
  {Inoue}}, \bibinfo {author} {\bibfnamefont {E.}~\bibnamefont {Waks}},\ and\
  \bibinfo {author} {\bibfnamefont {Y.}~\bibnamefont {Yamamoto}},\ }\bibfield
  {title} {\bibinfo {title} {Differential phase shift quantum key
  distribution},\ }\href {https://doi.org/10.1103/PhysRevLett.89.037902}
  {\bibfield  {journal} {\bibinfo  {journal} {Phys. Rev. Lett.}\ }\textbf
  {\bibinfo {volume} {89}},\ \bibinfo {pages} {037902} (\bibinfo {year}
  {2002})}\BibitemShut {NoStop}%
\bibitem [{\citenamefont {Inoue}\ \emph {et~al.}(2003)\citenamefont {Inoue},
  \citenamefont {Waks},\ and\ \citenamefont
  {Yamamoto}}]{inoue_differential-phase-shift_2003}%
  \BibitemOpen
  \bibfield  {author} {\bibinfo {author} {\bibfnamefont {K.}~\bibnamefont
  {Inoue}}, \bibinfo {author} {\bibfnamefont {E.}~\bibnamefont {Waks}},\ and\
  \bibinfo {author} {\bibfnamefont {Y.}~\bibnamefont {Yamamoto}},\ }\bibfield
  {title} {\bibinfo {title} {Differential-phase-shift quantum key distribution
  using coherent light},\ }\href {https://doi.org/10.1103/PhysRevA.68.022317}
  {\bibfield  {journal} {\bibinfo  {journal} {Phys. Rev. A}\ }\textbf {\bibinfo
  {volume} {68}},\ \bibinfo {pages} {022317} (\bibinfo {year}
  {2003})}\BibitemShut {NoStop}%
\bibitem [{\citenamefont {Stucki}\ \emph {et~al.}(2005)\citenamefont {Stucki},
  \citenamefont {Brunner}, \citenamefont {Gisin}, \citenamefont {Scarani},\
  and\ \citenamefont {Zbinden}}]{stucki_fast_2005}%
  \BibitemOpen
  \bibfield  {author} {\bibinfo {author} {\bibfnamefont {D.}~\bibnamefont
  {Stucki}}, \bibinfo {author} {\bibfnamefont {N.}~\bibnamefont {Brunner}},
  \bibinfo {author} {\bibfnamefont {N.}~\bibnamefont {Gisin}}, \bibinfo
  {author} {\bibfnamefont {V.}~\bibnamefont {Scarani}},\ and\ \bibinfo {author}
  {\bibfnamefont {H.}~\bibnamefont {Zbinden}},\ }\bibfield  {title} {\bibinfo
  {title} {Fast and simple one-way quantum key distribution},\ }\href
  {https://doi.org/10.1063/1.2126792} {\bibfield  {journal} {\bibinfo
  {journal} {Appl. Phys. Lett.}\ }\textbf {\bibinfo {volume} {87}},\ \bibinfo
  {pages} {194108} (\bibinfo {year} {2005})}\BibitemShut {NoStop}%
\bibitem [{\citenamefont {Sasaki}\ \emph {et~al.}(2014)\citenamefont {Sasaki},
  \citenamefont {Yamamoto},\ and\ \citenamefont
  {Koashi}}]{sasaki_practical_2014}%
  \BibitemOpen
  \bibfield  {author} {\bibinfo {author} {\bibfnamefont {T.}~\bibnamefont
  {Sasaki}}, \bibinfo {author} {\bibfnamefont {Y.}~\bibnamefont {Yamamoto}},\
  and\ \bibinfo {author} {\bibfnamefont {M.}~\bibnamefont {Koashi}},\
  }\bibfield  {title} {\bibinfo {title} {Practical quantum key distribution
  protocol without monitoring signal disturbance},\ }\href
  {https://doi.org/10.1038/nature13303} {\bibfield  {journal} {\bibinfo
  {journal} {Nature}\ }\textbf {\bibinfo {volume} {509}},\ \bibinfo {pages}
  {475} (\bibinfo {year} {2014})}\BibitemShut {NoStop}%
\bibitem [{\citenamefont {Takesue}\ \emph {et~al.}(2005)\citenamefont
  {Takesue}, \citenamefont {Diamanti}, \citenamefont {Honjo}, \citenamefont
  {Langrock}, \citenamefont {Fejer}, \citenamefont {Inoue},\ and\ \citenamefont
  {Yamamoto}}]{Takesue_differential_2005}%
  \BibitemOpen
  \bibfield  {author} {\bibinfo {author} {\bibfnamefont {H.}~\bibnamefont
  {Takesue}}, \bibinfo {author} {\bibfnamefont {E.}~\bibnamefont {Diamanti}},
  \bibinfo {author} {\bibfnamefont {T.}~\bibnamefont {Honjo}}, \bibinfo
  {author} {\bibfnamefont {C.}~\bibnamefont {Langrock}}, \bibinfo {author}
  {\bibfnamefont {M.~M.}\ \bibnamefont {Fejer}}, \bibinfo {author}
  {\bibfnamefont {K.}~\bibnamefont {Inoue}},\ and\ \bibinfo {author}
  {\bibfnamefont {Y.}~\bibnamefont {Yamamoto}},\ }\bibfield  {title} {\bibinfo
  {title} {Differential phase shift quantum key distribution experiment over
  105 km fibre},\ }\href {https://doi.org/10.1088/1367-2630/7/1/232} {\bibfield
   {journal} {\bibinfo  {journal} {New J. Phys.}\ }\textbf {\bibinfo {volume}
  {7}},\ \bibinfo {pages} {232} (\bibinfo {year} {2005})}\BibitemShut {NoStop}%
\bibitem [{\citenamefont {Diamanti}\ \emph {et~al.}(2006)\citenamefont
  {Diamanti}, \citenamefont {Takesue}, \citenamefont {Langrock}, \citenamefont
  {Fejer},\ and\ \citenamefont {Yamamoto}}]{Diamanti_100km_2006}%
  \BibitemOpen
  \bibfield  {author} {\bibinfo {author} {\bibfnamefont {E.}~\bibnamefont
  {Diamanti}}, \bibinfo {author} {\bibfnamefont {H.}~\bibnamefont {Takesue}},
  \bibinfo {author} {\bibfnamefont {C.}~\bibnamefont {Langrock}}, \bibinfo
  {author} {\bibfnamefont {M.~M.}\ \bibnamefont {Fejer}},\ and\ \bibinfo
  {author} {\bibfnamefont {Y.}~\bibnamefont {Yamamoto}},\ }\bibfield  {title}
  {\bibinfo {title} {100 km differential phase shift quantum key distribution
  experiment with low jitter up-conversion detectors},\ }\href@noop {}
  {\bibfield  {journal} {\bibinfo  {journal} {Opt. Express}\ }\textbf {\bibinfo
  {volume} {14}},\ \bibinfo {pages} {13073} (\bibinfo {year}
  {2006})}\BibitemShut {NoStop}%
\bibitem [{\citenamefont {Takesue}\ \emph {et~al.}(2007)\citenamefont
  {Takesue}, \citenamefont {Nam}, \citenamefont {Zhang}, \citenamefont
  {Hadfield}, \citenamefont {Honjo}, \citenamefont {Tamaki},\ and\
  \citenamefont {Yamamoto}}]{takesue_quantum_2007}%
  \BibitemOpen
  \bibfield  {author} {\bibinfo {author} {\bibfnamefont {H.}~\bibnamefont
  {Takesue}}, \bibinfo {author} {\bibfnamefont {S.~W.}\ \bibnamefont {Nam}},
  \bibinfo {author} {\bibfnamefont {Q.}~\bibnamefont {Zhang}}, \bibinfo
  {author} {\bibfnamefont {R.~H.}\ \bibnamefont {Hadfield}}, \bibinfo {author}
  {\bibfnamefont {T.}~\bibnamefont {Honjo}}, \bibinfo {author} {\bibfnamefont
  {K.}~\bibnamefont {Tamaki}},\ and\ \bibinfo {author} {\bibfnamefont
  {Y.}~\bibnamefont {Yamamoto}},\ }\bibfield  {title} {\bibinfo {title}
  {Quantum key distribution over a 40-{dB} channel loss using superconducting
  single-photon detectors},\ }\href {https://doi.org/10.1038/nphoton.2007.75}
  {\bibfield  {journal} {\bibinfo  {journal} {Nat. Photonics}\ }\textbf
  {\bibinfo {volume} {1}},\ \bibinfo {pages} {343} (\bibinfo {year}
  {2007})}\BibitemShut {NoStop}%
\bibitem [{\citenamefont {Sasaki}\ \emph {et~al.}(2011)\citenamefont {Sasaki},
  \citenamefont {Fujiwara}, \citenamefont {Ishizuka}, \citenamefont {Klaus},
  \citenamefont {Wakui}, \citenamefont {Takeoka}, \citenamefont {Miki},
  \citenamefont {Yamashita}, \citenamefont {Wang}, \citenamefont {Tanaka},
  \citenamefont {Yoshino}, \citenamefont {Nambu}, \citenamefont {Takahashi},
  \citenamefont {Tajima}, \citenamefont {Tomita}, \citenamefont {Domeki},
  \citenamefont {Hasegawa}, \citenamefont {Sakai}, \citenamefont {Kobayashi},
  \citenamefont {Asai}, \citenamefont {Shimizu}, \citenamefont {Tokura},
  \citenamefont {Tsurumaru}, \citenamefont {Matsui}, \citenamefont {Honjo},
  \citenamefont {Tamaki}, \citenamefont {Takesue}, \citenamefont {Tokura},
  \citenamefont {Dynes}, \citenamefont {Dixon}, \citenamefont {Sharpe},
  \citenamefont {Yuan}, \citenamefont {Shields}, \citenamefont {Uchikoga},
  \citenamefont {Legr\'{e}}, \citenamefont {Robyr}, \citenamefont {Trinkler},
  \citenamefont {Monat}, \citenamefont {Page}, \citenamefont {Ribordy},
  \citenamefont {Poppe}, \citenamefont {Allacher}, \citenamefont {Maurhart},
  \citenamefont {L\"{a}nger}, \citenamefont {Peev},\ and\ \citenamefont
  {Zeilinger}}]{Sasaki_field_2011}%
  \BibitemOpen
  \bibfield  {author} {\bibinfo {author} {\bibfnamefont {M.}~\bibnamefont
  {Sasaki}}, \bibinfo {author} {\bibfnamefont {M.}~\bibnamefont {Fujiwara}},
  \bibinfo {author} {\bibfnamefont {H.}~\bibnamefont {Ishizuka}}, \bibinfo
  {author} {\bibfnamefont {W.}~\bibnamefont {Klaus}}, \bibinfo {author}
  {\bibfnamefont {K.}~\bibnamefont {Wakui}}, \bibinfo {author} {\bibfnamefont
  {M.}~\bibnamefont {Takeoka}}, \bibinfo {author} {\bibfnamefont
  {S.}~\bibnamefont {Miki}}, \bibinfo {author} {\bibfnamefont {T.}~\bibnamefont
  {Yamashita}}, \bibinfo {author} {\bibfnamefont {Z.}~\bibnamefont {Wang}},
  \bibinfo {author} {\bibfnamefont {A.}~\bibnamefont {Tanaka}}, \bibinfo
  {author} {\bibfnamefont {K.}~\bibnamefont {Yoshino}}, \bibinfo {author}
  {\bibfnamefont {Y.}~\bibnamefont {Nambu}}, \bibinfo {author} {\bibfnamefont
  {S.}~\bibnamefont {Takahashi}}, \bibinfo {author} {\bibfnamefont
  {A.}~\bibnamefont {Tajima}}, \bibinfo {author} {\bibfnamefont
  {A.}~\bibnamefont {Tomita}}, \bibinfo {author} {\bibfnamefont
  {T.}~\bibnamefont {Domeki}}, \bibinfo {author} {\bibfnamefont
  {T.}~\bibnamefont {Hasegawa}}, \bibinfo {author} {\bibfnamefont
  {Y.}~\bibnamefont {Sakai}}, \bibinfo {author} {\bibfnamefont
  {H.}~\bibnamefont {Kobayashi}}, \bibinfo {author} {\bibfnamefont
  {T.}~\bibnamefont {Asai}}, \bibinfo {author} {\bibfnamefont {K.}~\bibnamefont
  {Shimizu}}, \bibinfo {author} {\bibfnamefont {T.}~\bibnamefont {Tokura}},
  \bibinfo {author} {\bibfnamefont {T.}~\bibnamefont {Tsurumaru}}, \bibinfo
  {author} {\bibfnamefont {M.}~\bibnamefont {Matsui}}, \bibinfo {author}
  {\bibfnamefont {T.}~\bibnamefont {Honjo}}, \bibinfo {author} {\bibfnamefont
  {K.}~\bibnamefont {Tamaki}}, \bibinfo {author} {\bibfnamefont
  {H.}~\bibnamefont {Takesue}}, \bibinfo {author} {\bibfnamefont
  {Y.}~\bibnamefont {Tokura}}, \bibinfo {author} {\bibfnamefont {J.~F.}\
  \bibnamefont {Dynes}}, \bibinfo {author} {\bibfnamefont {A.~R.}\ \bibnamefont
  {Dixon}}, \bibinfo {author} {\bibfnamefont {A.~W.}\ \bibnamefont {Sharpe}},
  \bibinfo {author} {\bibfnamefont {Z.~L.}\ \bibnamefont {Yuan}}, \bibinfo
  {author} {\bibfnamefont {A.~J.}\ \bibnamefont {Shields}}, \bibinfo {author}
  {\bibfnamefont {S.}~\bibnamefont {Uchikoga}}, \bibinfo {author}
  {\bibfnamefont {M.}~\bibnamefont {Legr\'{e}}}, \bibinfo {author}
  {\bibfnamefont {S.}~\bibnamefont {Robyr}}, \bibinfo {author} {\bibfnamefont
  {P.}~\bibnamefont {Trinkler}}, \bibinfo {author} {\bibfnamefont
  {L.}~\bibnamefont {Monat}}, \bibinfo {author} {\bibfnamefont {J.-B.}\
  \bibnamefont {Page}}, \bibinfo {author} {\bibfnamefont {G.}~\bibnamefont
  {Ribordy}}, \bibinfo {author} {\bibfnamefont {A.}~\bibnamefont {Poppe}},
  \bibinfo {author} {\bibfnamefont {A.}~\bibnamefont {Allacher}}, \bibinfo
  {author} {\bibfnamefont {O.}~\bibnamefont {Maurhart}}, \bibinfo {author}
  {\bibfnamefont {T.}~\bibnamefont {L\"{a}nger}}, \bibinfo {author}
  {\bibfnamefont {M.}~\bibnamefont {Peev}},\ and\ \bibinfo {author}
  {\bibfnamefont {A.}~\bibnamefont {Zeilinger}},\ }\bibfield  {title} {\bibinfo
  {title} {Field test of quantum key distribution in the tokyo qkd network},\
  }\href {https://doi.org/10.1364/OE.19.010387} {\bibfield  {journal} {\bibinfo
   {journal} {Opt. Express}\ }\textbf {\bibinfo {volume} {19}},\ \bibinfo
  {pages} {10387} (\bibinfo {year} {2011})}\BibitemShut {NoStop}%
\bibitem [{\citenamefont {Wen}\ \emph {et~al.}(2009)\citenamefont {Wen},
  \citenamefont {Tamaki},\ and\ \citenamefont
  {Yamamoto}}]{wen_unconditional_2009}%
  \BibitemOpen
  \bibfield  {author} {\bibinfo {author} {\bibfnamefont {K.}~\bibnamefont
  {Wen}}, \bibinfo {author} {\bibfnamefont {K.}~\bibnamefont {Tamaki}},\ and\
  \bibinfo {author} {\bibfnamefont {Y.}~\bibnamefont {Yamamoto}},\ }\bibfield
  {title} {\bibinfo {title} {Unconditional security of single-photon
  differential phase shift quantum key distribution},\ }\href
  {https://doi.org/10.1103/PhysRevLett.103.170503} {\bibfield  {journal}
  {\bibinfo  {journal} {Phys. Rev. Lett.}\ }\textbf {\bibinfo {volume} {103}},\
  \bibinfo {pages} {170503} (\bibinfo {year} {2009})}\BibitemShut {NoStop}%
\bibitem [{\citenamefont {Tamaki}\ \emph {et~al.}(2012)\citenamefont {Tamaki},
  \citenamefont {Koashi},\ and\ \citenamefont
  {Kato}}]{tamaki_2012_unconditional}%
  \BibitemOpen
  \bibfield  {author} {\bibinfo {author} {\bibfnamefont {K.}~\bibnamefont
  {Tamaki}}, \bibinfo {author} {\bibfnamefont {M.}~\bibnamefont {Koashi}},\
  and\ \bibinfo {author} {\bibfnamefont {G.}~\bibnamefont {Kato}},\ }\href@noop
  {} {\bibinfo {title} {Unconditional security of coherent-state-based
  differential phase shift quantum key distribution protocol with block-wise
  phase randomization}} (\bibinfo {year} {2012}),\ \Eprint
  {https://arxiv.org/abs/1208.1995} {arXiv:1208.1995} \BibitemShut
  {NoStop}%
\bibitem [{\citenamefont {Mizutani}\ \emph {et~al.}(2017)\citenamefont
  {Mizutani}, \citenamefont {Sasaki}, \citenamefont {Kato}, \citenamefont
  {Takeuchi},\ and\ \citenamefont {Tamaki}}]{Mizutani_information_2018}%
  \BibitemOpen
  \bibfield  {author} {\bibinfo {author} {\bibfnamefont {A.}~\bibnamefont
  {Mizutani}}, \bibinfo {author} {\bibfnamefont {T.}~\bibnamefont {Sasaki}},
  \bibinfo {author} {\bibfnamefont {G.}~\bibnamefont {Kato}}, \bibinfo {author}
  {\bibfnamefont {Y.}~\bibnamefont {Takeuchi}},\ and\ \bibinfo {author}
  {\bibfnamefont {K.}~\bibnamefont {Tamaki}},\ }\bibfield  {title} {\bibinfo
  {title} {Information-theoretic security proof of differential-phase-shift
  quantum key distribution protocol based on complementarity},\ }\href
  {https://doi.org/10.1088/2058-9565/aa8705} {\bibfield  {journal} {\bibinfo
  {journal} {Quantum Sci. Technol.}\ }\textbf {\bibinfo {volume} {3}},\
  \bibinfo {pages} {014003} (\bibinfo {year} {2017})}\BibitemShut {NoStop}%
\bibitem [{\citenamefont {Mizutani}\ \emph {et~al.}(2019)\citenamefont
  {Mizutani}, \citenamefont {Sasaki}, \citenamefont {Takeuchi}, \citenamefont
  {Tamaki},\ and\ \citenamefont {Koashi}}]{mizutani_quantum_2019}%
  \BibitemOpen
  \bibfield  {author} {\bibinfo {author} {\bibfnamefont {A.}~\bibnamefont
  {Mizutani}}, \bibinfo {author} {\bibfnamefont {T.}~\bibnamefont {Sasaki}},
  \bibinfo {author} {\bibfnamefont {Y.}~\bibnamefont {Takeuchi}}, \bibinfo
  {author} {\bibfnamefont {K.}~\bibnamefont {Tamaki}},\ and\ \bibinfo {author}
  {\bibfnamefont {M.}~\bibnamefont {Koashi}},\ }\bibfield  {title} {\bibinfo
  {title} {Quantum key distribution with simply characterized light sources},\
  }\href {https://doi.org/10.1038/s41534-019-0194-3} {\bibfield  {journal}
  {\bibinfo  {journal} {npj Quantum Inf.}\ }\textbf {\bibinfo {volume} {5}},\
  \bibinfo {pages} {87} (\bibinfo {year} {2019})}\BibitemShut {NoStop}%
\bibitem [{\citenamefont {Mizutani}(2020)}]{Mizutani_quantum_2020}%
  \BibitemOpen
  \bibfield  {author} {\bibinfo {author} {\bibfnamefont {A.}~\bibnamefont
  {Mizutani}},\ }\bibfield  {title} {\bibinfo {title} {Quantum key distribution
  with any two independent and identically distributed states},\ }\href
  {https://doi.org/10.1103/PhysRevA.102.022613} {\bibfield  {journal} {\bibinfo
   {journal} {Phys. Rev. A}\ }\textbf {\bibinfo {volume} {102}},\ \bibinfo
  {pages} {022613} (\bibinfo {year} {2020})}\BibitemShut {NoStop}%
\bibitem [{\citenamefont {Endo}\ \emph {et~al.}(2022)\citenamefont {Endo},
  \citenamefont {Sasaki}, \citenamefont {Takeoka}, \citenamefont {Fujiwara},
  \citenamefont {Koashi},\ and\ \citenamefont {Sasaki}}]{Endo_line_2022}%
  \BibitemOpen
  \bibfield  {author} {\bibinfo {author} {\bibfnamefont {H.}~\bibnamefont
  {Endo}}, \bibinfo {author} {\bibfnamefont {T.}~\bibnamefont {Sasaki}},
  \bibinfo {author} {\bibfnamefont {M.}~\bibnamefont {Takeoka}}, \bibinfo
  {author} {\bibfnamefont {M.}~\bibnamefont {Fujiwara}}, \bibinfo {author}
  {\bibfnamefont {M.}~\bibnamefont {Koashi}},\ and\ \bibinfo {author}
  {\bibfnamefont {M.}~\bibnamefont {Sasaki}},\ }\bibfield  {title} {\bibinfo
  {title} {Line-of-sight quantum key distribution with differential phase shift
  keying},\ }\href {https://doi.org/10.1088/1367-2630/ac5056} {\bibfield
  {journal} {\bibinfo  {journal} {New Journal of Physics}\ }\textbf {\bibinfo
  {volume} {24}},\ \bibinfo {pages} {025008} (\bibinfo {year}
  {2022})}\BibitemShut {NoStop}%
\bibitem [{\citenamefont {Mizutani}\ \emph {et~al.}(2023)\citenamefont
  {Mizutani}, \citenamefont {Takeuchi},\ and\ \citenamefont
  {Tamaki}}]{mizutani_finite_2023}%
  \BibitemOpen
  \bibfield  {author} {\bibinfo {author} {\bibfnamefont {A.}~\bibnamefont
  {Mizutani}}, \bibinfo {author} {\bibfnamefont {Y.}~\bibnamefont {Takeuchi}},\
  and\ \bibinfo {author} {\bibfnamefont {K.}~\bibnamefont {Tamaki}},\
  }\bibfield  {title} {\bibinfo {title} {Finite-key security analysis of
  differential-phase-shift quantum key distribution},\ }\href
  {https://doi.org/10.1103/PhysRevResearch.5.023132} {\bibfield  {journal}
  {\bibinfo  {journal} {Phys. Rev. Res.}\ }\textbf {\bibinfo {volume} {5}},\
  \bibinfo {pages} {023132} (\bibinfo {year} {2023})}\BibitemShut {NoStop}%
\bibitem [{\citenamefont {Sandfuchs}\ \emph {et~al.}(2023)\citenamefont
  {Sandfuchs}, \citenamefont {Haberland}, \citenamefont {Vilasini},\ and\
  \citenamefont {Wolf}}]{sandfuchs_2023_security}%
  \BibitemOpen
  \bibfield  {author} {\bibinfo {author} {\bibfnamefont {M.}~\bibnamefont
  {Sandfuchs}}, \bibinfo {author} {\bibfnamefont {M.}~\bibnamefont
  {Haberland}}, \bibinfo {author} {\bibfnamefont {V.}~\bibnamefont
  {Vilasini}},\ and\ \bibinfo {author} {\bibfnamefont {R.}~\bibnamefont
  {Wolf}},\ }\href@noop {} {\bibinfo {title} {Security of differential phase
  shift qkd from relativistic principles}} (\bibinfo {year} {2023}),\ \Eprint
  {https://arxiv.org/abs/2301.11340} {arXiv:2301.11340} \BibitemShut
  {NoStop}%
\bibitem [{\citenamefont {Peev}\ \emph {et~al.}(2009)\citenamefont {Peev},
  \citenamefont {Pacher}, \citenamefont {All\'eaume}, \citenamefont {Barreiro},
  \citenamefont {Bouda}, \citenamefont {Boxleitner}, \citenamefont
  {Debuisschert}, \citenamefont {Diamanti}, \citenamefont {Dianati},
  \citenamefont {Dynes}, \citenamefont {Fasel}, \citenamefont {Fossier},
  \citenamefont {F\"urst}, \citenamefont {Gautier}, \citenamefont {Gay},
  \citenamefont {Gisin}, \citenamefont {Grangier}, \citenamefont {Happe},
  \citenamefont {Hasani}, \citenamefont {Hentschel}, \citenamefont {H\"ubel},
  \citenamefont {Humer}, \citenamefont {L\"anger}, \citenamefont {Legr\'e},
  \citenamefont {Lieger}, \citenamefont {Lodewyck}, \citenamefont {Lor\"unser},
  \citenamefont {L\"utkenhaus}, \citenamefont {Marhold}, \citenamefont
  {Matyus}, \citenamefont {Maurhart}, \citenamefont {Monat}, \citenamefont
  {Nauerth}, \citenamefont {Page}, \citenamefont {Poppe}, \citenamefont
  {Querasser}, \citenamefont {Ribordy}, \citenamefont {Robyr}, \citenamefont
  {Salvail}, \citenamefont {Sharpe}, \citenamefont {Shields}, \citenamefont
  {Stucki}, \citenamefont {Suda}, \citenamefont {Tamas}, \citenamefont
  {Themel}, \citenamefont {Thew}, \citenamefont {Thoma}, \citenamefont
  {Treiber}, \citenamefont {Trinkler}, \citenamefont {Tualle-Brouri},
  \citenamefont {Vannel}, \citenamefont {Walenta}, \citenamefont {Weier},
  \citenamefont {Weinfurter}, \citenamefont {Wimberger}, \citenamefont {Yuan},
  \citenamefont {Zbinden},\ and\ \citenamefont {Zeilinger}}]{Peev_2009}%
  \BibitemOpen
  \bibfield  {author} {\bibinfo {author} {\bibfnamefont {M.}~\bibnamefont
  {Peev}}, \bibinfo {author} {\bibfnamefont {C.}~\bibnamefont {Pacher}},
  \bibinfo {author} {\bibfnamefont {R.}~\bibnamefont {All\'eaume}}, \bibinfo
  {author} {\bibfnamefont {C.}~\bibnamefont {Barreiro}}, \bibinfo {author}
  {\bibfnamefont {J.}~\bibnamefont {Bouda}}, \bibinfo {author} {\bibfnamefont
  {W.}~\bibnamefont {Boxleitner}}, \bibinfo {author} {\bibfnamefont
  {T.}~\bibnamefont {Debuisschert}}, \bibinfo {author} {\bibfnamefont
  {E.}~\bibnamefont {Diamanti}}, \bibinfo {author} {\bibfnamefont
  {M.}~\bibnamefont {Dianati}}, \bibinfo {author} {\bibfnamefont {J.~F.}\
  \bibnamefont {Dynes}}, \bibinfo {author} {\bibfnamefont {S.}~\bibnamefont
  {Fasel}}, \bibinfo {author} {\bibfnamefont {S.}~\bibnamefont {Fossier}},
  \bibinfo {author} {\bibfnamefont {M.}~\bibnamefont {F\"urst}}, \bibinfo
  {author} {\bibfnamefont {J.-D.}\ \bibnamefont {Gautier}}, \bibinfo {author}
  {\bibfnamefont {O.}~\bibnamefont {Gay}}, \bibinfo {author} {\bibfnamefont
  {N.}~\bibnamefont {Gisin}}, \bibinfo {author} {\bibfnamefont
  {P.}~\bibnamefont {Grangier}}, \bibinfo {author} {\bibfnamefont
  {A.}~\bibnamefont {Happe}}, \bibinfo {author} {\bibfnamefont
  {Y.}~\bibnamefont {Hasani}}, \bibinfo {author} {\bibfnamefont
  {M.}~\bibnamefont {Hentschel}}, \bibinfo {author} {\bibfnamefont
  {H.}~\bibnamefont {H\"ubel}}, \bibinfo {author} {\bibfnamefont
  {G.}~\bibnamefont {Humer}}, \bibinfo {author} {\bibfnamefont
  {T.}~\bibnamefont {L\"anger}}, \bibinfo {author} {\bibfnamefont
  {M.}~\bibnamefont {Legr\'e}}, \bibinfo {author} {\bibfnamefont
  {R.}~\bibnamefont {Lieger}}, \bibinfo {author} {\bibfnamefont
  {J.}~\bibnamefont {Lodewyck}}, \bibinfo {author} {\bibfnamefont
  {T.}~\bibnamefont {Lor\"unser}}, \bibinfo {author} {\bibfnamefont
  {N.}~\bibnamefont {L\"utkenhaus}}, \bibinfo {author} {\bibfnamefont
  {A.}~\bibnamefont {Marhold}}, \bibinfo {author} {\bibfnamefont
  {T.}~\bibnamefont {Matyus}}, \bibinfo {author} {\bibfnamefont
  {O.}~\bibnamefont {Maurhart}}, \bibinfo {author} {\bibfnamefont
  {L.}~\bibnamefont {Monat}}, \bibinfo {author} {\bibfnamefont
  {S.}~\bibnamefont {Nauerth}}, \bibinfo {author} {\bibfnamefont {J.-B.}\
  \bibnamefont {Page}}, \bibinfo {author} {\bibfnamefont {A.}~\bibnamefont
  {Poppe}}, \bibinfo {author} {\bibfnamefont {E.}~\bibnamefont {Querasser}},
  \bibinfo {author} {\bibfnamefont {G.}~\bibnamefont {Ribordy}}, \bibinfo
  {author} {\bibfnamefont {S.}~\bibnamefont {Robyr}}, \bibinfo {author}
  {\bibfnamefont {L.}~\bibnamefont {Salvail}}, \bibinfo {author} {\bibfnamefont
  {A.~W.}\ \bibnamefont {Sharpe}}, \bibinfo {author} {\bibfnamefont {A.~J.}\
  \bibnamefont {Shields}}, \bibinfo {author} {\bibfnamefont {D.}~\bibnamefont
  {Stucki}}, \bibinfo {author} {\bibfnamefont {M.}~\bibnamefont {Suda}},
  \bibinfo {author} {\bibfnamefont {C.}~\bibnamefont {Tamas}}, \bibinfo
  {author} {\bibfnamefont {T.}~\bibnamefont {Themel}}, \bibinfo {author}
  {\bibfnamefont {R.~T.}\ \bibnamefont {Thew}}, \bibinfo {author}
  {\bibfnamefont {Y.}~\bibnamefont {Thoma}}, \bibinfo {author} {\bibfnamefont
  {A.}~\bibnamefont {Treiber}}, \bibinfo {author} {\bibfnamefont
  {P.}~\bibnamefont {Trinkler}}, \bibinfo {author} {\bibfnamefont
  {R.}~\bibnamefont {Tualle-Brouri}}, \bibinfo {author} {\bibfnamefont
  {F.}~\bibnamefont {Vannel}}, \bibinfo {author} {\bibfnamefont
  {N.}~\bibnamefont {Walenta}}, \bibinfo {author} {\bibfnamefont
  {H.}~\bibnamefont {Weier}}, \bibinfo {author} {\bibfnamefont
  {H.}~\bibnamefont {Weinfurter}}, \bibinfo {author} {\bibfnamefont
  {I.}~\bibnamefont {Wimberger}}, \bibinfo {author} {\bibfnamefont {Z.~L.}\
  \bibnamefont {Yuan}}, \bibinfo {author} {\bibfnamefont {H.}~\bibnamefont
  {Zbinden}},\ and\ \bibinfo {author} {\bibfnamefont {A.}~\bibnamefont
  {Zeilinger}},\ }\bibfield  {title} {\bibinfo {title} {The secoqc quantum key
  distribution network in vienna},\ }\href
  {https://doi.org/10.1088/1367-2630/11/7/075001} {\bibfield  {journal}
  {\bibinfo  {journal} {New J. Phys.}\ }\textbf {\bibinfo {volume} {11}},\
  \bibinfo {pages} {075001} (\bibinfo {year} {2009})}\BibitemShut {NoStop}%
\bibitem [{\citenamefont {Stucki}\ \emph
  {et~al.}(2009{\natexlab{a}})\citenamefont {Stucki}, \citenamefont {Barreiro},
  \citenamefont {Fasel}, \citenamefont {Gautier}, \citenamefont {Gay},
  \citenamefont {Gisin}, \citenamefont {Thew}, \citenamefont {Thoma},
  \citenamefont {Trinkler}, \citenamefont {Vannel},\ and\ \citenamefont
  {Zbinden}}]{stucki_continuous_2009}%
  \BibitemOpen
  \bibfield  {author} {\bibinfo {author} {\bibfnamefont {D.}~\bibnamefont
  {Stucki}}, \bibinfo {author} {\bibfnamefont {C.}~\bibnamefont {Barreiro}},
  \bibinfo {author} {\bibfnamefont {S.}~\bibnamefont {Fasel}}, \bibinfo
  {author} {\bibfnamefont {J.-D.}\ \bibnamefont {Gautier}}, \bibinfo {author}
  {\bibfnamefont {O.}~\bibnamefont {Gay}}, \bibinfo {author} {\bibfnamefont
  {N.}~\bibnamefont {Gisin}}, \bibinfo {author} {\bibfnamefont
  {R.}~\bibnamefont {Thew}}, \bibinfo {author} {\bibfnamefont {Y.}~\bibnamefont
  {Thoma}}, \bibinfo {author} {\bibfnamefont {P.}~\bibnamefont {Trinkler}},
  \bibinfo {author} {\bibfnamefont {F.}~\bibnamefont {Vannel}},\ and\ \bibinfo
  {author} {\bibfnamefont {H.}~\bibnamefont {Zbinden}},\ }\bibfield  {title}
  {\bibinfo {title} {Continuous high speed coherent one-way quantum key
  distribution},\ }\href {https://doi.org/10.1364/OE.17.013326} {\bibfield
  {journal} {\bibinfo  {journal} {Opt. Express}\ }\textbf {\bibinfo {volume}
  {17}},\ \bibinfo {pages} {13326} (\bibinfo {year}
  {2009}{\natexlab{a}})}\BibitemShut {NoStop}%
\bibitem [{\citenamefont {Stucki}\ \emph
  {et~al.}(2009{\natexlab{b}})\citenamefont {Stucki}, \citenamefont {Walenta},
  \citenamefont {Vannel}, \citenamefont {Thew}, \citenamefont {Gisin},
  \citenamefont {Zbinden}, \citenamefont {Gray}, \citenamefont {Towery},\ and\
  \citenamefont {Ten}}]{stucki_high_2009}%
  \BibitemOpen
  \bibfield  {author} {\bibinfo {author} {\bibfnamefont {D.}~\bibnamefont
  {Stucki}}, \bibinfo {author} {\bibfnamefont {N.}~\bibnamefont {Walenta}},
  \bibinfo {author} {\bibfnamefont {F.}~\bibnamefont {Vannel}}, \bibinfo
  {author} {\bibfnamefont {R.~T.}\ \bibnamefont {Thew}}, \bibinfo {author}
  {\bibfnamefont {N.}~\bibnamefont {Gisin}}, \bibinfo {author} {\bibfnamefont
  {H.}~\bibnamefont {Zbinden}}, \bibinfo {author} {\bibfnamefont
  {S.}~\bibnamefont {Gray}}, \bibinfo {author} {\bibfnamefont {C.~R.}\
  \bibnamefont {Towery}},\ and\ \bibinfo {author} {\bibfnamefont
  {S.}~\bibnamefont {Ten}},\ }\bibfield  {title} {\bibinfo {title} {High rate,
  long-distance quantum key distribution over 250 km of ultra low loss
  fibres},\ }\href {https://doi.org/10.1088/1367-2630/11/7/075003} {\bibfield
  {journal} {\bibinfo  {journal} {New J. Phys.}\ }\textbf {\bibinfo {volume}
  {11}},\ \bibinfo {pages} {075003} (\bibinfo {year}
  {2009}{\natexlab{b}})}\BibitemShut {NoStop}%
\bibitem [{\citenamefont {Walenta}\ \emph {et~al.}(2014)\citenamefont
  {Walenta}, \citenamefont {Burg}, \citenamefont {Caselunghe}, \citenamefont
  {Constantin}, \citenamefont {Gisin}, \citenamefont {Guinnard}, \citenamefont
  {Houlmann}, \citenamefont {Junod}, \citenamefont {Korzh}, \citenamefont
  {Kulesza}, \citenamefont {Legr\'e}, \citenamefont {Lim}, \citenamefont
  {Lunghi}, \citenamefont {Monat}, \citenamefont {Portmann}, \citenamefont
  {Soucarros}, \citenamefont {Thew}, \citenamefont {Trinkler}, \citenamefont
  {Trolliet}, \citenamefont {Vannel},\ and\ \citenamefont
  {Zbinden}}]{walenta_fast_2014}%
  \BibitemOpen
  \bibfield  {author} {\bibinfo {author} {\bibfnamefont {N.}~\bibnamefont
  {Walenta}}, \bibinfo {author} {\bibfnamefont {A.}~\bibnamefont {Burg}},
  \bibinfo {author} {\bibfnamefont {D.}~\bibnamefont {Caselunghe}}, \bibinfo
  {author} {\bibfnamefont {J.}~\bibnamefont {Constantin}}, \bibinfo {author}
  {\bibfnamefont {N.}~\bibnamefont {Gisin}}, \bibinfo {author} {\bibfnamefont
  {O.}~\bibnamefont {Guinnard}}, \bibinfo {author} {\bibfnamefont
  {R.}~\bibnamefont {Houlmann}}, \bibinfo {author} {\bibfnamefont
  {P.}~\bibnamefont {Junod}}, \bibinfo {author} {\bibfnamefont
  {B.}~\bibnamefont {Korzh}}, \bibinfo {author} {\bibfnamefont
  {N.}~\bibnamefont {Kulesza}}, \bibinfo {author} {\bibfnamefont
  {M.}~\bibnamefont {Legr\'e}}, \bibinfo {author} {\bibfnamefont {C.~W.}\
  \bibnamefont {Lim}}, \bibinfo {author} {\bibfnamefont {T.}~\bibnamefont
  {Lunghi}}, \bibinfo {author} {\bibfnamefont {L.}~\bibnamefont {Monat}},
  \bibinfo {author} {\bibfnamefont {C.}~\bibnamefont {Portmann}}, \bibinfo
  {author} {\bibfnamefont {M.}~\bibnamefont {Soucarros}}, \bibinfo {author}
  {\bibfnamefont {R.~T.}\ \bibnamefont {Thew}}, \bibinfo {author}
  {\bibfnamefont {P.}~\bibnamefont {Trinkler}}, \bibinfo {author}
  {\bibfnamefont {G.}~\bibnamefont {Trolliet}}, \bibinfo {author}
  {\bibfnamefont {F.}~\bibnamefont {Vannel}},\ and\ \bibinfo {author}
  {\bibfnamefont {H.}~\bibnamefont {Zbinden}},\ }\bibfield  {title} {\bibinfo
  {title} {A fast and versatile quantum key distribution system with hardware
  key distillation and wavelength multiplexing},\ }\href
  {https://doi.org/10.1088/1367-2630/16/1/013047} {\bibfield  {journal}
  {\bibinfo  {journal} {New J. Phys.}\ }\textbf {\bibinfo {volume} {16}},\
  \bibinfo {pages} {013047} (\bibinfo {year} {2014})}\BibitemShut {NoStop}%
\bibitem [{\citenamefont {Korzh}\ \emph {et~al.}(2015)\citenamefont {Korzh},
  \citenamefont {Lim}, \citenamefont {Houlmann}, \citenamefont {Gisin},
  \citenamefont {Li}, \citenamefont {Nolan}, \citenamefont {Sanguinetti},
  \citenamefont {Thew},\ and\ \citenamefont {Zbinden}}]{korzh_provably_2015}%
  \BibitemOpen
  \bibfield  {author} {\bibinfo {author} {\bibfnamefont {B.}~\bibnamefont
  {Korzh}}, \bibinfo {author} {\bibfnamefont {C.~C.~W.}\ \bibnamefont {Lim}},
  \bibinfo {author} {\bibfnamefont {R.}~\bibnamefont {Houlmann}}, \bibinfo
  {author} {\bibfnamefont {N.}~\bibnamefont {Gisin}}, \bibinfo {author}
  {\bibfnamefont {M.~J.}\ \bibnamefont {Li}}, \bibinfo {author} {\bibfnamefont
  {D.}~\bibnamefont {Nolan}}, \bibinfo {author} {\bibfnamefont
  {B.}~\bibnamefont {Sanguinetti}}, \bibinfo {author} {\bibfnamefont
  {R.}~\bibnamefont {Thew}},\ and\ \bibinfo {author} {\bibfnamefont
  {H.}~\bibnamefont {Zbinden}},\ }\bibfield  {title} {\bibinfo {title}
  {Provably secure and practical quantum key distribution over 307 km of
  optical fibre},\ }\href {https://doi.org/10.1038/nphoton.2014.327} {\bibfield
   {journal} {\bibinfo  {journal} {Nat. Photonics}\ }\textbf {\bibinfo {volume}
  {9}},\ \bibinfo {pages} {163} (\bibinfo {year} {2015})}\BibitemShut {NoStop}%
\bibitem [{\citenamefont {Sibson}\ \emph
  {et~al.}(2017{\natexlab{a}})\citenamefont {Sibson}, \citenamefont {Erven},
  \citenamefont {Godfrey}, \citenamefont {Miki}, \citenamefont {Yamashita},
  \citenamefont {Fujiwara}, \citenamefont {Sasaki}, \citenamefont {Terai},
  \citenamefont {Tanner}, \citenamefont {Natarajan}, \citenamefont {Hadfield},
  \citenamefont {O'Brien},\ and\ \citenamefont
  {Thompson}}]{sibson_chip-based_2017}%
  \BibitemOpen
  \bibfield  {author} {\bibinfo {author} {\bibfnamefont {P.}~\bibnamefont
  {Sibson}}, \bibinfo {author} {\bibfnamefont {C.}~\bibnamefont {Erven}},
  \bibinfo {author} {\bibfnamefont {M.}~\bibnamefont {Godfrey}}, \bibinfo
  {author} {\bibfnamefont {S.}~\bibnamefont {Miki}}, \bibinfo {author}
  {\bibfnamefont {T.}~\bibnamefont {Yamashita}}, \bibinfo {author}
  {\bibfnamefont {M.}~\bibnamefont {Fujiwara}}, \bibinfo {author}
  {\bibfnamefont {M.}~\bibnamefont {Sasaki}}, \bibinfo {author} {\bibfnamefont
  {H.}~\bibnamefont {Terai}}, \bibinfo {author} {\bibfnamefont {M.~G.}\
  \bibnamefont {Tanner}}, \bibinfo {author} {\bibfnamefont {C.~M.}\
  \bibnamefont {Natarajan}}, \bibinfo {author} {\bibfnamefont {R.~H.}\
  \bibnamefont {Hadfield}}, \bibinfo {author} {\bibfnamefont {J.~L.}\
  \bibnamefont {O'Brien}},\ and\ \bibinfo {author} {\bibfnamefont {M.~G.}\
  \bibnamefont {Thompson}},\ }\bibfield  {title} {\bibinfo {title} {Chip-based
  quantum key distribution},\ }\href {https://doi.org/10.1038/ncomms13984}
  {\bibfield  {journal} {\bibinfo  {journal} {Nat. Commun.}\ }\textbf {\bibinfo
  {volume} {8}},\ \bibinfo {pages} {13984} (\bibinfo {year}
  {2017}{\natexlab{a}})}\BibitemShut {NoStop}%
\bibitem [{\citenamefont {Sibson}\ \emph
  {et~al.}(2017{\natexlab{b}})\citenamefont {Sibson}, \citenamefont {Kennard},
  \citenamefont {Stanisic}, \citenamefont {Erven}, \citenamefont {O'Brien},\
  and\ \citenamefont {Thompson}}]{sibson_integrated_2017}%
  \BibitemOpen
  \bibfield  {author} {\bibinfo {author} {\bibfnamefont {P.}~\bibnamefont
  {Sibson}}, \bibinfo {author} {\bibfnamefont {J.~E.}\ \bibnamefont {Kennard}},
  \bibinfo {author} {\bibfnamefont {S.}~\bibnamefont {Stanisic}}, \bibinfo
  {author} {\bibfnamefont {C.}~\bibnamefont {Erven}}, \bibinfo {author}
  {\bibfnamefont {J.~L.}\ \bibnamefont {O'Brien}},\ and\ \bibinfo {author}
  {\bibfnamefont {M.~G.}\ \bibnamefont {Thompson}},\ }\bibfield  {title}
  {\bibinfo {title} {Integrated silicon photonics for high-speed quantum key
  distribution},\ }\href {https://doi.org/10.1364/OPTICA.4.000172} {\bibfield
  {journal} {\bibinfo  {journal} {Optica}\ }\textbf {\bibinfo {volume} {4}},\
  \bibinfo {pages} {172} (\bibinfo {year} {2017}{\natexlab{b}})}\BibitemShut
  {NoStop}%
\bibitem [{\citenamefont {Roberts}\ \emph {et~al.}(2017)\citenamefont
  {Roberts}, \citenamefont {Lucamarini}, \citenamefont {Dynes}, \citenamefont
  {Savory}, \citenamefont {Yuan},\ and\ \citenamefont
  {Shields}}]{roberts_modulator-free_2017}%
  \BibitemOpen
  \bibfield  {author} {\bibinfo {author} {\bibfnamefont {G.~L.}\ \bibnamefont
  {Roberts}}, \bibinfo {author} {\bibfnamefont {M.}~\bibnamefont {Lucamarini}},
  \bibinfo {author} {\bibfnamefont {J.~F.}\ \bibnamefont {Dynes}}, \bibinfo
  {author} {\bibfnamefont {S.~J.}\ \bibnamefont {Savory}}, \bibinfo {author}
  {\bibfnamefont {Z.~L.}\ \bibnamefont {Yuan}},\ and\ \bibinfo {author}
  {\bibfnamefont {A.~J.}\ \bibnamefont {Shields}},\ }\bibfield  {title}
  {\bibinfo {title} {Modulator-free coherent-one-way quantum key distribution: Modulator-free coherent-oneway quantum key distribution},\ }\href {https://doi.org/10.1002/lpor.201700067} {\bibfield
  {journal} {\bibinfo  {journal} {Laser Photonics Rev.}\ }\textbf {\bibinfo
  {volume} {11}},\ \bibinfo {pages} {1700067} (\bibinfo {year}
  {2017})}\BibitemShut {NoStop}%
\bibitem [{\citenamefont {Dai}\ \emph {et~al.}(2020)\citenamefont {Dai},
  \citenamefont {Zhang}, \citenamefont {Fu}, \citenamefont {Zheng},\ and\
  \citenamefont {Yang}}]{dai_pass-block_2020}%
  \BibitemOpen
  \bibfield  {author} {\bibinfo {author} {\bibfnamefont {J.}~\bibnamefont
  {Dai}}, \bibinfo {author} {\bibfnamefont {L.}~\bibnamefont {Zhang}}, \bibinfo
  {author} {\bibfnamefont {X.}~\bibnamefont {Fu}}, \bibinfo {author}
  {\bibfnamefont {X.}~\bibnamefont {Zheng}},\ and\ \bibinfo {author}
  {\bibfnamefont {L.}~\bibnamefont {Yang}},\ }\bibfield  {title} {\bibinfo
  {title} {Pass-block architecture for distributed-phase-reference quantum key
  distribution using silicon photonics},\ }\href
  {https://doi.org/10.1364/OL.388654} {\bibfield  {journal} {\bibinfo
  {journal} {Opt. Lett.}\ }\textbf {\bibinfo {volume} {45}},\ \bibinfo {pages}
  {2014} (\bibinfo {year} {2020})}\BibitemShut {NoStop}%
\bibitem [{\citenamefont {Branciard}\ \emph {et~al.}(2008)\citenamefont
  {Branciard}, \citenamefont {Gisin},\ and\ \citenamefont
  {Scarani}}]{branciard_upper_2008}%
  \BibitemOpen
  \bibfield  {author} {\bibinfo {author} {\bibfnamefont {C.}~\bibnamefont
  {Branciard}}, \bibinfo {author} {\bibfnamefont {N.}~\bibnamefont {Gisin}},\
  and\ \bibinfo {author} {\bibfnamefont {V.}~\bibnamefont {Scarani}},\
  }\bibfield  {title} {\bibinfo {title} {Upper bounds for the security of two
  distributed-phase reference protocols of quantum cryptography},\ }\href
  {https://doi.org/10.1088/1367-2630/10/1/013031} {\bibfield  {journal}
  {\bibinfo  {journal} {New J. Phys.}\ }\textbf {\bibinfo {volume} {10}},\
  \bibinfo {pages} {013031} (\bibinfo {year} {2008})}\BibitemShut {NoStop}%
\bibitem [{\citenamefont {Moroder}\ \emph {et~al.}(2012)\citenamefont
  {Moroder}, \citenamefont {Curty}, \citenamefont {Lim}, \citenamefont {Thinh},
  \citenamefont {Zbinden},\ and\ \citenamefont
  {Gisin}}]{moroder_security_2012}%
  \BibitemOpen
  \bibfield  {author} {\bibinfo {author} {\bibfnamefont {T.}~\bibnamefont
  {Moroder}}, \bibinfo {author} {\bibfnamefont {M.}~\bibnamefont {Curty}},
  \bibinfo {author} {\bibfnamefont {C.~C.~W.}\ \bibnamefont {Lim}}, \bibinfo
  {author} {\bibfnamefont {L.~P.}\ \bibnamefont {Thinh}}, \bibinfo {author}
  {\bibfnamefont {H.}~\bibnamefont {Zbinden}},\ and\ \bibinfo {author}
  {\bibfnamefont {N.}~\bibnamefont {Gisin}},\ }\bibfield  {title} {\bibinfo
  {title} {Security of distributed-phase-reference quantum key distribution},\ }\href {https://doi.org/10.1103/PhysRevLett.109.260501}
  {\bibfield  {journal} {\bibinfo  {journal} {Phys. Rev. Lett.}\ }\textbf
  {\bibinfo {volume} {109}},\ \bibinfo {pages} {260501} (\bibinfo {year}
  {2012})}\BibitemShut {NoStop}%
\bibitem [{\citenamefont {Gonz\'alez-Payo}\ \emph {et~al.}(2020)\citenamefont
  {Gonz\'alez-Payo}, \citenamefont {Tr\'enyi}, \citenamefont {Wang},\ and\
  \citenamefont {Curty}}]{gonzalez-payo_upper_2020}%
  \BibitemOpen
  \bibfield  {author} {\bibinfo {author} {\bibfnamefont {J.}~\bibnamefont
  {Gonz\'alez-Payo}}, \bibinfo {author} {\bibfnamefont {R.}~\bibnamefont
  {Tr\'enyi}}, \bibinfo {author} {\bibfnamefont {W.}~\bibnamefont {Wang}},\
  and\ \bibinfo {author} {\bibfnamefont {M.}~\bibnamefont {Curty}},\ }\bibfield
   {title} {\bibinfo {title} {Upper security bounds for
coherent-one-way quantum key distribution},\ }\href
  {https://doi.org/10.1103/PhysRevLett.125.260510} {\bibfield  {journal}
  {\bibinfo  {journal} {Phys. Rev. Lett.}\ }\textbf {\bibinfo {volume} {125}},\
  \bibinfo {pages} {260510} (\bibinfo {year} {2020})}\BibitemShut {NoStop}%
\bibitem [{\citenamefont {Tr\'enyi}\ and\ \citenamefont
  {Curty}(2021)}]{trenyi_zero-error_2021}%
  \BibitemOpen
  \bibfield  {author} {\bibinfo {author} {\bibfnamefont {R.}~\bibnamefont
  {Tr\'enyi}}\ and\ \bibinfo {author} {\bibfnamefont {M.}~\bibnamefont
  {Curty}},\ }\bibfield  {title} {\bibinfo {title} {Zero-error attack against
  coherent-one-way quantum key distribution},\ }\href
  {https://doi.org/10.1088/1367-2630/ac1e41} {\bibfield  {journal} {\bibinfo
  {journal} {New J. Phys.}\ }\textbf {\bibinfo {volume} {23}},\ \bibinfo
  {pages} {093005} (\bibinfo {year} {2021})}\BibitemShut {NoStop}%
\bibitem [{\citenamefont {Wang}\ \emph
  {et~al.}(2019{\natexlab{a}})\citenamefont {Wang}, \citenamefont
  {Primaatmaja}, \citenamefont {Lavie}, \citenamefont {Varvitsiotis},\ and\
  \citenamefont {Lim}}]{wang_characterising_2019}%
  \BibitemOpen
  \bibfield  {author} {\bibinfo {author} {\bibfnamefont {Y.}~\bibnamefont
  {Wang}}, \bibinfo {author} {\bibfnamefont {I.~W.}\ \bibnamefont
  {Primaatmaja}}, \bibinfo {author} {\bibfnamefont {E.}~\bibnamefont {Lavie}},
  \bibinfo {author} {\bibfnamefont {A.}~\bibnamefont {Varvitsiotis}},\ and\
  \bibinfo {author} {\bibfnamefont {C.~C.~W.}\ \bibnamefont {Lim}},\ }\bibfield
   {title} {\bibinfo {title} {Characterising the correlations of
  prepare-and-measure quantum networks},\ }\href
  {https://doi.org/10.1038/s41534-019-0133-3} {\bibfield  {journal} {\bibinfo
  {journal} {npj Quantum Inf.}\ }\textbf {\bibinfo {volume} {5}},\ \bibinfo
  {pages} {17} (\bibinfo {year} {2019}{\natexlab{a}})}\BibitemShut {NoStop}%
\bibitem [{\citenamefont {De~Marco}\ \emph {et~al.}(2021)\citenamefont
  {De~Marco}, \citenamefont {Woodward}, \citenamefont {Roberts}, \citenamefont
  {Para\"iso}, \citenamefont {Roger}, \citenamefont {Sanzaro}, \citenamefont
  {Lucamarini}, \citenamefont {Yuan},\ and\ \citenamefont
  {Shields}}]{de_marco_real-time_2021}%
  \BibitemOpen
  \bibfield  {author} {\bibinfo {author} {\bibfnamefont {I.}~\bibnamefont
  {De~Marco}}, \bibinfo {author} {\bibfnamefont {R.~I.}\ \bibnamefont
  {Woodward}}, \bibinfo {author} {\bibfnamefont {G.~L.}\ \bibnamefont
  {Roberts}}, \bibinfo {author} {\bibfnamefont {T.~K.}\ \bibnamefont
  {Para\"iso}}, \bibinfo {author} {\bibfnamefont {T.}~\bibnamefont {Roger}},
  \bibinfo {author} {\bibfnamefont {M.}~\bibnamefont {Sanzaro}}, \bibinfo
  {author} {\bibfnamefont {M.}~\bibnamefont {Lucamarini}}, \bibinfo {author}
  {\bibfnamefont {Z.}~\bibnamefont {Yuan}},\ and\ \bibinfo {author}
  {\bibfnamefont {A.~J.}\ \bibnamefont {Shields}},\ }\bibfield  {title}
  {\bibinfo {title} {Real-time operation of a multi-rate, multi-protocol
  quantum key distribution transmitter},\ }\href
  {https://doi.org/10.1364/OPTICA.423517} {\bibfield  {journal} {\bibinfo
  {journal} {Optica}\ }\textbf {\bibinfo {volume} {8}},\ \bibinfo {pages} {911}
  (\bibinfo {year} {2021})}\BibitemShut {NoStop}%
\bibitem [{\citenamefont {Lavie}\ and\ \citenamefont
  {Lim}(2022)}]{Lavie-improved-2022}%
  \BibitemOpen
  \bibfield  {author} {\bibinfo {author} {\bibfnamefont {E.}~\bibnamefont
  {Lavie}}\ and\ \bibinfo {author} {\bibfnamefont {C.~C.-W.}\ \bibnamefont
  {Lim}},\ }\bibfield  {title} {\bibinfo {title} {Improved coherent one-way
  quantum key distribution for high-loss channels},\ }\href
  {https://doi.org/10.1103/PhysRevApplied.18.064053} {\bibfield  {journal}
  {\bibinfo  {journal} {Phys. Rev. Appl.}\ }\textbf {\bibinfo {volume} {18}},\
  \bibinfo {pages} {064053} (\bibinfo {year} {2022})}\BibitemShut {NoStop}%
\bibitem [{\citenamefont {Gao}\ \emph {et~al.}(2022)\citenamefont {Gao},
  \citenamefont {Xie}, \citenamefont {Gu}, \citenamefont {Liu}, \citenamefont
  {Weng}, \citenamefont {Li}, \citenamefont {Yin},\ and\ \citenamefont
  {Chen}}]{gao_simple_2022}%
  \BibitemOpen
  \bibfield  {author} {\bibinfo {author} {\bibfnamefont {R.-Q.}\ \bibnamefont
  {Gao}}, \bibinfo {author} {\bibfnamefont {Y.-M.}\ \bibnamefont {Xie}},
  \bibinfo {author} {\bibfnamefont {J.}~\bibnamefont {Gu}}, \bibinfo {author}
  {\bibfnamefont {W.-B.}\ \bibnamefont {Liu}}, \bibinfo {author} {\bibfnamefont
  {C.-X.}\ \bibnamefont {Weng}}, \bibinfo {author} {\bibfnamefont {B.-H.}\
  \bibnamefont {Li}}, \bibinfo {author} {\bibfnamefont {H.-L.}\ \bibnamefont
  {Yin}},\ and\ \bibinfo {author} {\bibfnamefont {Z.-B.}\ \bibnamefont
  {Chen}},\ }\bibfield  {title} {\bibinfo {title} {Simple security proof of
  coherent-one-way quantum key distribution},\ }\href
  {https://doi.org/10.1364/OE.461669} {\bibfield  {journal} {\bibinfo
  {journal} {Opt. Express}\ }\textbf {\bibinfo {volume} {30}},\ \bibinfo
  {pages} {23783} (\bibinfo {year} {2022})}\BibitemShut {NoStop}%
\bibitem [{\citenamefont {Tomamichel}\ and\ \citenamefont
  {Renner}(2011)}]{tomamichel_uncertainty_2011}%
  \BibitemOpen
  \bibfield  {author} {\bibinfo {author} {\bibfnamefont {M.}~\bibnamefont
  {Tomamichel}}\ and\ \bibinfo {author} {\bibfnamefont {R.}~\bibnamefont
  {Renner}},\ }\bibfield  {title} {\bibinfo {title} {Uncertainty relation for smooth entropies},\ }\href
  {https://doi.org/10.1103/PhysRevLett.106.110506} {\bibfield  {journal}
  {\bibinfo  {journal} {Phys. Rev. Lett.}\ }\textbf {\bibinfo {volume} {106}},\
  \bibinfo {pages} {110506} (\bibinfo {year} {2011})}\BibitemShut {NoStop}%
\bibitem [{\citenamefont {Tomamichel}\ \emph {et~al.}(2012)\citenamefont
  {Tomamichel}, \citenamefont {Lim}, \citenamefont {Gisin},\ and\ \citenamefont
  {Renner}}]{tomamichel_tight_2012}%
  \BibitemOpen
  \bibfield  {author} {\bibinfo {author} {\bibfnamefont {M.}~\bibnamefont
  {Tomamichel}}, \bibinfo {author} {\bibfnamefont {C.~C.~W.}\ \bibnamefont
  {Lim}}, \bibinfo {author} {\bibfnamefont {N.}~\bibnamefont {Gisin}},\ and\
  \bibinfo {author} {\bibfnamefont {R.}~\bibnamefont {Renner}},\ }\bibfield
  {title} {\bibinfo {title} {Tight finite-key analysis for quantum
  cryptography},\ }\href {https://doi.org/10.1038/ncomms1631} {\bibfield
  {journal} {\bibinfo  {journal} {Nat. Commun.}\ }\textbf {\bibinfo {volume}
  {3}},\ \bibinfo {pages} {634} (\bibinfo {year} {2012})}\BibitemShut {NoStop}%
\bibitem [{\citenamefont {M\"uller-Quade}\ and\ \citenamefont
  {Renner}(2009)}]{Müller-Quade_2009}%
  \BibitemOpen
  \bibfield  {author} {\bibinfo {author} {\bibfnamefont {J.}~\bibnamefont
  {M\"uller-Quade}}\ and\ \bibinfo {author} {\bibfnamefont {R.}~\bibnamefont
  {Renner}},\ }\bibfield  {title} {\bibinfo {title} {Composability in quantum
  cryptography},\ }\href {https://doi.org/10.1088/1367-2630/11/8/085006}
  {\bibfield  {journal} {\bibinfo  {journal} {New J. Phys.}\ }\textbf {\bibinfo
  {volume} {11}},\ \bibinfo {pages} {085006} (\bibinfo {year}
  {2009})}\BibitemShut {NoStop}%
\bibitem [{\citenamefont {Wang}\ \emph {et~al.}(2016)\citenamefont {Wang},
  \citenamefont {Bao}, \citenamefont {Zhou}, \citenamefont {Jiang},\ and\
  \citenamefont {Li}}]{wang_tight_2016}%
  \BibitemOpen
  \bibfield  {author} {\bibinfo {author} {\bibfnamefont {Y.}~\bibnamefont
  {Wang}}, \bibinfo {author} {\bibfnamefont {W.-S.}\ \bibnamefont {Bao}},
  \bibinfo {author} {\bibfnamefont {C.}~\bibnamefont {Zhou}}, \bibinfo {author}
  {\bibfnamefont {M.-S.}\ \bibnamefont {Jiang}},\ and\ \bibinfo {author}
  {\bibfnamefont {H.-W.}\ \bibnamefont {Li}},\ }\bibfield  {title} {\bibinfo
  {title} {Tight finite-key analysis of a practical decoy-state quantum key
  distribution with unstable sources},\ }\href
  {https://doi.org/10.1103/PhysRevA.94.032335} {\bibfield  {journal} {\bibinfo
  {journal} {Phys. Rev. A}\ }\textbf {\bibinfo {volume} {94}},\ \bibinfo
  {pages} {032335} (\bibinfo {year} {2016})}\BibitemShut {NoStop}%
\bibitem [{\citenamefont {Wang}\ \emph
  {et~al.}(2019{\natexlab{b}})\citenamefont {Wang}, \citenamefont {Bao},
  \citenamefont {Zhou}, \citenamefont {Jiang},\ and\ \citenamefont
  {Li}}]{wang_finite_2019}%
  \BibitemOpen
  \bibfield  {author} {\bibinfo {author} {\bibfnamefont {Y.}~\bibnamefont
  {Wang}}, \bibinfo {author} {\bibfnamefont {W.-S.}\ \bibnamefont {Bao}},
  \bibinfo {author} {\bibfnamefont {C.}~\bibnamefont {Zhou}}, \bibinfo {author}
  {\bibfnamefont {M.-S.}\ \bibnamefont {Jiang}},\ and\ \bibinfo {author}
  {\bibfnamefont {H.-W.}\ \bibnamefont {Li}},\ }\bibfield  {title} {\bibinfo
  {title} {Finite-key analysis of practical decoy-state
  measurement-device-independent quantum key distribution with unstable
  sources},\ }\href {https://doi.org/10.1364/JOSAB.36.000B83} {\bibfield
  {journal} {\bibinfo  {journal} {J. Opt. Soc. Am. B}\ }\textbf {\bibinfo
  {volume} {36}},\ \bibinfo {pages} {B83} (\bibinfo {year}
  {2019}{\natexlab{b}})}\BibitemShut {NoStop}%
\bibitem [{\citenamefont {Renner}(2008)}]{renner_security_2008}%
  \BibitemOpen
  \bibfield  {author} {\bibinfo {author} {\bibfnamefont {R.}~\bibnamefont
  {Renner}},\ }\bibfield  {title} {\bibinfo {title} {Security of quantum key
  distribution},\ }\href {https://doi.org/10.1142/S0219749908003256} {\bibfield
   {journal} {\bibinfo  {journal} {Int. J. Quantum Inf.}\ }\textbf {\bibinfo
  {volume} {06}},\ \bibinfo {pages} {1} (\bibinfo {year} {2008})}\BibitemShut
  {NoStop}%
\bibitem [{\citenamefont {Kato}(2020)}]{kato_concentration_2020}%
  \BibitemOpen
  \bibfield  {author} {\bibinfo {author} {\bibfnamefont {G.}~\bibnamefont
  {Kato}},\ }\href@noop {} {\bibinfo {title} {Concentration inequality using
  unconfirmed knowledge}} (\bibinfo {year} {2020}),\ \Eprint
  {https://arxiv.org/abs/2002.04357} {arXiv:2002.04357} \BibitemShut
  {NoStop}%
\bibitem [{\citenamefont {Azuma}(1967)}]{azuma_weighted_1967}%
  \BibitemOpen
  \bibfield  {author} {\bibinfo {author} {\bibfnamefont {K.}~\bibnamefont
  {Azuma}},\ }\bibfield  {title} {\bibinfo {title} {Weighted sums of certain
  dependent random variables},\ }\href
  {https://projecteuclid.org/journals/tohoku-mathematical-journal/volume-19/issue-3/Weighted-sums-of-certain-dependent-random-variables/10.2748/tmj/1178243286.full}
  {\bibfield  {journal} {\bibinfo  {journal} {Tohoku Math. J.}\ }\textbf
  {\bibinfo {volume} {19}} (\bibinfo {year} {1967})}\BibitemShut {NoStop}%
\bibitem [{\citenamefont {Konig}\ \emph {et~al.}(2009)\citenamefont {Konig},
  \citenamefont {Renner},\ and\ \citenamefont
  {Schaffner}}]{konig_operational_2009}%
  \BibitemOpen
  \bibfield  {author} {\bibinfo {author} {\bibfnamefont {R.}~\bibnamefont
  {Konig}}, \bibinfo {author} {\bibfnamefont {R.}~\bibnamefont {Renner}},\ and\
  \bibinfo {author} {\bibfnamefont {C.}~\bibnamefont {Schaffner}},\ }\bibfield
  {title} {\bibinfo {title} {The operational meaning of min- and max-entropy},\ }\href {https://doi.org/10.1109/TIT.2009.2025545}
  {\bibfield  {journal} {\bibinfo  {journal} {IEEE Trans. Inf. Theory}\
  }\textbf {\bibinfo {volume} {55}},\ \bibinfo {pages} {4337} (\bibinfo {year}
  {2009})}\BibitemShut {NoStop}%
\bibitem [{\citenamefont {Renes}\ and\ \citenamefont
  {Renner}(2012)}]{renes_one-shot_2012}%
  \BibitemOpen
  \bibfield  {author} {\bibinfo {author} {\bibfnamefont {J.~M.}\ \bibnamefont
  {Renes}}\ and\ \bibinfo {author} {\bibfnamefont {R.}~\bibnamefont {Renner}},\
  }\bibfield  {title} {\bibinfo {title} {One-shot classical data compression with quantum side information and the distillation of common randomness or secret keys},\ }\href
  {https://doi.org/10.1109/TIT.2011.2177589} {\bibfield  {journal} {\bibinfo
  {journal} {IEEE Trans. Inf. Theory}\ }\textbf {\bibinfo {volume} {58}},\
  \bibinfo {pages} {1985} (\bibinfo {year} {2012})}\BibitemShut {NoStop}%
\bibitem [{\citenamefont {Yin}\ and\ \citenamefont
  {Chen}(2019)}]{yin_finite-key_2019}%
  \BibitemOpen
  \bibfield  {author} {\bibinfo {author} {\bibfnamefont {H.-L.}\ \bibnamefont
  {Yin}}\ and\ \bibinfo {author} {\bibfnamefont {Z.-B.}\ \bibnamefont {Chen}},\
  }\bibfield  {title} {\bibinfo {title} {Finite-key analysis for twin-field
  quantum key distribution with composable security},\ }\href
  {https://doi.org/10.1038/s41598-019-53435-4} {\bibfield  {journal} {\bibinfo
  {journal} {Sci. Rep.}\ }\textbf {\bibinfo {volume} {9}},\ \bibinfo {pages}
  {17113} (\bibinfo {year} {2019})}\BibitemShut {NoStop}%
\bibitem [{\citenamefont {Curr\'as-Lorenzo}\ \emph {et~al.}(2021)\citenamefont
  {Curr\'as-Lorenzo}, \citenamefont {Navarrete}, \citenamefont {Azuma},
  \citenamefont {Kato}, \citenamefont {Curty},\ and\ \citenamefont
  {Razavi}}]{curras-lorenzo_tight_2021}%
  \BibitemOpen
  \bibfield  {author} {\bibinfo {author} {\bibfnamefont {G.}~\bibnamefont
  {Curr\'as-Lorenzo}}, \bibinfo {author} {\bibfnamefont {{\'A}.}~\bibnamefont
  {Navarrete}}, \bibinfo {author} {\bibfnamefont {K.}~\bibnamefont {Azuma}},
  \bibinfo {author} {\bibfnamefont {G.}~\bibnamefont {Kato}}, \bibinfo {author}
  {\bibfnamefont {M.}~\bibnamefont {Curty}},\ and\ \bibinfo {author}
  {\bibfnamefont {M.}~\bibnamefont {Razavi}},\ }\bibfield  {title} {\bibinfo
  {title} {Tight finite-key security for twin-field quantum key distribution},\
  }\href {https://doi.org/10.1038/s41534-020-00345-3} {\bibfield  {journal}
  {\bibinfo  {journal} {npj Quantum Inf.}\ }\textbf {\bibinfo {volume} {7}},\
  \bibinfo {pages} {22} (\bibinfo {year} {2021})}\BibitemShut {NoStop}%
\end{thebibliography}

%

\end{document}